\documentclass[twocolumn]{aastex631}

\usepackage{CJK}
\usepackage{multirow}
\usepackage{makecell}
\usepackage{booktabs}
\usepackage{graphbox}
\usepackage{amsmath}
\usepackage{ulem}

\newcommand{\getsf}{\texttt{getsf}}
\newcommand{\unpbcor}{\texttt{unpbcor}}
\newcommand{\pbcor}{\texttt{pbcor}}

\newcommand{\massrate}{$M_{\odot}$\,yr$^{-1}$}

\newcommand{\hii}{H\textsc{ii}}
\newcommand{\msun}{$ M_\odot$}
\newcommand{\lsun}{$ L_\odot$}
\newcommand{\kms}{km\,s$^{-1}$}
\newcommand{\jybeam}{Jy\,beam$^{-1}$}
\newcommand{\mjybeam}{mJy\,beam$^{-1}$}

\newcommand{\degree}{$^{\circ}$}
\newcommand{\parcsec}{\mbox{$.\!\!\arcsec$}}
\newcommand{\ssstyle}{\scriptscriptstyle}

\newcommand{\htcn}{H$^{13}$CN}

\newcommand{\chtoh}{CH$_3$OH}

\newcommand{\htcs}{H$_2$CS}

\newcommand{\chtocho}{CH$_3$OCHO}

\shorttitle{Dynamic View of Protocluster Evolution}
\shortauthors{Xu et al.}

\begin{document}

\title{The ALMA Survey of Star Formation and Evolution in Massive Protoclusters with Blue Profiles (ASSEMBLE): Core Growth, Cluster Contraction, and Primordial Mass Segregation}

\correspondingauthor{Ke Wang}
\email{kwang.astro@pku.edu.cn}

\author[0000-0001-5950-1932]{Fengwei Xu}
\affiliation{Kavli Institute for Astronomy and Astrophysics, Peking University, Beijing 100871, People's Republic of China}
\affiliation{Department of Astronomy, School of Physics, Peking University, Beijing, 100871, People's Republic of China}

\author[0000-0002-7237-3856]{Ke Wang}
\affiliation{Kavli Institute for Astronomy and Astrophysics, Peking University, Beijing 100871, People's Republic of China}

\author[0000-0002-5286-2564]{Tie Liu}
\affiliation{Shanghai Astronomical Observatory, Chinese Academy of Sciences, 80 Nandan Road, Shanghai 200030, People's Republic of China}

\author[0000-0001-9160-2944]{Mengyao Tang}
\affiliation{Institute of Astrophysics, School of Physics and Electronic Science, Chuxiong Normal University, Chuxiong 675000, People's Republic of China}


\author[0000-0001-5175-1777]{Neal J. Evans II}
\affiliation{Department of Astronomy, The University of Texas at Austin, 2515 Speedway, Stop C1400, Austin, TX 78712-1205, USA}

\author[0000-0002-9569-9234]{Aina Palau}
\affiliation{Instituto de Radioastronom\'ia y Astrof\'isica, Universidad Nacional Aut\'onoma de M\'exico, Antigua Carretera a P\'atzcuaro 8701, Ex-Hda. San Jos\'e de la Huerta, 58089 Morelia, Michoac\'an, M\'exico}

\author[0000-0002-6752-6061]{Kaho Morii}
\affiliation{Department of Astronomy, Graduate School of Science, The University of Tokyo, 7-3-1 Hongo, Bunkyo-ku, Tokyo 113-0033, Japan}
\affiliation{National Astronomical Observatory of Japan, National Institutes of Natural Sciences, 2-21-1 Osawa, Mitaka, Tokyo 181-8588, Japan}

\author[0000-0002-3938-4393]{Jinhua He}
\affiliation{Yunnan Observatories, Chinese Academy of Sciences, 396 Yangfangwang,
Guandu District, Kunming, 650216, People's Republic of China}
\affiliation{Chinese Academy of Sciences South AmericaCenterfor Astronomy,National
Astronomical Observatories, CAS, Beijing 100101, People's Republic of China}
\affiliation{Departamento de Astronom\'{\i}a, Universidad de Chile, Las Condes, 7591245 Santiago, Chile}

\author[0000-0002-7125-7685]{Patricio Sanhueza}
\affiliation{National Astronomical Observatory of Japan, National Institutes of Natural Sciences, 2-21-1 Osawa, Mitaka, Tokyo 181-8588, Japan}
\affiliation{Astronomical Science Program, The Graduate University for Advanced Studies, SOKENDAI, 2-21-1 Osawa, Mitaka, Tokyo 181-8588, Japan}

\author[0000-0003-3343-9645]{Hong-Li Liu}
\affiliation{Department of Astronomy, Yunnan University, Kunming 650091, People's Republic of China}

\author[0000-0003-2300-8200]{Amelia Stutz}
\affiliation{Departamento de Astronom\'ia, Universidad de Concepci\'on, Casilla 160-C, Concepci\'on, Chile}

\author[0000-0003-2384-6589]{Qizhou Zhang}
\affiliation{Center for Astrophysics $|$ Harvard \& Smithsonian, 60 Garden Street, Cambridge, MA 02138, USA}

\author[0000-0002-5435-925X]{Xi Chen}
\affiliation{Center for Astrophysics, Guangzhou University, Guangzhou 510006, People's Republic of China}
\affiliation{Shanghai Astronomical Observatory, Chinese Academy of Sciences, 80 Nandan Road, Shanghai 200030, People's Republic of China}
\affiliation{Peng Cheng Lab, Shenzhen, 518066, People's Republic of China}

\author[0000-0001-8077-7095]{Pak Shing Li}
\affiliation{Shanghai Astronomical Observatory, Chinese Academy of Sciences, 80 Nandan Road, Shanghai 200030, People's Republic of China}

\author[0000-0003-4714-0636]{Gilberto C. Gómez}
\affiliation{Instituto de Radioastronom\'ia y Astrof\'isica, Universidad Nacional Aut\'onoma de M\'exico, Antigua Carretera a P\'atzcuaro 8701, Ex-Hda. San Jos\'e de la Huerta, 58089 Morelia, Michoac\'an, M\'exico}

\author[0000-0002-1424-3543]{Enrique Vázquez-Semadeni}
\affiliation{Instituto de Radioastronom\'ia y Astrof\'isica, Universidad Nacional Aut\'onoma de M\'exico, Antigua Carretera a P\'atzcuaro 8701, Ex-Hda. San Jos\'e de la Huerta, 58089 Morelia, Michoac\'an, M\'exico}

\author[0000-0003-1275-5251]{Shanghuo Li}
\affiliation{Max Planck Institute for Astronomy, Königstuhl 17, D-69117 Heidelberg, Germany}

\author[0000-0001-7573-0145]{Xiaofeng Mai}
\affiliation{Shanghai Astronomical Observatory, Chinese Academy of Sciences, 80 Nandan Road, Shanghai 200030, People's Republic of China}

\author[0000-0003-2619-9305]{Xing Lu}
\affiliation{Shanghai Astronomical Observatory, Chinese Academy of Sciences, 80 Nandan Road, Shanghai 200030, People's Republic of China}

\author[0000-0002-5789-7504]{Meizhu Liu}
\affiliation{Department of Astronomy, Yunnan University, Kunming 650091, People's Republic of China}

\author[0009-0009-8154-4205]{Li Chen}
\affiliation{Department of Astronomy, Yunnan University, Kunming 650091, People's Republic of China}

\author[0000-0001-5710-6509]{Chuanshou Li}
\affiliation{Department of Astronomy, Yunnan University, Kunming 650091, People's Republic of China}

\author[0000-0001-8277-1367]{Hongqiong Shi}
\affiliation{Department of Astronomy, Yunnan University, Kunming 650091, People's Republic of China}

\author[0000-0003-4659-1742]{Zhiyuan Ren}
\affiliation{National Astronomical Observatories, Chinese Academy of Sciences, Beijing 100101, People's Republic of China}

\author[0000-0003-3010-7661]{Di Li}
\affiliation{National Astronomical Observatories, CAS, 20A, Datun Road, Chaoyang District, Beijing, 100101, People's Republic of China}
\affiliation{Research Center for Intelligent Computing Platforms, Zhejiang Laboratory, Hangzhou 311100, People's Republic of China}
\affiliation{NAOC-UKZN Computational Astrophysics Centre (NUCAC), University of KwaZulu-Natal, Durban, 4000, South Africa}

\author[0000-0003-1649-7958]{Guido Garay}
\affiliation{Departamento de Astronom\'{\i}a, Universidad de Chile, Las Condes, 7591245 Santiago, Chile}

\author[0000-0002-9574-8454]{Leonardo Bronfman}
\affiliation{Departamento de Astronom\'{\i}a, Universidad de Chile, Las Condes, 7591245 Santiago, Chile}

\author[0000-0001-6725-0483]{Lokesh Dewangan}
\affiliation{Physical Research Laboratory, Navrangpura, Ahmedabad-380 009, India}

\author[0000-0002-5809-4834]{Mika Juvela}
\affiliation{Department of Physics, University of Helsinki, PO Box 64, FI-00014 Helsinki, Finland}

\author[0000-0002-3179-6334]{Chang Won Lee}
\affiliation{Korea Astronomy and Space Science Institute, 776 Daedeok-daero, Yuseong-gu, Daejeon 34055, Republic of Korea}
\affiliation{University of Science and Technology, 217 Gajeong-ro, Yuseong-gu, Daejeon 34113, Republic of Korea}

\author[0000-0002-9836-0279]{S. Zhang}
\affiliation{Kavli Institute for Astronomy and Astrophysics, Peking University, Beijing 100871, People's Republic of China}

\author[0000-0003-0355-6875]{Nannan Yue}
\affiliation{Kavli Institute for Astronomy and Astrophysics, Peking University, Beijing 100871, People's Republic of China}

\author{Chao Wang}
\affiliation{Kavli Institute for Astronomy and Astrophysics, Peking University, Beijing 100871, People's Republic of China}
\affiliation{Department of Astronomy, School of Physics, Peking University, Beijing, 100871, People's Republic of China}

\author[0000-0002-8727-0868]{Yifei Ge}
\affiliation{Kavli Institute for Astronomy and Astrophysics, Peking University, Beijing 100871, People's Republic of China}
\affiliation{Department of Astronomy, School of Physics, Peking University, Beijing, 100871, People's Republic of China}

\author[0000-0001-9822-7817]{Wenyu Jiao}
\affiliation{Kavli Institute for Astronomy and Astrophysics, Peking University, Beijing 100871, People's Republic of China}
\affiliation{Department of Astronomy, School of Physics, Peking University, Beijing, 100871, People's Republic of China}
\affiliation{Max Planck Institute for Astronomy, Königstuhl 17, D-69117 Heidelberg, Germany}

\author[0000-0003-4506-3171]{Qiuyi Luo}
\affiliation{Shanghai Astronomical Observatory, Chinese Academy of Sciences, 80 Nandan Road, Shanghai 200030, People's Republic of China}

\author{J.-W. Zhou}
\affiliation{Max-Planck-Institut für Radioastronomie, Auf dem Hügel 69, 53121 Bonn, Germany}

\author[0000-0002-8149-8546]{Ken'ichi Tatematsu}
\affiliation{National Astronomical Observatory of Japan, National Institutes of Natural Sciences, 2-21-1 Osawa, Mitaka, Tokyo 181-8588, Japan}
\affiliation{Astronomical Science Program, The Graduate University for Advanced Studies, SOKENDAI, 2-21-1 Osawa, Mitaka, Tokyo 181-8588, Japan}

\author[0000-0002-9875-7436]{James O. Chibueze}
\affiliation{Department of Mathematical Sciences, University of South Africa, Cnr Christian de Wet Rd and Pioneer Avenue, Florida Park, 1709, Roodepoort, South Africa}
\affiliation{Centre for Space Research, Physics Department, North-West University, Potchefstroom 2520, South Africa}
\affiliation{Department of Physics and Astronomy, Faculty of Physical Sciences, University of Nigeria, Carver Building, 1 University Road, Nsukka 410001, Nigeria}

\author[0009-0003-2243-7983]{Keyun Su}
\affiliation{Kavli Institute for Astronomy and Astrophysics, Peking University, Beijing 100871, People's Republic of China}

\author[0000-0002-9796-1507]{Shenglan Sun}
\affiliation{Kavli Institute for Astronomy and Astrophysics, Peking University, Beijing 100871, People's Republic of China}

\author[0000-0002-1469-6323]{I. Ristorcelli}
\affiliation{Univ. Toulouse, CNRS, IRAP, 9 Av. du colonel Roche, BP 44346, 31028, Toulouse, France}

\author[0000-0002-5310-4212]{L. Viktor Toth}
\affiliation{Eötvös University Budapest, Pazmany P. s. 1/A, H-1117 Budapest, Hungary}
\affiliation{University of Debrecen, Faculty of Science and Technology, Egyetem tér 1, H-4032 Debrecen, Hungary}




\begin{abstract}

The ALMA Survey of Star Formation and Evolution in Massive Protoclusters with Blue Profiles (ASSEMBLE) aims to investigate the process of mass assembly and its connection to high-mass star formation theories in protoclusters in a dynamic view. We observed 11 massive ($M_{\rm clump}\gtrsim10^3$\,\msun), luminous ($L_{\rm bol}\gtrsim10^4$\,\lsun), and blue-profile (infall signature) clumps by ALMA with resolution of $\sim$2200--5500\,au (median value of 3500\,au) at 350\,GHz (870\,$\mu$m). 248 dense cores were identified, including 106 cores showing protostellar signatures and 142 prestellar core candidates. Compared to early-stage infrared dark clouds (IRDCs) by ASHES, the core mass and surface density within the ASSEMBLE clumps exhibited significant increment, suggesting concurrent core accretion during the evolution of the clumps. The maximum mass of prestellar cores was found to be 2 times larger than that in IRDCs, indicating that evolved protoclusters have the potential to harbor massive prestellar cores. The mass relation between clumps and their most massive core (MMCs) is observed in ASSEMBLE but not in IRDCs, which is suggested to be regulated by multiscale mass accretion. The mass correlation between the core clusters and their MMCs has a steeper slope compared to that observed in stellar clusters, which can be due to fragmentation of the MMC and stellar multiplicity. We observe a decrease in core separation and an increase in central concentration as protoclusters evolve. We confirm primordial mass segregation in the ASSEMBLE protoclusters, possibly resulting from gravitational concentration and/or gas accretion.

\end{abstract}



\section{Introduction} \label{sec:intro}

Observations suggest that massive stars form either in bound clusters \citep{Lada2003Cluster,Longmore2011G8,Longmore2014YMC,Motte2018Review} or in large-scale hierarchically structured associations \citep{Ward2018Nocluster}. However, the process of stellar mass assembly, which includes fragmentation and accretion, remains poorly understood. This is a critical step in determining important parameters such as the number of massive stars and their final stellar mass. Also, it is important to note that the fragmentation of molecular gas and core accretion are both time dependent, as the instantaneous physical conditions in the cloud vary during ongoing star formation and feedback (e.g., radiation and outflow). So, it is challenging to pinpoint the physical conditions that give rise to the fragmentation and accretion observed at present. 

Over the past decades, researchers have focused on massive clumps associated with infrared dark clouds (IRDCs), which are believed to harbor the earliest stage of massive star and cluster formation \citep[e.g.][]{Rathborne2006IRDCs,Rathborne2007IRDCs,Chambers2009IRDCs,Zhang2009Fragmentation, Zhang2011IRDC30, Wang2011G28, Sanhueza2012IRDCs, Wang2014Snake, Zhang2015G28, Yuan2017HMSC,Pillai2019,HuangBo2023}. Despite their large reservoir of molecular gas at high densities $>10^4$\,cm$^{-3}$, IRDC clumps show few signs of star formation. For example, only 12\% in a sample of 140 IRDCs has water masers \citep{Wang2006H2Omaser}. Moreover, IRDCs have consistently lower gas temperatures and line widths, with studies in NH$_3$ finding temperatures of $\lesssim15$\,K \citep{Pillai2006Ammonia,  Ragan2011IRDCs,Wang2012Heating,Wang2014Snake,Xie2021Temperature} and linewidths of $\lesssim2$\,\kms~averaged over a spatial scale of 1\,pc \citep{Wang2008G28Ammonia,Ragan2011IRDCs,Ragan2012IRDCs}. Both two parameters are lower than those observed in high-mass protostellar objects (HMPOs) with temperature of $\sim20$\,K and linewidths of $\sim2$\,\kms~\citep{Molinari1996preUCHII,Sridharan2002HMPO,Wu2006HMPO,Longmore2007HMC,Urquhart2011MSX}, and those observed in ultra-compact \hii~(UC\hii) regions with $>25$\,K and $\gtrsim3$\,\kms~\citep{Churchwell1990Ammonia,Harju1993Ammonia,Molinari1996preUCHII,Sridharan2002HMPO}. Therefore, there is a clear evolutionary sequence from IRDCs to HMPOs and then to UC\hii~regions, which sets the basis for a time-dependent study of massive star formation. 

Taking advantage of the low contamination from stellar feedback in IRDCs, great efforts have been made to investigate the initial conditions of massive star formation therein. For example, \citet{Zhang2009Fragmentation} first conducted arcsec resolution studies of the IRDC G28.34+0.06 with the Submillimetre Array (SMA) and found that dense cores giving rise to massive stars are much more massive than the thermal Jeans mass of the clump. This discovery challenges the notion in the ``competitive accretion'' model that massive stars should arise from cores of thermal Jeans mass \citep{Bonnell2001Simulation}. The larger core mass in the fragments demands either additional support from turbulence and magnetic fields \citep{Wang2012Heating} or a continuous accretion onto the core \citep{VS2023Simulation}. On the other hand, observations also find that the mass of these cores does not contain sufficient material to form a massive star \citep{Sanhueza2017G28,Sanhueza2019ASHES,Morii2023ASHES}, and the cores typically continue to fragment when observed at higher angular resolution \citep{Wang2011G28,Wang2014Snake,Zhang2015G28,Olguin2021DIHCA-I,Olguin2022DIHCA-II}, or at slightly later evolutionary stages \citep[e.g.,][]{Palau2015Fragmentation,Beuther2018Fragmentation}. Therefore, the idea of monolithic collapse \citep{Mckee2003TurbulentCore} for massive star formation does not match the observations. On the simulation side, recent work by \citet{Pelkonen2021MNRAS} has shed light on the inadequacies of both core collapse and competitive accretion scenarios. Their findings reveal a lack of a direct correlation between the progenitor core mass and the final stellar mass for individual stars, as well as a lack of an increase in accretion rate with core mass. 

However, \citet{Padoan2020ApJ} suggested a scenario where massive stars are assembled by large-scale, converging, inertial flows that naturally occur in supersonic turbulence. Very recently, \citet{He2023Simulation} performed high resolution up to $\sim7$\,au and found that gas should be continuously supplied from larger scales beyond the mass reservoir of the core. Such a continuous mass accretion is observed directly \citep{Dewangan2022W42,Redaelli2022G14,Xu2023SDC335} or indirectly \citep{Contreras2018Infall}. More recent observations of IRDCs with the Atacama Large Millimeter/submillimeter Array (ALMA) routinely reach a mass sensitivity far below the thermal Jeans mass and detect a large population of low-mass cores in the clumps that are compatible with the thermal Jeans mass \citep{Svoboda2019HMSC,Sanhueza2019ASHES,Morii2023ASHES}. These cores may form low-mass stars in a cluster. To summarize, these observations point to a picture of massive star formation in which dense cores continue to gain material from the parental molecular clump, while the embedded protostar undergoes accretion \citep[see review in Section 1.1 in][]{Xu2023SDC335}. 
 
Mass assembly is a dynamic process that occurs over time after all, and it is essential to compare the predictions of theoretical models and numerical simulations with observations of massive clumps at a broad range of evolutionary stages to understand high-mass star and cluster formation. While the state-of-the-art understanding of massive star formation suggests gas transfers along filamentary structures to feed the massive dense cores where protostars grow in mass \citep{Gomez2014Simulation,Motte2018Review,NR2022Simulation,Xu2023SDC335}, observational evidence is required to provide more straightforward constraints on the physical processes during protocluster evolution, which will yield a time-tracked understanding of high-mass star and cluster formation. 

Therefore, we conduct the ALMA Survey of Star Formation and Evolution in Massive Protoclusters with Blue Profiles (ASSEMBLE), designed to study mass assembly systematically, including fragmentation and accretion, and its connection to high-mass star formation theories. The survey aims at providing a ``dynamic'' view from two main perspectives: 1) answering a series of kinematics questions such as when infall starts and stops, how gas transfers inwards, and where infalling gas goes; 2) and unveiling the evolution of key physical parameters in the protoclusters since the sample in the survey provides more evolved protoclusters compared to early-stage IRDCs. The first idea is reflected in our sample selection that all the 11 massive clumps are chosen from pilot single-dish surveys with evident blue profiles indicating global infall motions and rapid mass assembly. The sample also benefits from synergy with ALMA Three-millimeter Observations of Massive Star-forming regions \citep[ATOMS;][]{Liu2020ATOMS-I}, supporting gas kinematics analyses \citep{Xu2023SDC335}. The second idea is to compare the ASSEMBLE results with those in early-stage IRDCs reported by \citet{Sanhueza2019ASHES,Morii2023ASHES} as well as \citet{Svoboda2019HMSC}. \citet{Sanhueza2019ASHES,Morii2023ASHES} are both included in series work ``The ALMA Survey of 70\,$\mu$m Dark High-mass Clumps in Early Stages'' (ASHES hereafter), which focus on a pilot sample of 12 \citep[ASHES Pilots;][]{Sanhueza2019ASHES} and a total sample of 39 \citep[ASHES Totals;][]{Morii2023ASHES} of carefully chosen IRDCs, respectively. The mean temperature of these IRDCs is $\sim$15\,K, with a range of 9 to 23\,K, and the luminosity-to-mass ratio ranges from 0.1 to 1\,$L_\odot/M_\odot$, supporting the idea that these clumps host the early stages of massive star formation \citep{Morii2023ASHES}. 

In this paper, we present comprehensive analyses of dust continuum emission from a carefully selected sample comprising 11 massive protocluster clumps that exhibit evidence of gas infall. Our study focuses on investigating the physical properties and evolution of cores within these clumps, including their mass, spatial distribution, and comparison with earlier stages. The paper is structured as follows. Section\,\ref{sec:sample} describes the criteria used for the selection of our sample. Section\,\ref{sec:observations} provides a summary of the observation setups and details the data reduction process. In Section\,\ref{sec:result}, we present the fundamental results derived from the ASSEMBLE data.
Section\,\ref{sec:discuss} offers in-depth discussions on the implications and significance of the observed results. To gain further insights into protocluster evolution, Section\,\ref{sec:evolution} presents comparative analyses with the ASHES data and contributes to the development of a comprehensive understanding of the protocluster evolution. Finally, in Section\,\ref{sec:conclusion}, we summarize the key findings and provide future prospects.

\section{Sample Selection} \label{sec:sample}


\begin{deluxetable*}{cccccccccccccc}
\rotate
\tabletypesize{\small}
\tablewidth{0pt}
\linespread{1.2}
\tablecaption{Physical Properties of the ASSEMBLE Clumps \label{tab:sample}}
\tablehead{
\multicolumn{2}{c}{ASSEMBLE Clump} & \multicolumn{2}{c}{Position} & \colhead{V$_{\rm lsr}$} & \colhead{$\delta V$} & \colhead{Dist.} & \colhead{$R_{\rm cl}$\tablenotemark{c}} & \colhead{$T_{\rm dust}$} & \colhead{$\log(M_{\rm cl})$} & \colhead{$\log(L_{\rm bol})$} & \colhead{$L/M$} & \colhead{$\Sigma_{\rm cl}$} & \multirow{2}{*}{\makecell[c]{Color \\ ID}} \\
\cmidrule(r){1-2} \cmidrule(r){3-4}
\colhead{IRAS\tablenotemark{a}} & \colhead{AGAL\tablenotemark{b}} & \colhead{$\alpha$(J2000)} & \colhead{$\delta$(J2000)} & \colhead{(\kms)} & \colhead{} & \colhead{(kpc)} & \colhead{(pc [\arcsec])} & \colhead{(K)} & \colhead{(\msun)} & \colhead{(\lsun)} & \colhead{\lsun/\msun} & \colhead{(g\,cm$^{-2}$)} & \colhead{}
}
\colnumbers
\startdata
I14382-6017 & G316.139-00.506 & 14:42:02.76 & -60:30:35.1 & -60.17 & -0.60 & $3.62_{-0.57}^{+0.68}$ & 0.49(28) & 28.0 & 3.65 & 5.20 & 35 & 1.21 & \includegraphics[scale=0.08]{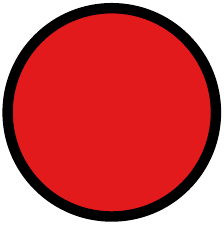} \\
I14498-5856 & G318.049+00.086 & 14:53:42.81 & -59:08:56.5 & -49.82 & -0.66 & $2.90_{-0.47}^{+0.52}$ & 0.23(16) & 26.7 & 3.01 & 4.43 & 26 & 1.29 & \includegraphics[scale=0.08]{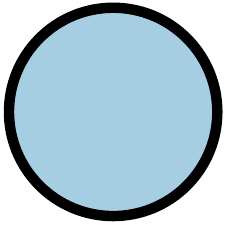} \\
I15520-5234 & G328.809+00.632 & 15:55:48.84 & -52:43:06.2 & -41.67 & -0.46 & $2.50_{-0.40}^{+0.38}$ & 0.16(13) & 32.2 & 3.23 & 5.13 & 79 & 4.17 & \includegraphics[scale=0.08]{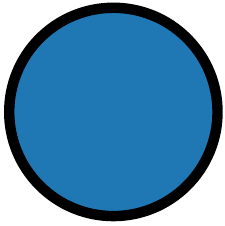} \\
I15596-5301 & G329.406-00.459 & 16:03:32.29 & -53:09:28.1 & -74.20 & -0.61 & $4.21_{-0.40}^{+0.41}$ & 0.36(17) & 28.5 & 3.93 & 5.47 & 34 & 4.40 & \includegraphics[scale=0.08]{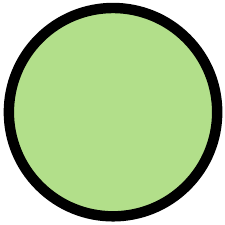} \\
I16060-5146 & G330.954-00.182 & 16:09:52.85 & -51:54:54.7 & -90.44 & -0.56 & $5.08_{-0.44}^{+0.54}$ & 0.30(12) & 32.2 & 3.95 & 5.78 & 67 & 6.50 & \includegraphics[scale=0.08]{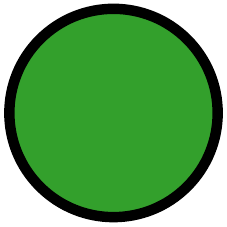}\\
I16071-5142 & G331.133-00.244 & 16:11:00.01 & -51:50:21.6 & -86.80 & -0.47 & $4.84_{-0.35}^{+0.55}$ & 0.41(17) & 23.9 & 3.68 & 4.78 & 12 & 1.87 & \includegraphics[scale=0.08]{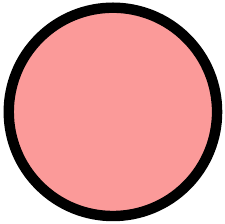} \\
I16076-5134 & G331.279-00.189 & 16:11:27.12 & -51:41:56.9 & -88.37 & -0.75 & $4.99_{-0.45}^{+0.45}$ & 0.52(21) & 30.1 & 3.57 & 5.27 & 49 & 0.92 & \includegraphics[scale=0.08]{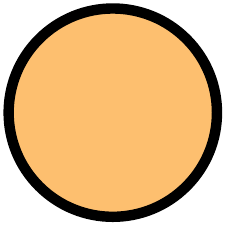} \\
I16272-4837 & G335.586-00.291 & 16:30:59.08 & -48:43:53.3 & -46.57 & -0.40 & $2.96_{-0.30}^{+0.41}$ & 0.22(15) & 23.1 & 3.24 & 4.29 & 11 & 2.34 & \includegraphics[scale=0.08]{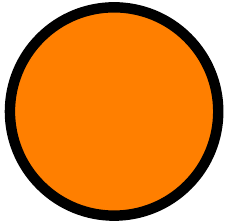} \\
I16351-4722 & G337.406-00.402 & 16:38:50.98 & -47:27:57.8 & -40.21 & -0.76 & $2.80_{-0.40}^{+0.35}$ & 0.21(15) & 30.4 & 3.22 & 4.88 & 46 & 2.57 & \includegraphics[scale=0.08]{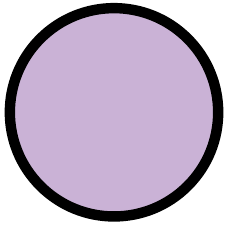} \\
I17204-3636 & G351.041-00.336 & 17:23:50.32 & -36:38:58.1 & -17.74 & -0.83 & $2.76_{-0.69}^{+0.60}$ & 0.26(19) & 25.8 & 2.88 & 4.17 & 19 & 0.77 & \includegraphics[scale=0.08]{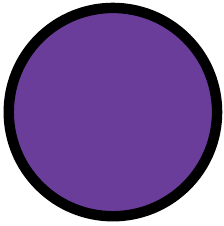} \\
I17220-3609 & G351.581-00.352 & 17:25:24.99 & -36:12:45.1 & -94.94 & -0.77 & $7.54_{-0.18}^{+0.31}$ & 0.79(21) & 25.4 & 4.35 & 5.66 & 20 & 2.38 & \includegraphics[scale=0.08]{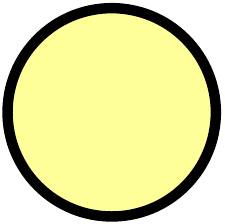} \\
\enddata
\tablecomments{The IRAS and ATLASGAL names of the ASSEMBLE clumps are listed in (1)--(2). Field centers of the ALMA mosaic observations are listed in (3)--(4), which are not always consistent with the continuum peak but cover most of the emission of the clump. Kinematic properties in (5)--(6) are derived from the CO\,(4-3) and C$^{17}$O\,(3-2) lines in APEX observations \citep{Yue2021APEX}. The distance and its uncertainties are estimated using the latest rotation curve model of the Milky Way \citep{Reid2019Distance} and listed in (7). Clump radius ($R_{\rm cl}$) is derived from 2D Gaussian fitting, shown in (8). Clump properties including dust temperature ($T_{\rm dust}$), mass ($M_{\rm cl}$), bolometric luminosity ($L_{\rm cl}$), and luminosity-to-mass ratio ($L/M$) are retrieved from \citet{Urquhart2018ATLASGAL}, which are listed in (9)--(12). The clump surface density, in column (13), is calculated by $\Sigma_{\rm cl} = M_{\rm cl} / \pi R_{\rm cl}^2$. Identical color, in column (14), is used to distinguish in statistical plots hereafter.
}
\tablenotetext{a}{The ``IRAS'' is replaced by ``I'' in short.}
\tablenotetext{b}{The ATLASGAL name from \citet{Urquhart2018ATLASGAL}.}
\tablenotetext{c}{Effective clump radius by Source Extractor \citep{Urquhart2014CSC}.}
\end{deluxetable*}

\begin{figure*}[!ht]
\centering
\includegraphics[width=0.92\linewidth]{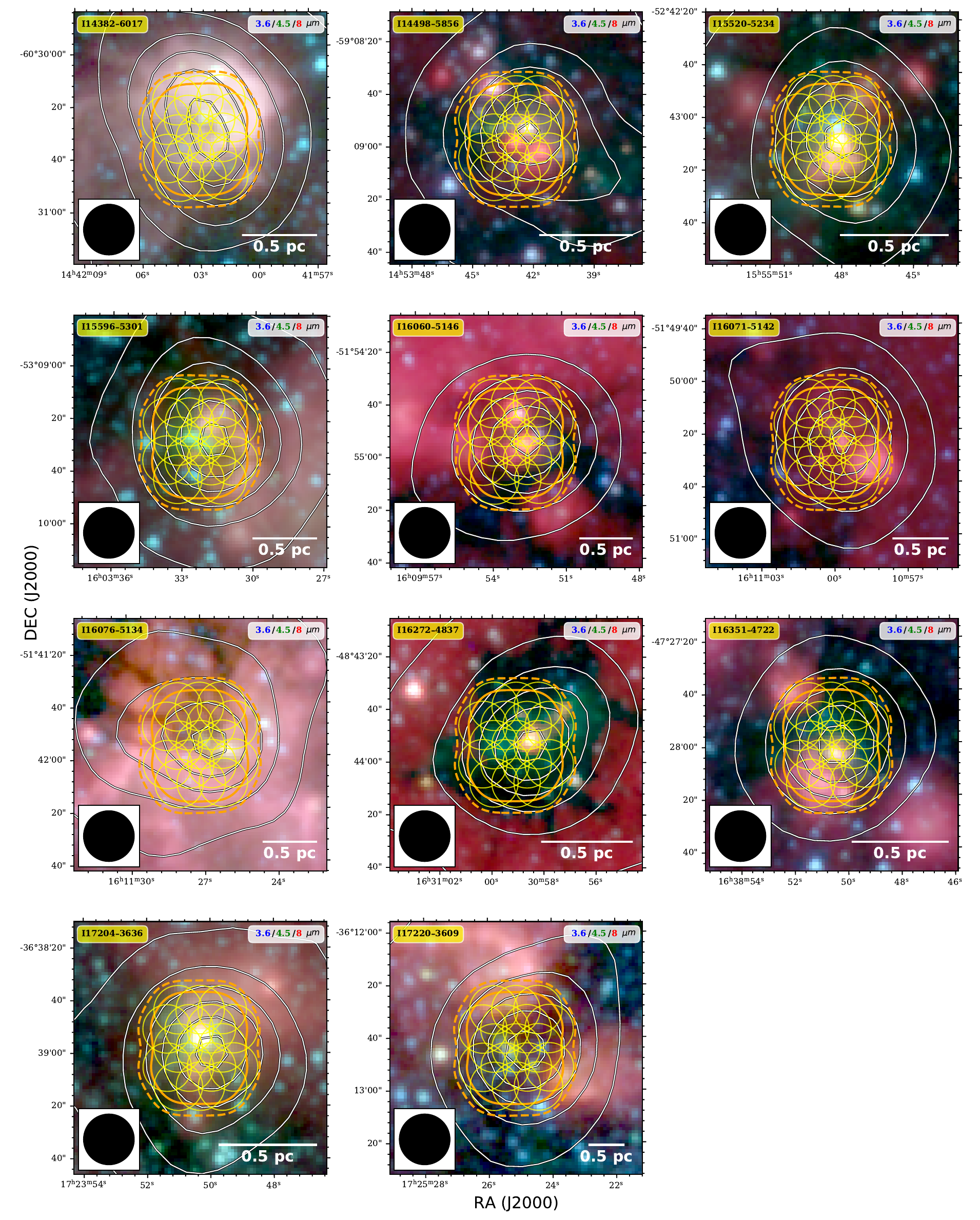}
\caption{Showcase of the ASSEMBLE sample. Background shows the \textit{Spitzer} infrared three-color map (blue: 3.6\,$\mu$m; green: 4.5\,$\mu$m; red: 8\,$\mu$m). White contours are ATLASGAL 870\,$\mu$m continuum emission, with levels starting from $5\sigma$ increasing in steps of $f(n)=3\times n^p + 2$ where $n=1,2,3,...N$. The beam of the ATLASGAL continuum map is 19\parcsec2 as shown on the bottom left. The ALMA mosaicked pointings are shown with yellow circles and the primary beam responses of 0.5 and 0.2 are outlined by orange solid and dashed lines respectively. The scale bar of 0.5\,pc is labelled in the bottom right corner. \label{fig:infrared}}
\end{figure*}

\subsection{Massive Clumps with Infall Motion} \label{sample:selection}

The ASSEMBLE sample, consisting of 11 carefully selected sources, owes its creation to advanced observational tools such as \textit{IRAS}, \textit{Spitzer}, and \textit{Herschel} satellites, as well as various ground-based surveys focusing on dust continuum and molecular lines. \citet{Bronfman1996CS} conducted a comprehensive and homogeneous CS\,(2-1) line survey of 1427 bright IRAS point sources in the Galactic plane candidates that were suspected to harbor UC\hii~regions. Subsequently, \citet{Faundez2004SIMBA} conducted a follow-up survey of 146 sources suspected of hosting high-mass star formation regions (bright CS, (2-1) emission of $T_b>2$, K, indicative of reasonably dense gas), using 1.2\,mm continuum emission. The same set of 146 high-mass star-forming clumps was then surveyed by \citet{Liu2016ASTE}, using HCN\,(4-3) and CS\,(7-6) lines with the 10-m Atacama Submillimeter Telescope Experiment (ASTE) telescope. With the most reliable tracer of infall motions HCN\,(4-3) lines \citep[Xu et al. in submission]{Chira2014Infall}, they identified 30 infall candidates based on the ``blue profiles''.

Out of the 30 infall candidates, 18 are further confirmed by HCN\,(3-2) and CO\,(4-3) lines observed with the Atacama Pathfinder Experiment (APEX) 12-m telescope \citep{Yue2021APEX}. Furthermore, the 18 sources were found to have virial parameters below 2, indicating that they're likely undergoing global collapse. All the 18 sources are covered by both the APEX Telescope Large Area Survey of the Galaxy \citep[ATLASGAL;][]{Urquhart2018ATLASGAL} and the \textit{Herschel} Infrared Galactic Plane Survey \citep[Hi-GAL;][]{Molinari2010HiGAL,Elia2017HiGAL,Elia2021HiGAL}, allowing well-constrained estimates of clump mass and luminosity from the infrared SED fitting \citep{Urquhart2018ATLASGAL}. Given ASSEMBLE's goal of investigating massive and luminous star-forming clumps, the study adopts additional selection criteria that the clump should be massive and luminous. 

To summarize, the sample of 11 ASSEMBLE targets meets the following key criteria: (1) the CS\,(2-1) emission has a brightness temperature of $T_{\rm b} > 2$\,K; (2) the HCN (3-2), HCN\,(4-3), and CO\,(4-3) lines exhibit ``blue profiles''; (3) the clump masses range from $8\times10^2$ to $2\times10^4$\,\msun, with a median value of $\sim4\times10^3$\,\msun; and (4) the bolometric luminosities range from $1\times10^4$ to $6\times10^5$\,\lsun, with a median value of $\sim1\times10^5$\,\lsun. 

\subsection{Physical Properties of Selected Sample} \label{sample:property}

Table\,\ref{tab:sample} presents the basic properties of the ASSEMBLE sample, including the clump kinematic properties, distances, and physical characteristics. The velocity in the local standard of rest ($V_{\rm lsr}$) was determined from the C$^{17}$O\,(3-2) lines in the APEX observations \citep{Yue2021APEX}, which is listed in column (5). The line asymmetric parameter ($\delta V \equiv (V_{\rm CO}-V_{\rm C^{17}O})/\Delta V_{\rm C^{17}O}$) in column (6) defines the line as having a blue profile. The kinematic distance as well as its upper and lower uncertainties is estimated using the latest rotation curve model of the Milky Way \citep{Reid2019Distance} and is listed in column (7). The clump radius is derived from 2D Gaussian fitting and is listed in column (8). The radius is derived from the 2D Gaussian fitting, the same method as adopted in \citet{Sanhueza2019ASHES} to better compare with. The dust temperature ($T_{\rm dust}$), clump mass ($M_{\rm cl}$), bolometric luminosity ($L_{\rm bol}$), and luminosity-to-mass ratio ($L/M$) are obtained from the far-IR (70--870\,$\mu$m by Herschel and ATLASGAL survey) SED fitting \citep[Table\,5 in][]{Urquhart2018ATLASGAL} and are listed in columns (9)--(12), respectively. The clump surface density, $\Sigma_{\rm cl} = M_{\rm cl}/\pi R_{\rm cl}^2$, is listed in column (13). It is noteworthy that all of the ASSEMBLE clumps have a surface density of $\Sigma_{\rm cl}\gtrsim1$\,g\,cm$^{-2}$, significantly surpassing the threshold (0.05\,g\,cm$^{-2}$) for high-mass star formation proposed by \citet{Urquhart2014CSC} and \citet{He2015Infall}, which further justifies our sample selection. 

The background \textit{Spitzer} three-color composite map (blue: 3.6\,$\mu$m; green: 4.5\,$\mu$m; red: 8\,$\mu$m) in Figure\,\ref{fig:infrared} displays the infrared environment. All the 11 targets exhibit bright infrared sources indicating active massive star formation, although the $L_{\rm bol}/M_{\rm cl}$ derived from Table\,\ref{tab:sample} columns (10)--(11), varies from 12 to 80. The differences in the $L_{\rm bol}/M_{\rm cl}$ suggest potential variations in the evolutionary stages among the samples. For instance, I16272-4837 \citep[also known as SDC335;][]{Peretto2009SDC} with the value of 12 is in an early stage of high-mass star formation embedded in a typical IRDC \citep{Xu2023SDC335}. Another example is I15520-5234, where extended radio ($\nu=8.64$\,GHz) continuum emission indicates evolved UC\hii~regions \citep[see Fig. 4 of][]{Ellingsen2005I15520}. Accordingly, I15520-5234 has the highest value of $L_{\rm bol}/M_{\rm cl}\simeq80$. 

In Appendix\,\ref{app:radio}, we present additional information on the radio emission derived from the MeerKAT Galactic Plane Survey 1.28\,GHz data \citep[Goedhart et al. in prep.]{Padmanabh2023MeerKAT}. All of the protoclusters included in our study exhibit embedded radio emission, which can originate from UC\hii~regions, or extended radio emission, which may arise from radio jets or extended \hii~regions. For instance, in the case of I14382-6017, we observe cometary radio emission that exhibits a spatial correlation with the 8\,$\mu$m emission, outlining the extended shell of the \hii~region. However, at an early stage, I16272-4837 displays two radio point sources associated with two UC\hii~regions \citep{Avison2015SDC335}. Additionally, I17720-3609 exhibits northward extended radio emission that is linked to a blue-shifted outflow \citep{Baug2020Outflow, Baug2021Outflow}.


\section{Observations and Data Reduction} \label{sec:observations}


\begin{deluxetable*}{cccccccc}
\rotate 
\tabletypesize{\small}
\tablewidth{0pt} 
\linespread{1.2}
\tablecaption{ALMA Observational Parameters 
\label{tab:almaobs}}
\tablehead{
\multirow{2}{*}{\makecell[c]{ASSEMBLE \\ Clump}} & \colhead{On-source time} & Line-free bandwidth\tablenotemark{a} & \colhead{RMS Noise} & \colhead{Beam Sizes} & \colhead{MRS\tablenotemark{b}} & \colhead{Phase Calibrators} & \colhead{Flux \& Bandpass Calibrators}\\
& \colhead{(mins/pointing)} & \colhead{(MHz [percent])} & \colhead{(mJy\,beam$^{-1}$)} & \colhead{($\arcsec\times\arcsec$)} & \colhead{($\arcsec$)} & \colhead{} & \colhead{}
}
\colnumbers
\startdata
I14382-6017 & 3.2 & 3569 (76.1\%)  & 1.0 &1.175$\arcsec$$\times$0.793$\arcsec$ & 8.433 & J1524-5903 &J1427-4206 \\
I14498-5856 & 3.2 & 1407 (30.0\%)  & 1.0 &1.065$\arcsec$$\times$0.785$\arcsec$ & 8.179 & J1524-5903 &J1427-4206 \\
I15520-5234 & 3.7 & 1249 (26.6\%)  & 1.5 &0.852$\arcsec$$\times$0.699$\arcsec$ & 9.197 & J1650-5044 &J1924-2914,J1427-4206,J1517-2422 \\
I15596-5301 & 3.7 & 1347 (28.7\%)  & 0.5 &0.824$\arcsec$$\times$0.691$\arcsec$ & 9.071 & J1650-5044 &J1924-2914,J1427-4206,J1517-2422 \\
I16060-5146 & 3.7 & 58 (1.2\%)     & 2.5 &0.820$\arcsec$$\times$0.681$\arcsec$ & 8.953 & J1650-5044 &J1924-2914,J1427-4206,J1517-2422 \\
I16071-5142 & 3.7 & 70 (1.5\%)     & 1.3 &0.810$\arcsec$$\times$0.666$\arcsec$ & 8.974 & J1650-5044 &J1924-2914,J1427-4206,J1517-2422 \\
I16076-5134 & 3.7 & 354 (7.5\%)    & 0.6 &0.734$\arcsec$$\times$0.628$\arcsec$ & 8.755 & J1650-5044 &J1924-2914,J1427-4206,J1517-2422 \\
I16272-4837 & 3.7 & 72 (1.5\%)     & 1.4 &0.799$\arcsec$$\times$0.662$\arcsec$ & 
8.450 & J1650-5044 &J1924-2914,J1427-4206,J1517-2422 \\
I16351-4722 & 3.7 & 131 (2.6\%)    & 1.2 &0.771$\arcsec$$\times$0.651$\arcsec$ & 8.221 & J1650-5044 &J1924-2914,J1427-4206,J1517-2422 \\
I17204-3636 & 2.7 & 2432 (51.9\%)  & 0.5 &0.790$\arcsec$$\times$0.660$\arcsec$ & 7.246 & J1733-3722 &J1924-2914 \\
I17220-3609 & 2.7 & 320 (6.8\%)    & 1.7 &0.813$\arcsec$$\times$0.653$\arcsec$ & 
7.237 & J1733-3722 &J1924-2914 \\
\enddata
\tablecomments{The ASSEMBLE target names are listed in (1).}
\tablenotetext{a}{The bandwidth used for continuum subtraction as well as the percentage to total bandwidth.}
\tablenotetext{b}{Maximum recoverable scale.}
\end{deluxetable*}

\subsection{ALMA Band-7 Observing Strategy} \label{obs:strategy}

The 17-pointing mosaicked observations were carried out with ALMA using the 12-m array in Band 7 during Cycle 5 (Project ID: 2017.1.00545.S; PI: Tie Liu) from 18 May to 23 May 2018. The observations have been divided into 6 executions; 48 antennas were used to obtain a total of 1128 baselines with lengths ranging from 15 to 313.7 meters across all the executions.

The mosaicked observing fields of ALMA are designed to cover the densest part of the massive clumps traced by the ATLASGAL 870\,$\mu$m continuum emission. The fields are outlined by the yellow dashed loops in Figure\,\ref{fig:infrared}. The field center of each mosaicking field is shown in the column (2)--(3) of Table\,\ref{tab:sample}. Each mosaicked field has a uniform size of $\sim46$\arcsec~and a sky coverage of $\sim0.58$\,arcmin$^2$. The on-source time per pointing is 2.7--3.7\,minutes, which is listed in column (2) of Table\,\ref{tab:almaobs}.

The four spectral windows (SPWs) numbered 31, 29, 25, and 27 are centered at frequencies of 343.2, 345.1, 354.4, and 356.7\,GHz, respectively. SPWs 25 and 27 possess a bandwidth of 468.75\,MHz and spectral resolution of 0.24\,\kms, which are specifically designed to observe the HCN\,(4-3) and HCO$^+$\,(4-3) strong lines, respectively. These lines serve as reliable tracers for infall and outflow \citep[e.g.,][]{Chira2014Infall}. On the other hand, SPWs 29 and 31 have a bandwidth of 1875\,MHz and a spectral resolution of 0.98\,\kms, which are intended to cover a wide range of spectral lines. The CO\,(3-2) line serves as outflow tracer according to \citet{Sanhueza2010G34,Wang2011G28, Baug2020Outflow,Baug2021Outflow}. High-density tracers, such as \htcn\,(4-3) and CS\,(7-6), can determine the core velocity. Additionally, some sulfur-bearing molecules, such as \htcs~and SO$_2$, can serve as tracers of rotational envelopes, while shock tracers include SO\,$^3\Sigma$\,($8_8-7_7$) and \chtoh\,($13_{1,12}-13_{0,13}$). The hot-core molecular lines, such as \htcs, \chtocho, and CH$_3$COCH$_3$, have a sufficient number of transitions to facilitate rotation-temperature and chemical abundance studies. A summary of the target spectral lines can be found in Table\,A1 of \citet{Xu2023SDC335}.

\subsection{ALMA Data Calibration and Imaging} \label{obs:reduction}

The pipeline provided by the ALMA observatory was utilized to perform data calibration in CASA \citep{McMullin2007CASA} version 5.1.15. The phase, flux, and bandpass calibrators are listed in columns (7)--(8) of Table\,\ref{tab:almaobs}. The imaging was conducted through the \texttt{TCLEAN} task in CASA 5.3. To aggregate the continuum emission, line-free channels were meticulously selected by visual inspection, with the bandwidth and its percentage of total bandwidth listed in column (3). A total of three rounds of phase self-calibration and one round of amplitude self-calibration were run to enhance the dynamic range of the image. For self-calibration, antenna DA47 was designated as the reference antenna. During imaging, the deconvolution was set as ``hogbom'' while the weighting parameter was set as ``briggs'' with a robust value of 0.5 to balance sensitivity and angular resolution. The primary beam correction is conducted with \texttt{pblimit=0.2}. Following self-calibration, the sensitivity and dynamic range of the final continuum image were significantly improved, as indicated in column (4), ranging from 0.5--1.7\,\mjybeam~with a mean value of $\sim1$\,\mjybeam. The beam size (i.e., angular resolution) with 0\parcsec8--1\parcsec2 and maximum recoverable scale (MRS) with 7\parcsec2--9\parcsec2 are presented in columns (5)--(6).


\section{Results} \label{sec:result}

\subsection{Dust Continuum Emission} \label{result:dust}

Figure\,\ref{fig:continuum} presents the ALMA 870\,$\mu$m dust continuum images without primary beam correction for a uniform rms noise. As a comparison, the dust continuum emission at the same wavelength from the single-dish survey ATLASGAL \citep{Schuller2009ATLASGAL} with a beam size of 19\parcsec2 is overlaid as black contours. In all of the 11 targets, the small-scale structures resolved by ALMA show a good spatial correlation with the large-scale structures seen by ATLASGAL. In other words, the dense structures surviving in the interferometric ``filtering-out'' effect are mostly distributed in the densest part of the clump. However, the small-scale structures present various morphologies: some present elongated filaments (e.g., I14382-6017 and I16071-5142); some have centrally concentrated cores (e.g. I16272-4837); some have spiral-like dust arms (e.g., I15596-5301 and I16060-5146).

\subsection{Core Extraction and Catalog} \label{result:extractcores}

\begin{figure*}[!ht]
\centering
\includegraphics[width=0.5\linewidth]{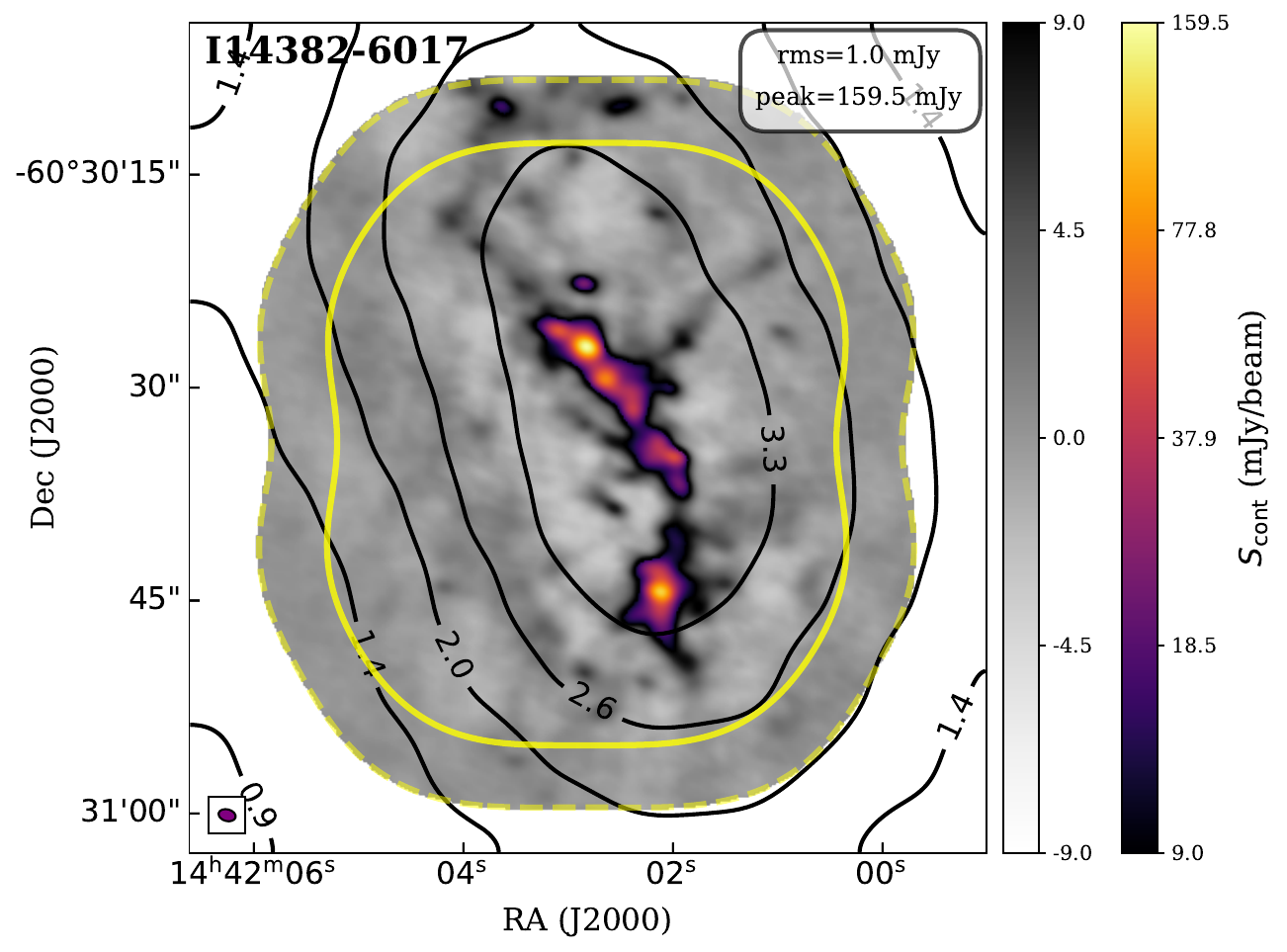}
\includegraphics[width=0.39\linewidth]{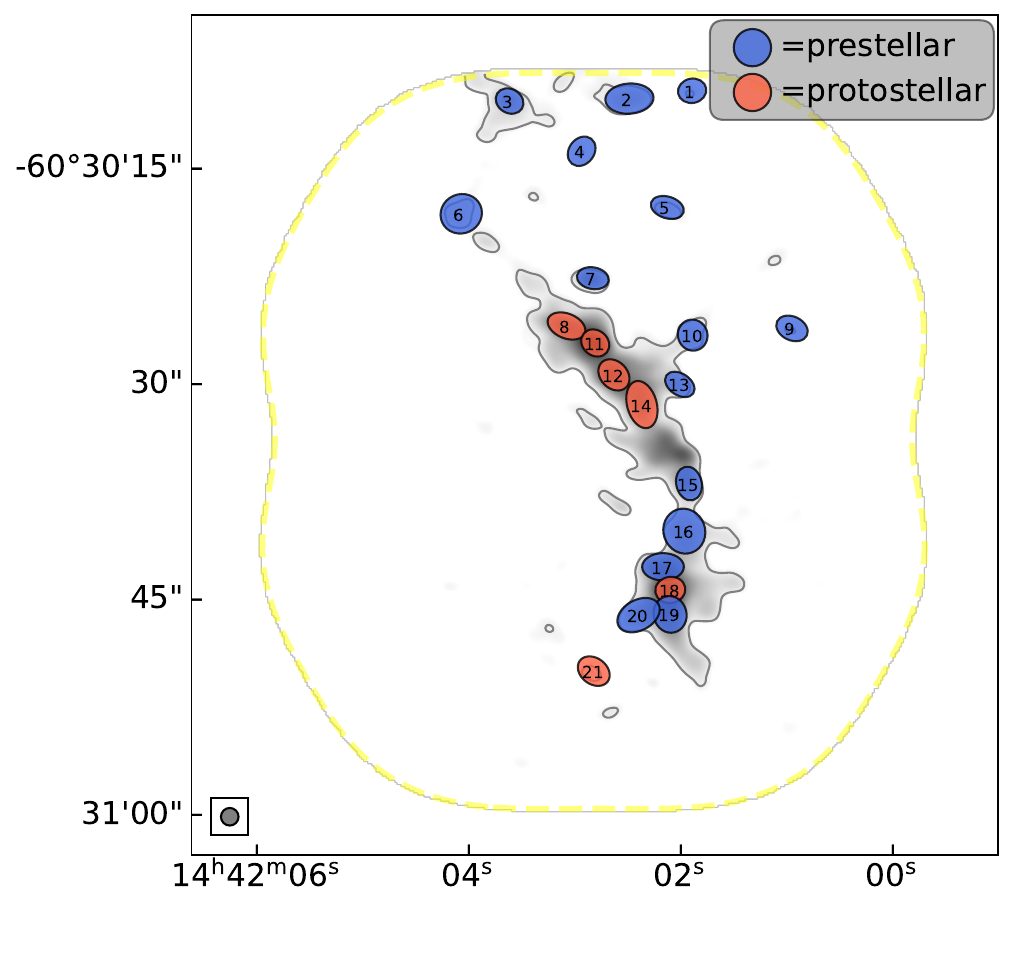}
\includegraphics[width=0.5\linewidth]{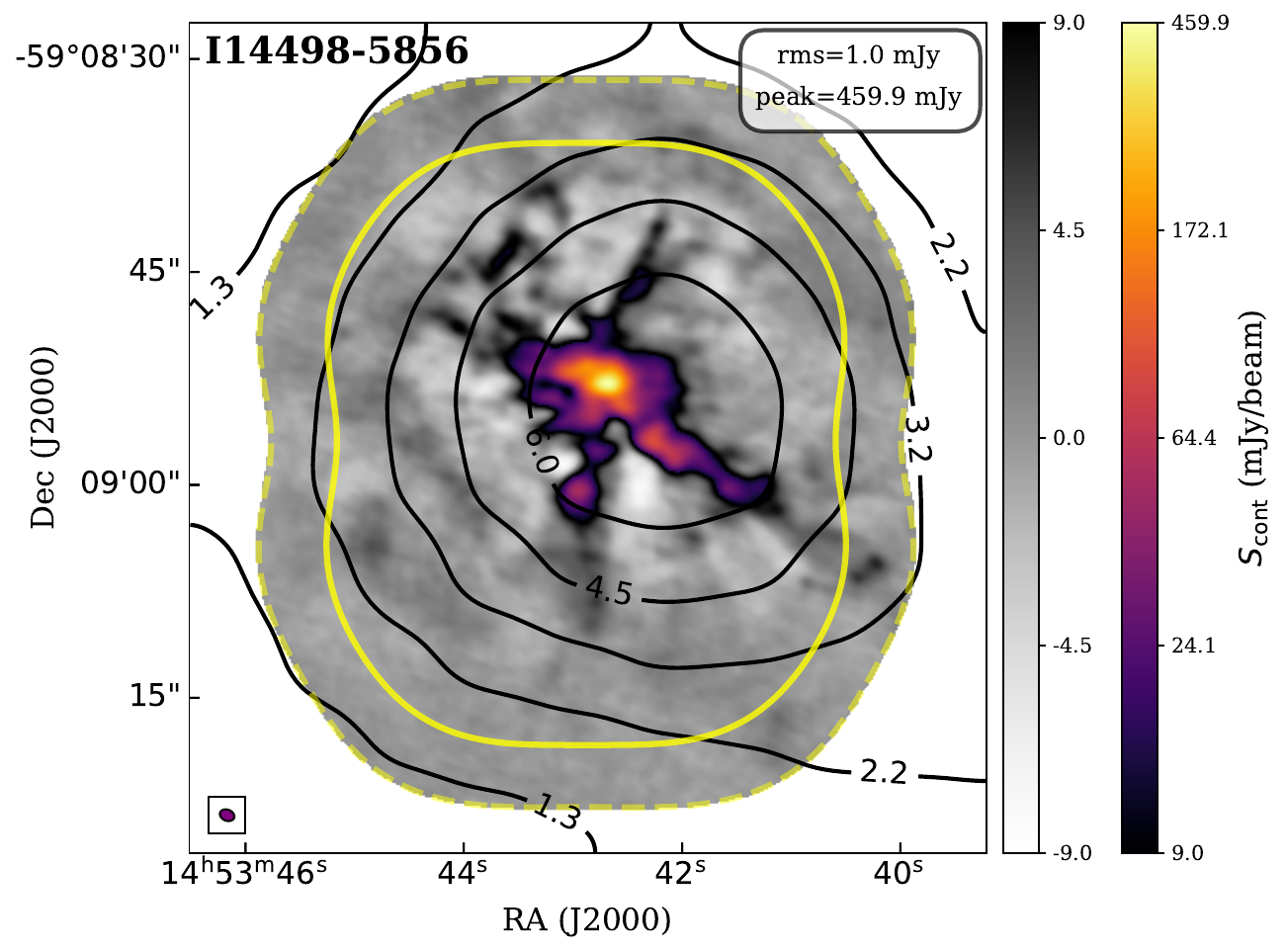}
\includegraphics[width=0.39\linewidth]{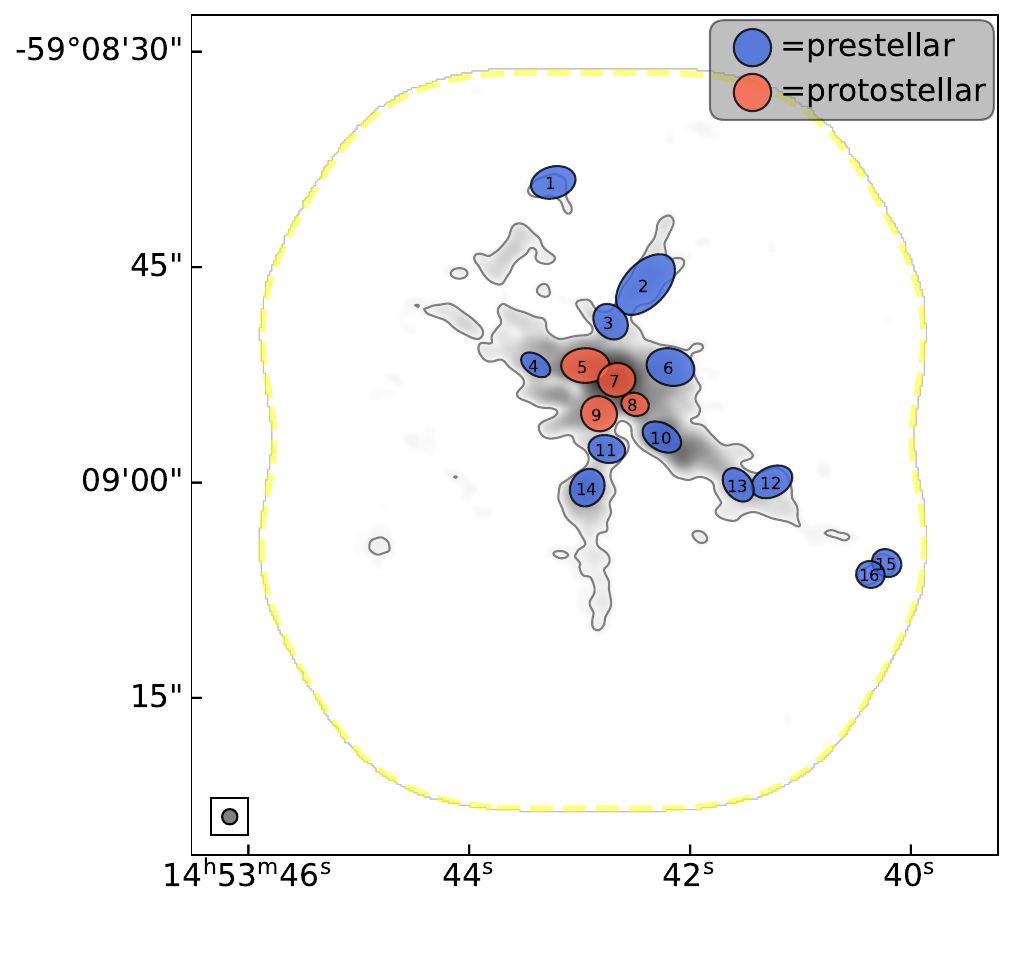}
\caption{The ALMA 870\,$\mu$m dust continuum emission without primary beam correction as well as extracted cores for two ASSEMBLE clumps (I14382-6017 and I14498-5856). The ALMA mosaicked primary beam responses of 0.5 and 0.2 are outlined by yellow solid and dashed lines respectively. Only the primary beam response of 0.2 is shown on the right panel. The beam size of each continuum image is shown in the bottom left corner. \textit{Left}: the background color map shows the ALMA 870\,$\mu$m emission with two colorbars, the first one (grayscale) showing -9 to +9 times the rms noise on a linear scale, then a second one (color-scheme) showing the range +9 times the rms noise to the peak value of the image in an arcsinh stretch. The rms noise and peak intensity are given on the top right. The black contours are from the ATLASGAL 870\,$\mu$m continuum emission, with power-law levels that start at $5\sigma$ and end at $I_\mathrm{peak}$, increasing in steps following the power law $f(n)=3\times n^p + 2$ where $n=1,2,3,...N$ and $p$ is determined from $D=3\times N^p+2$ ($D=I_\mathrm{peak}/\sigma$: the dynamic range; $N=8$: the number of contour levels). The values of each contour level are labeled with a unit of \jybeam. \textit{Right}: the background gray-scale map shows the arcsinh-stretch part in the left panel, outlined by the $5\sigma$ contour. The ALMA continuum emission map is smoothed to a circular beam with a size equal to the major axis of the original beam. The cores extracted by \getsf~algorithm are presented by red / blue ellipses, as well as black IDs, with numbers in order from North to South. The red and blue ones represent protostellar and prestellar cores defined in Section\,\ref{result:coreclass}. \label{fig:continuum}}
\end{figure*}
\addtocounter{figure}{-1}

\begin{figure*}[!ht]
\centering
\includegraphics[width=0.5\linewidth]{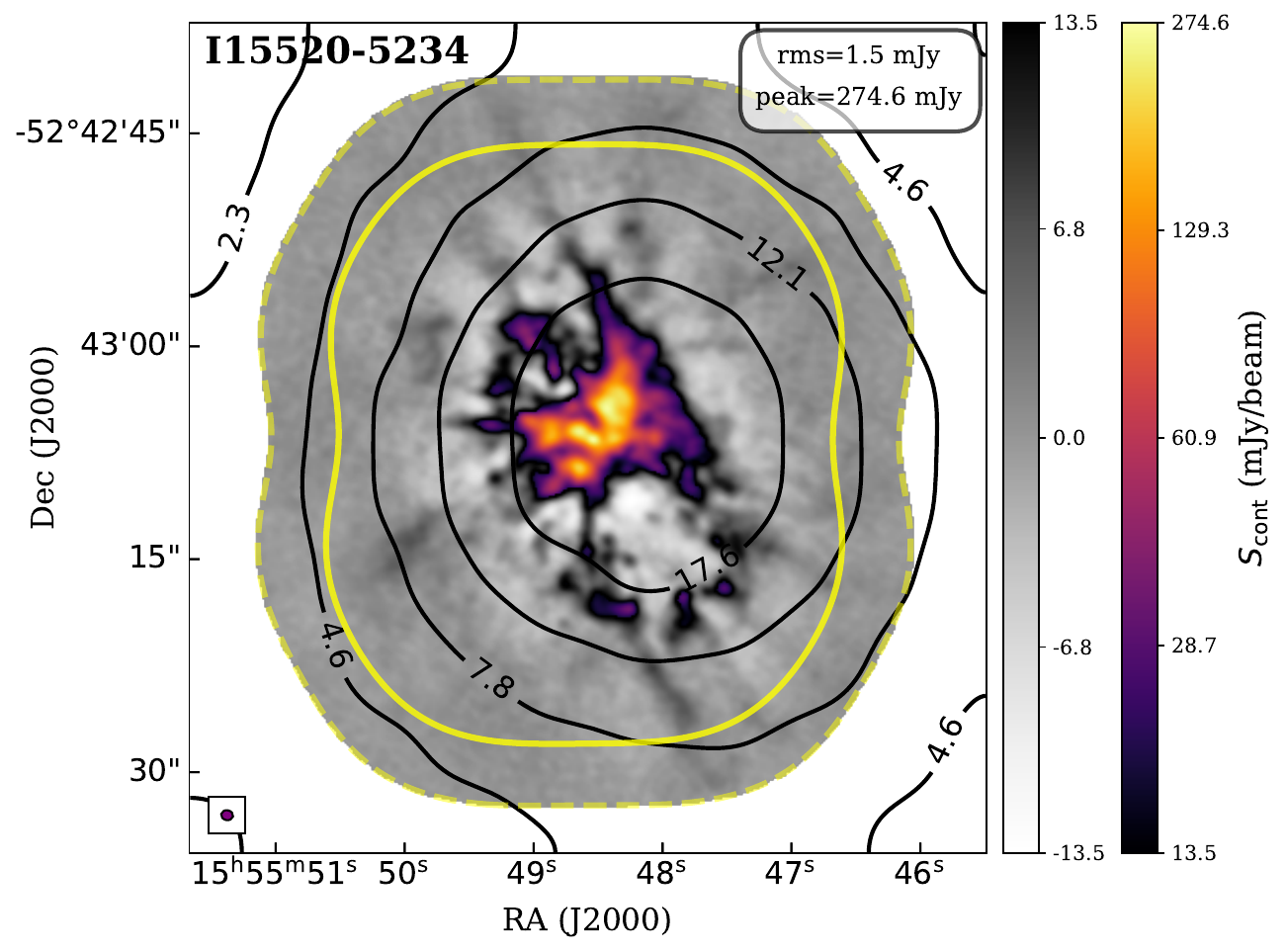}
\includegraphics[width=0.39\linewidth]{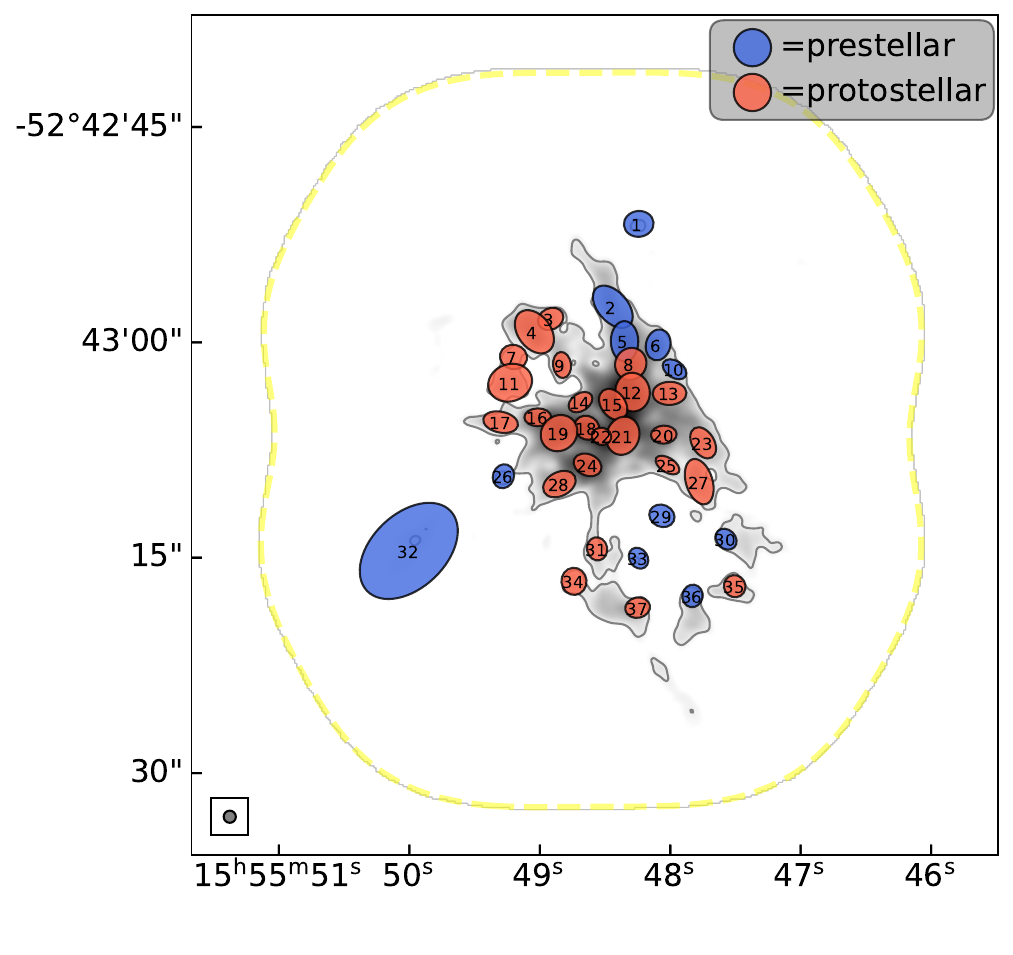}
\includegraphics[width=0.5\linewidth]{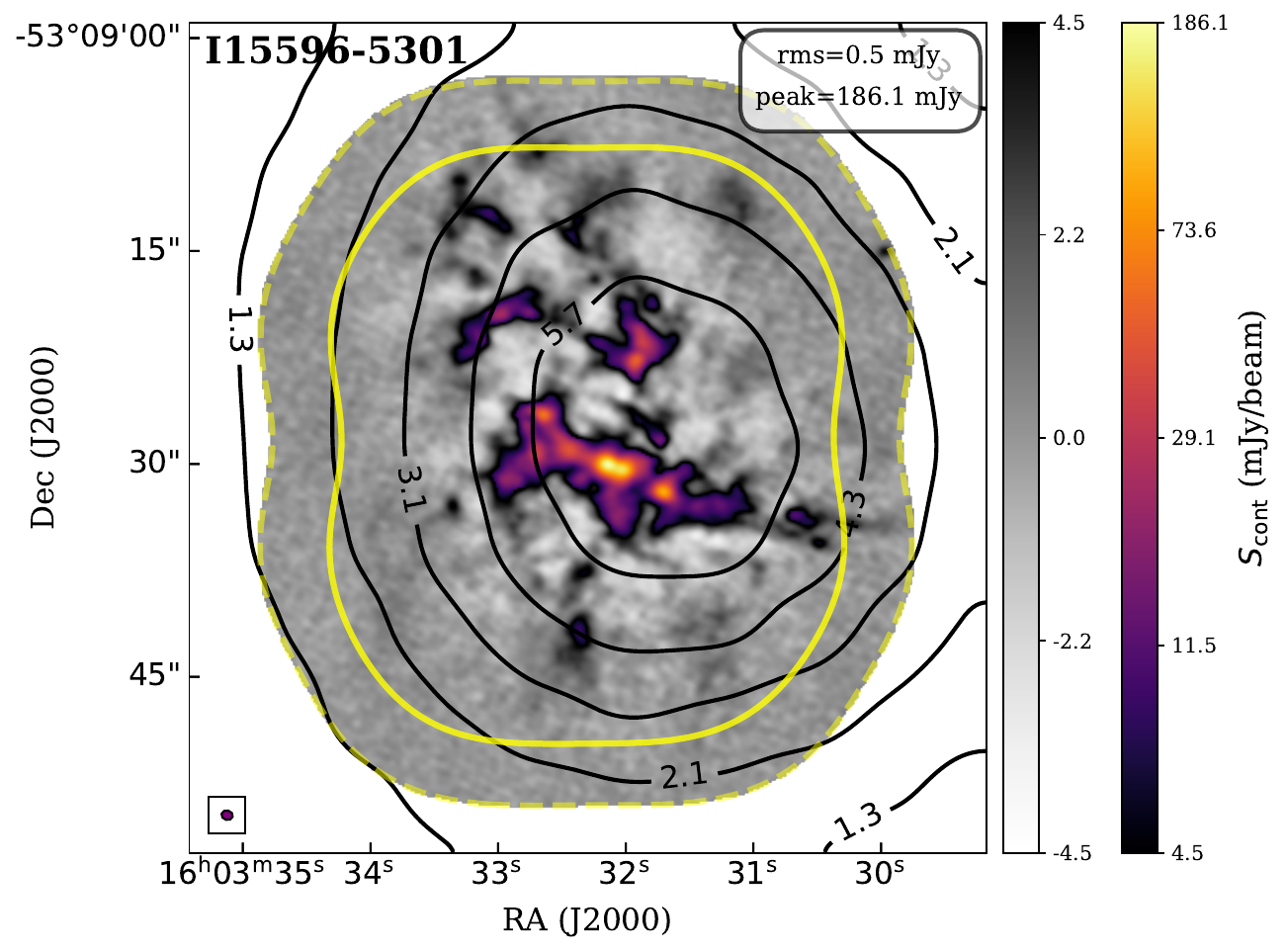}
\includegraphics[width=0.39\linewidth]{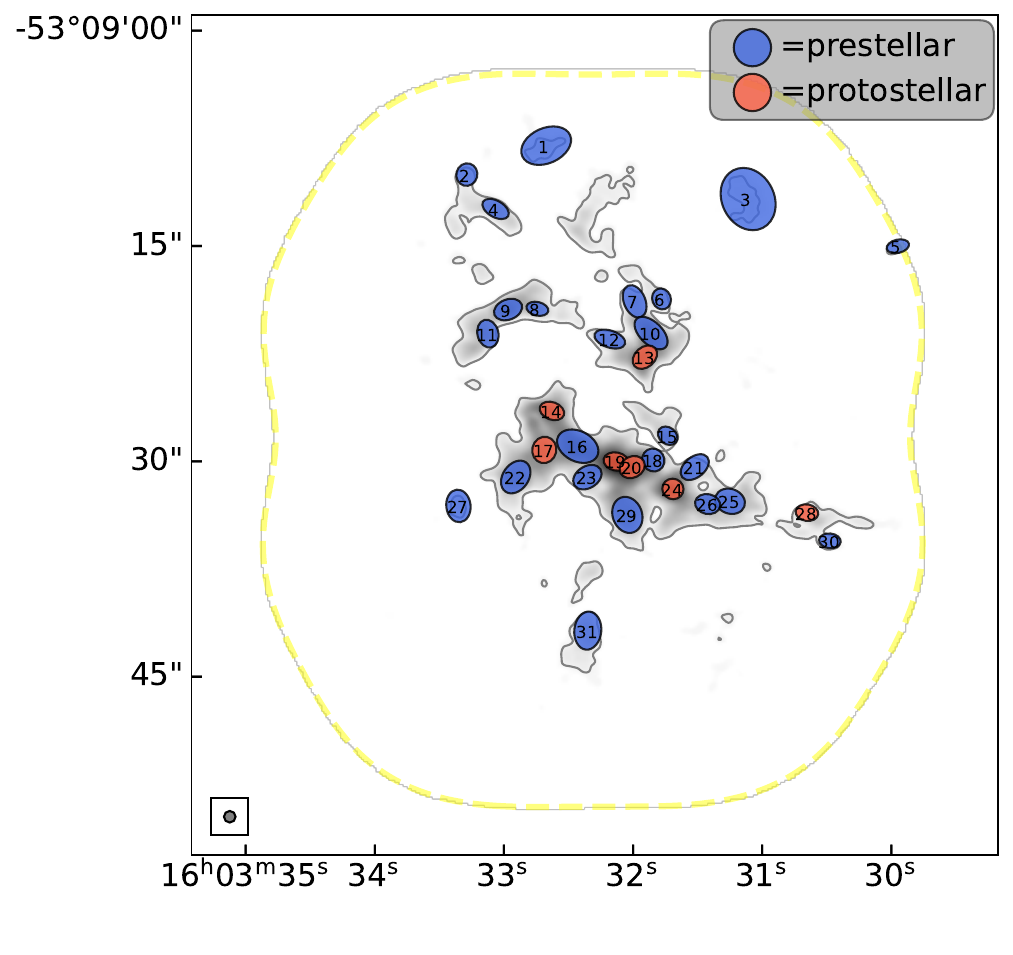}
\includegraphics[width=0.5\linewidth]{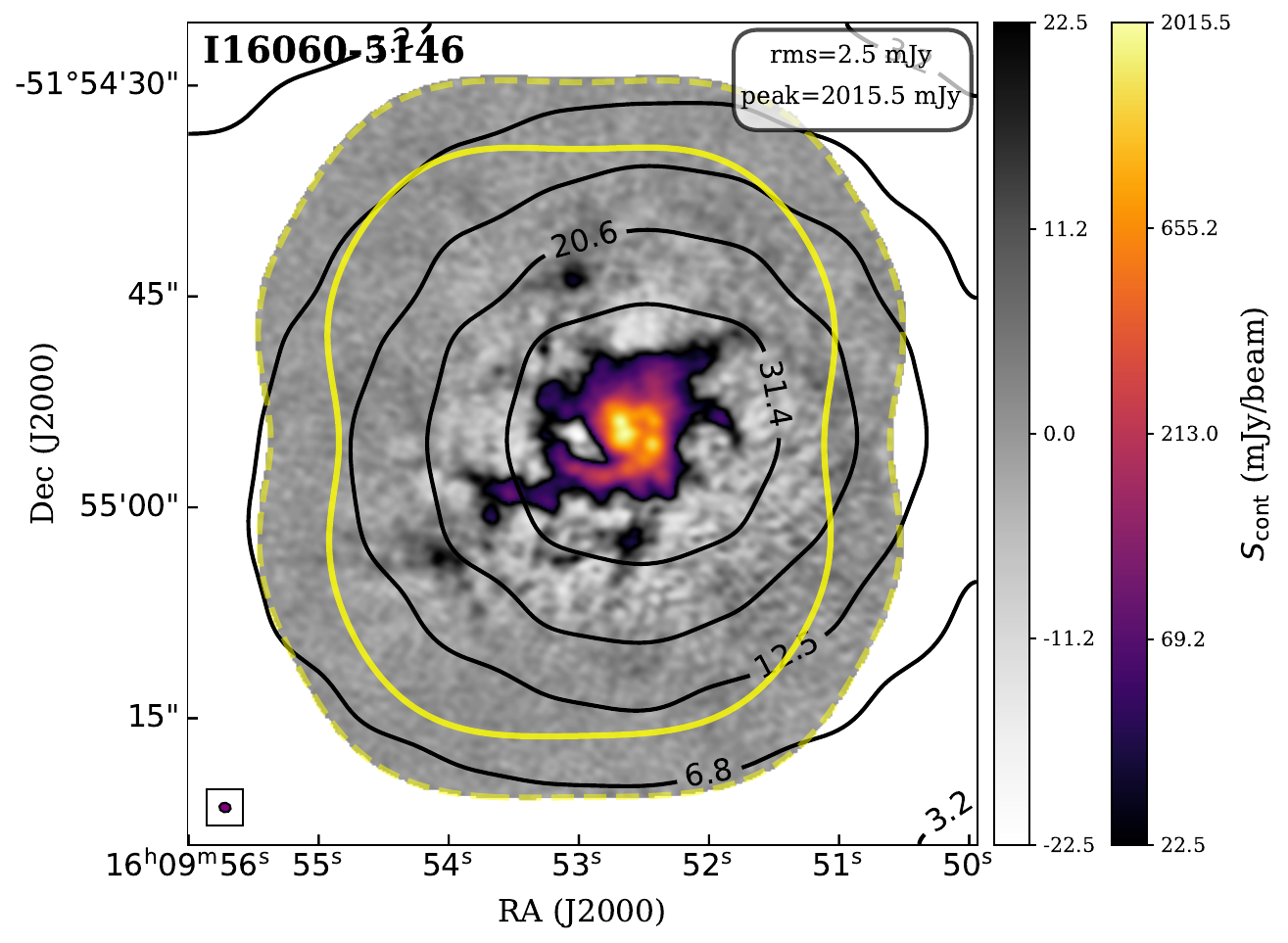}
\includegraphics[width=0.39\linewidth]{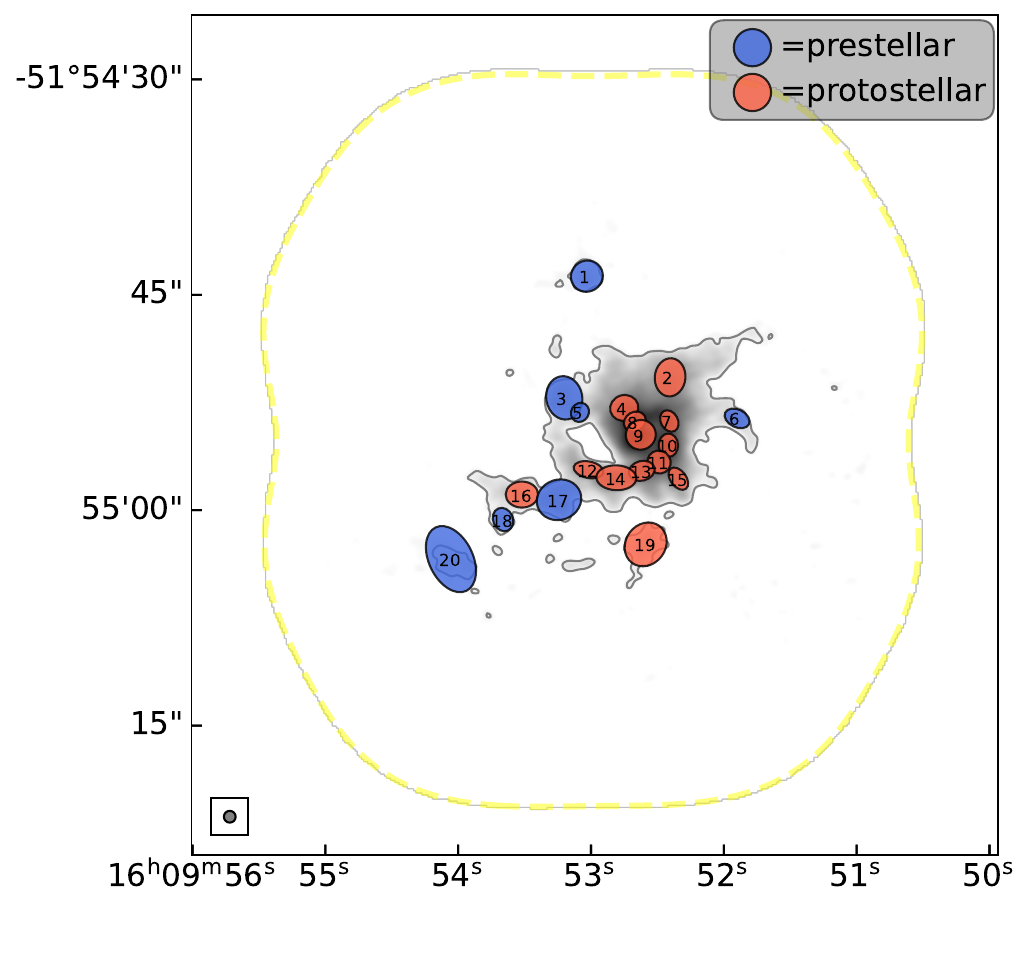}
\caption{Continued for three ASSEMBLE clumps (I15520-5234, I15596-5301, and I16060-5146).}
\end{figure*}
\addtocounter{figure}{-1}

\begin{figure*}[!ht]
\centering
\includegraphics[width=0.5\linewidth]{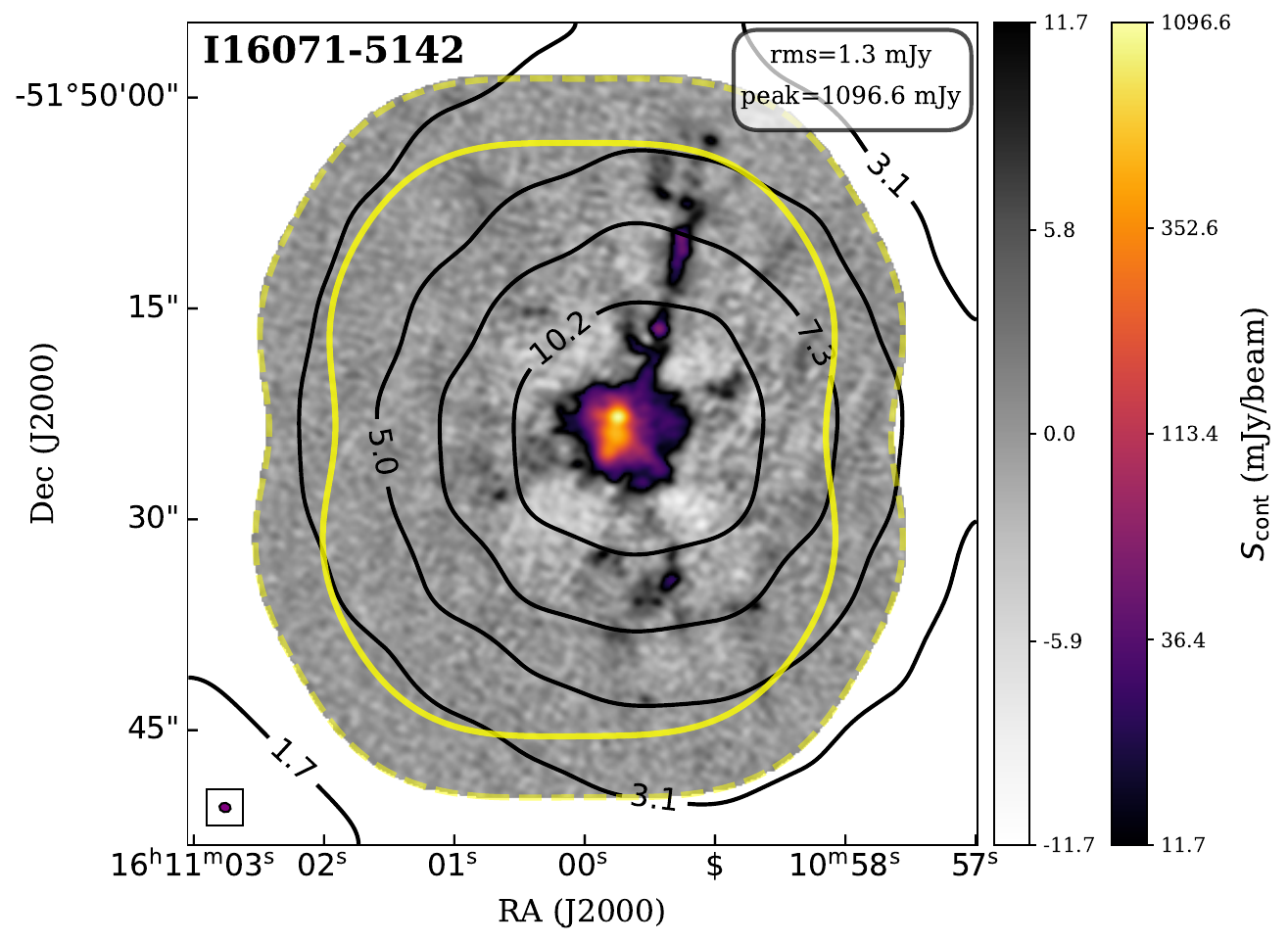}
\includegraphics[width=0.39\linewidth]{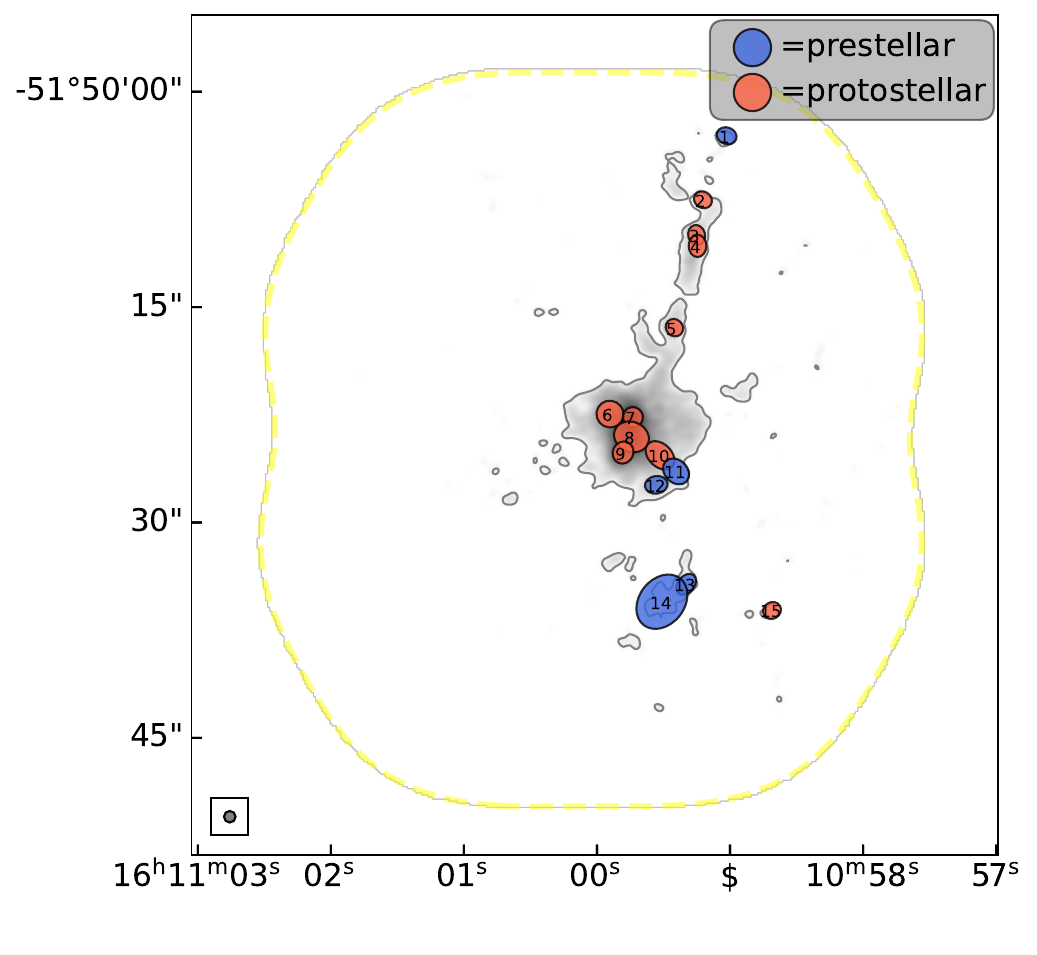}
\includegraphics[width=0.5\linewidth]{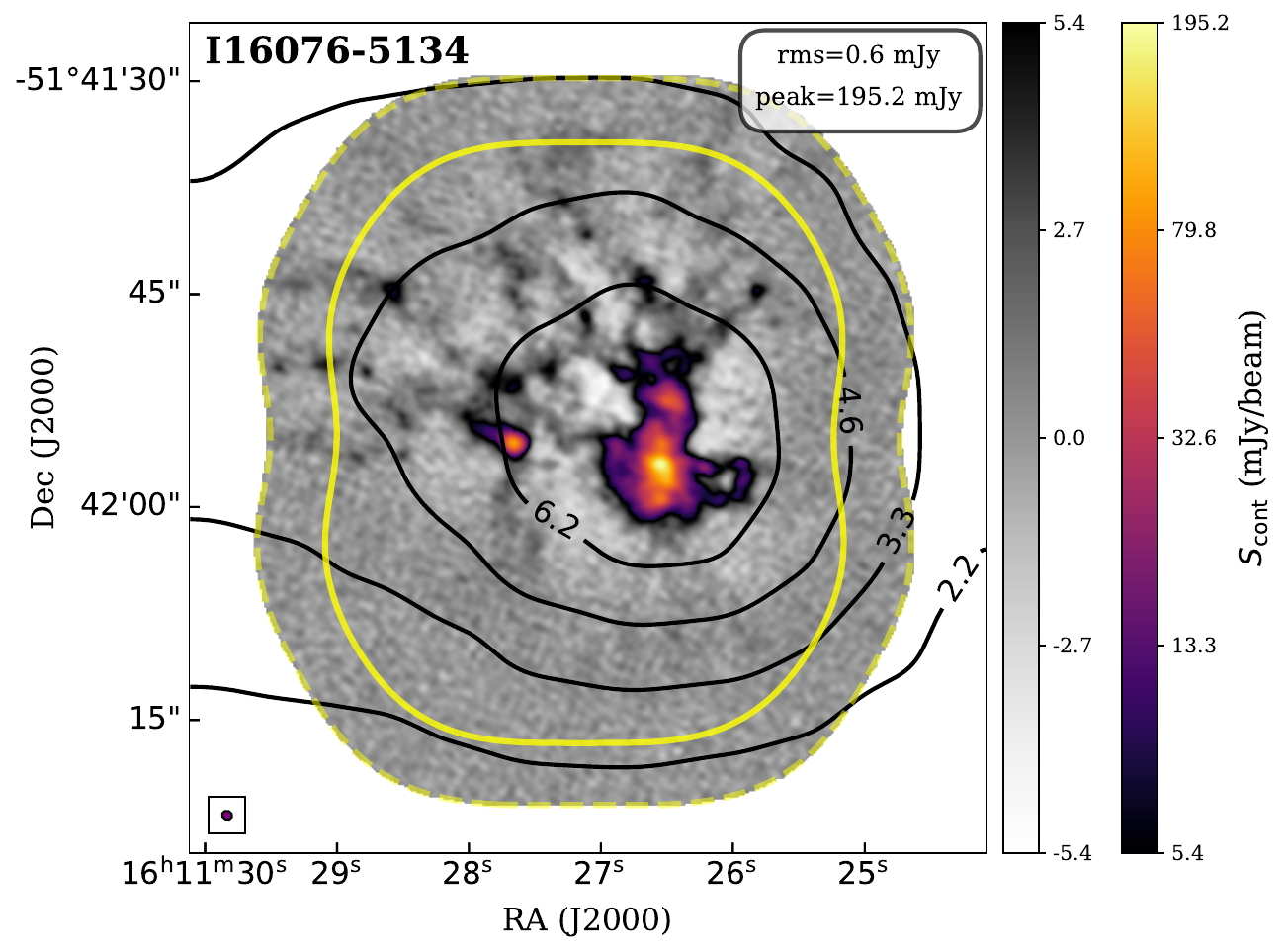}
\includegraphics[width=0.39\linewidth]{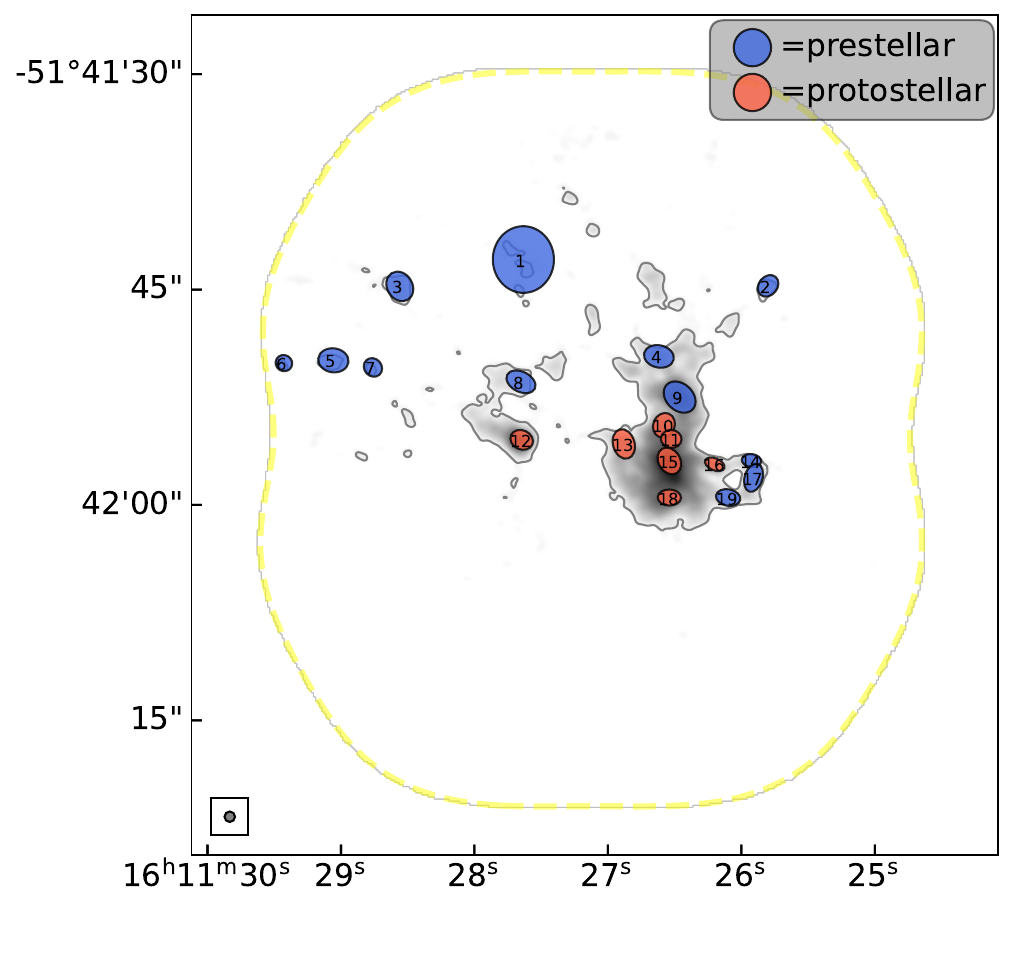}
\includegraphics[width=0.50\linewidth]{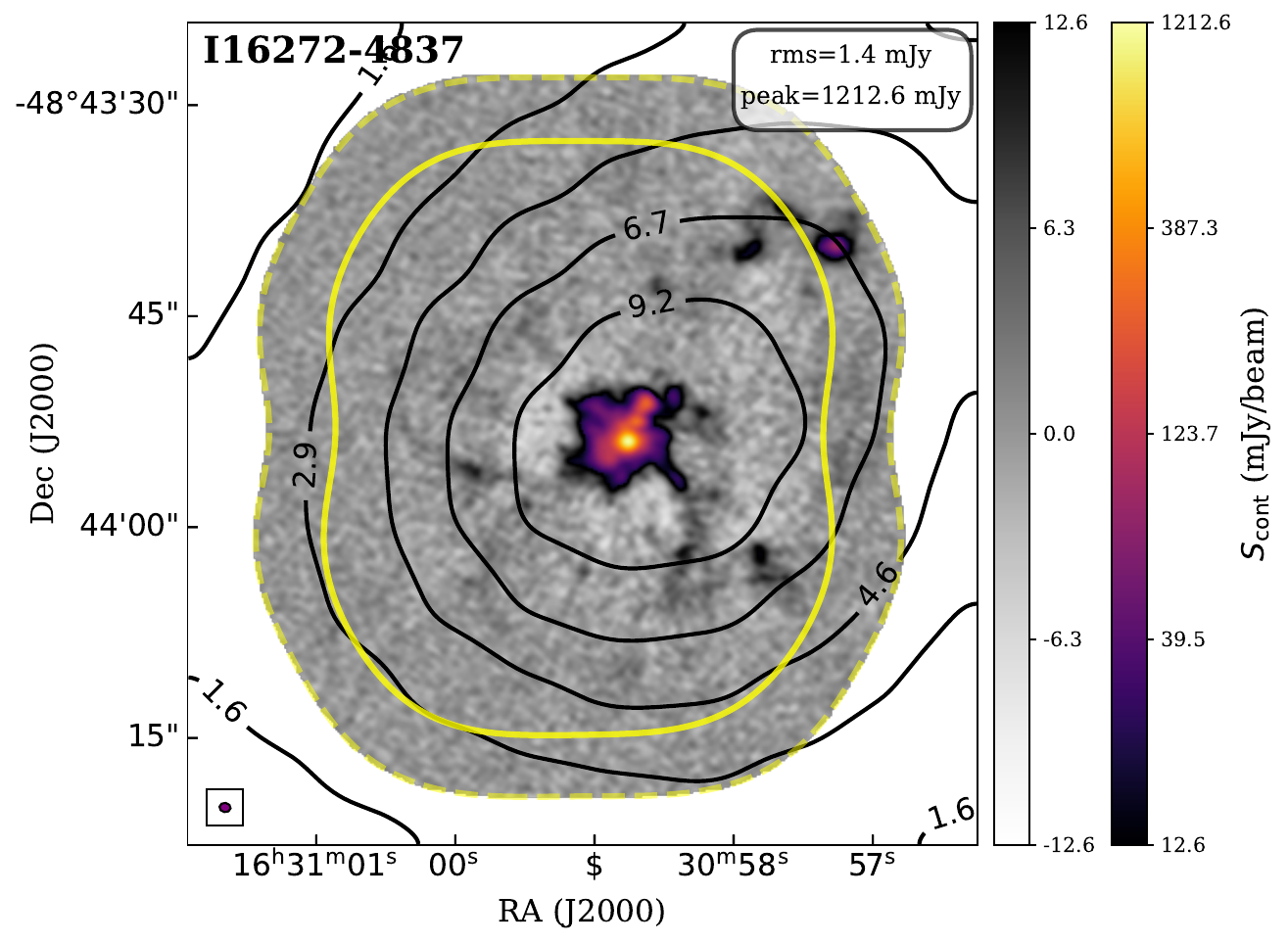}
\includegraphics[width=0.39\linewidth]{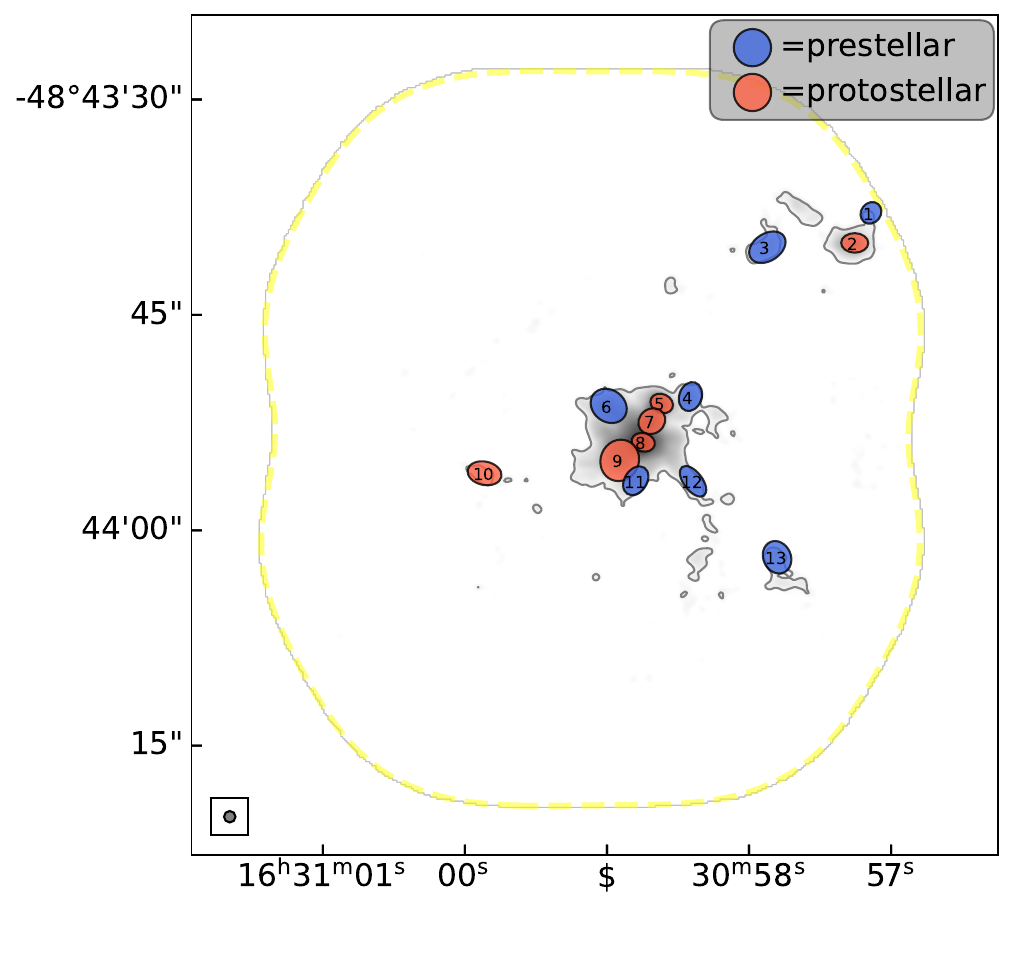}
\caption{Continued for three ASSEMBLE clumps (I16071-5142, I16076-5134, and I16272-4837).}
\end{figure*}
\addtocounter{figure}{-1}

\begin{figure*}[!ht]
\centering
\includegraphics[width=0.5\linewidth]{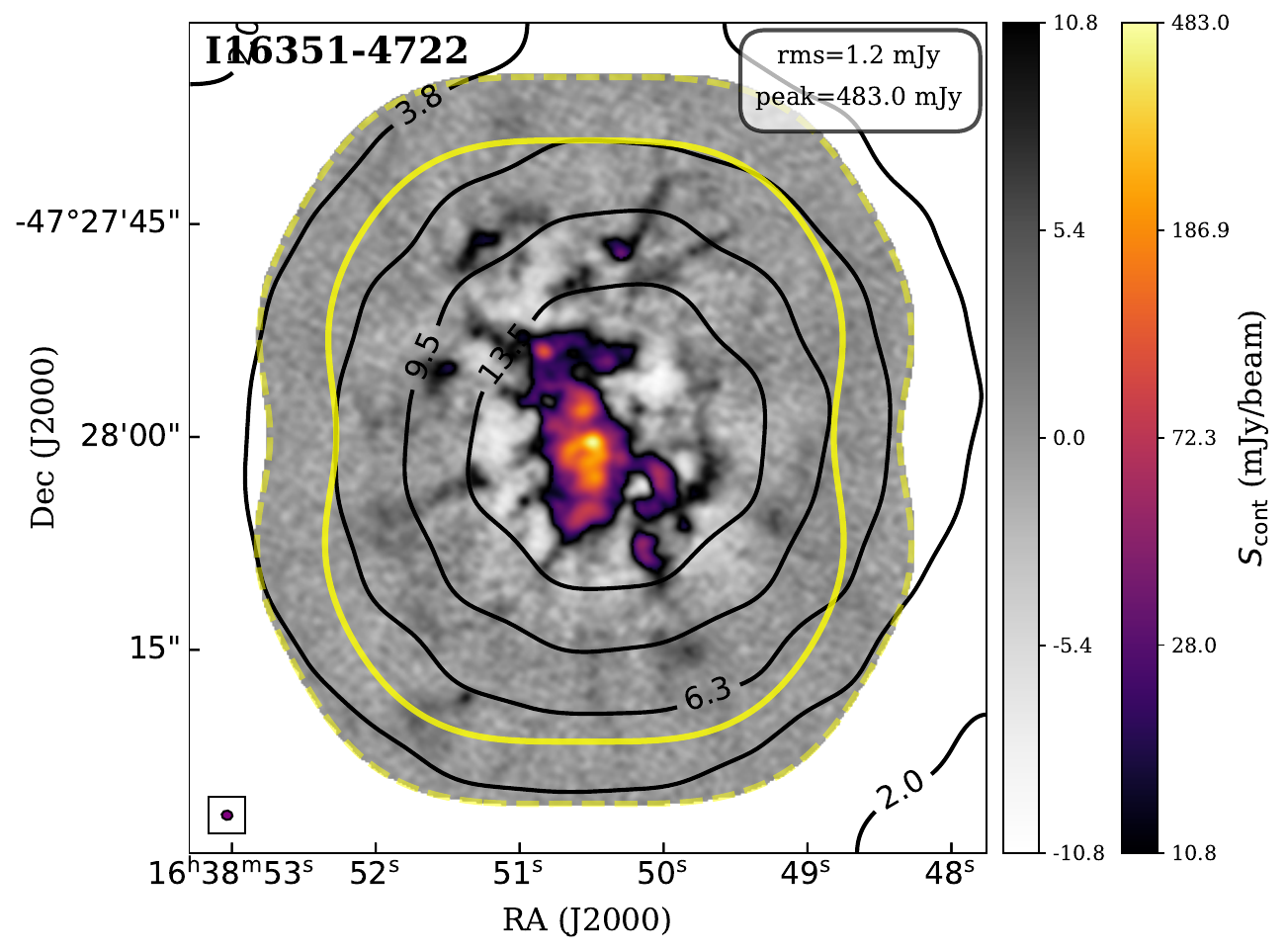}
\includegraphics[width=0.39\linewidth]{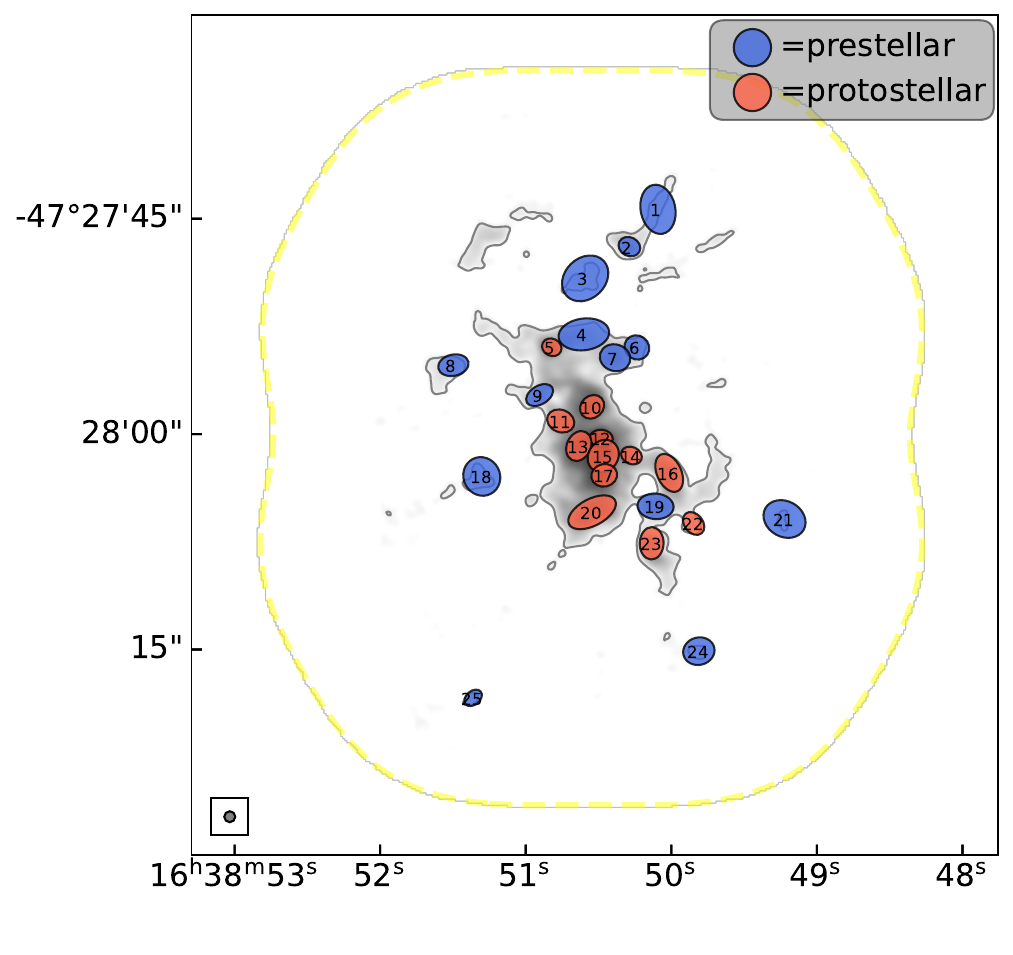}
\includegraphics[width=0.5\linewidth]{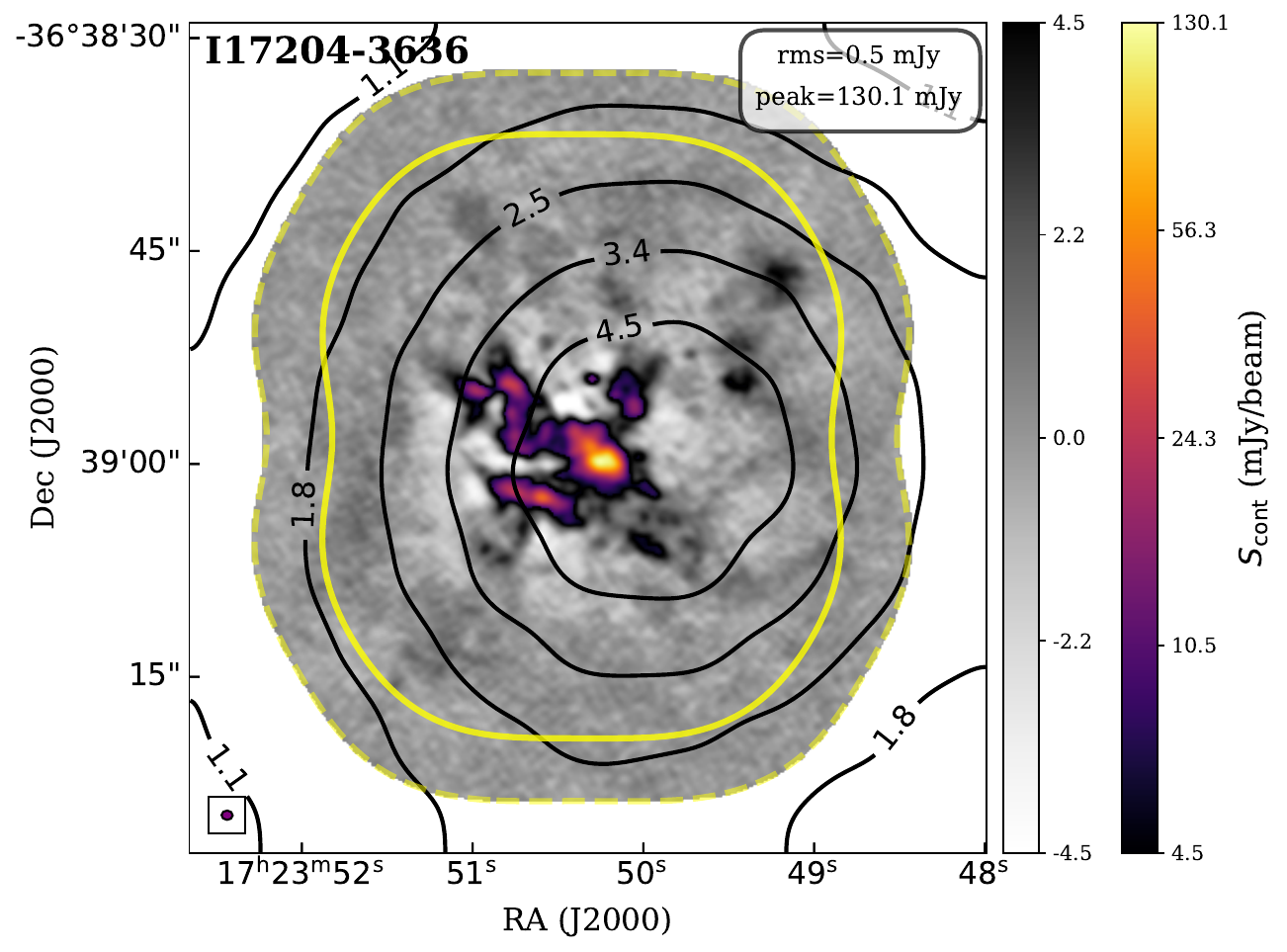}
\includegraphics[width=0.39\linewidth]{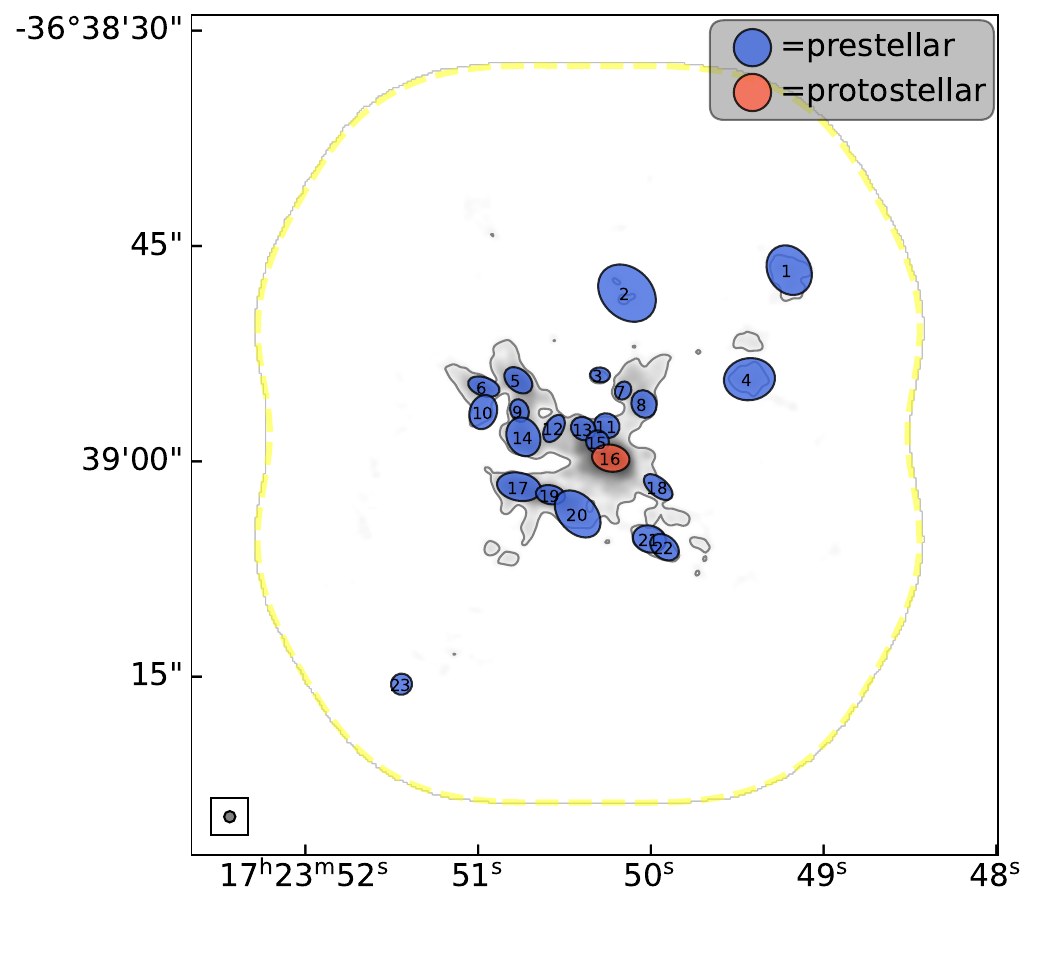}
\includegraphics[width=0.5\linewidth]{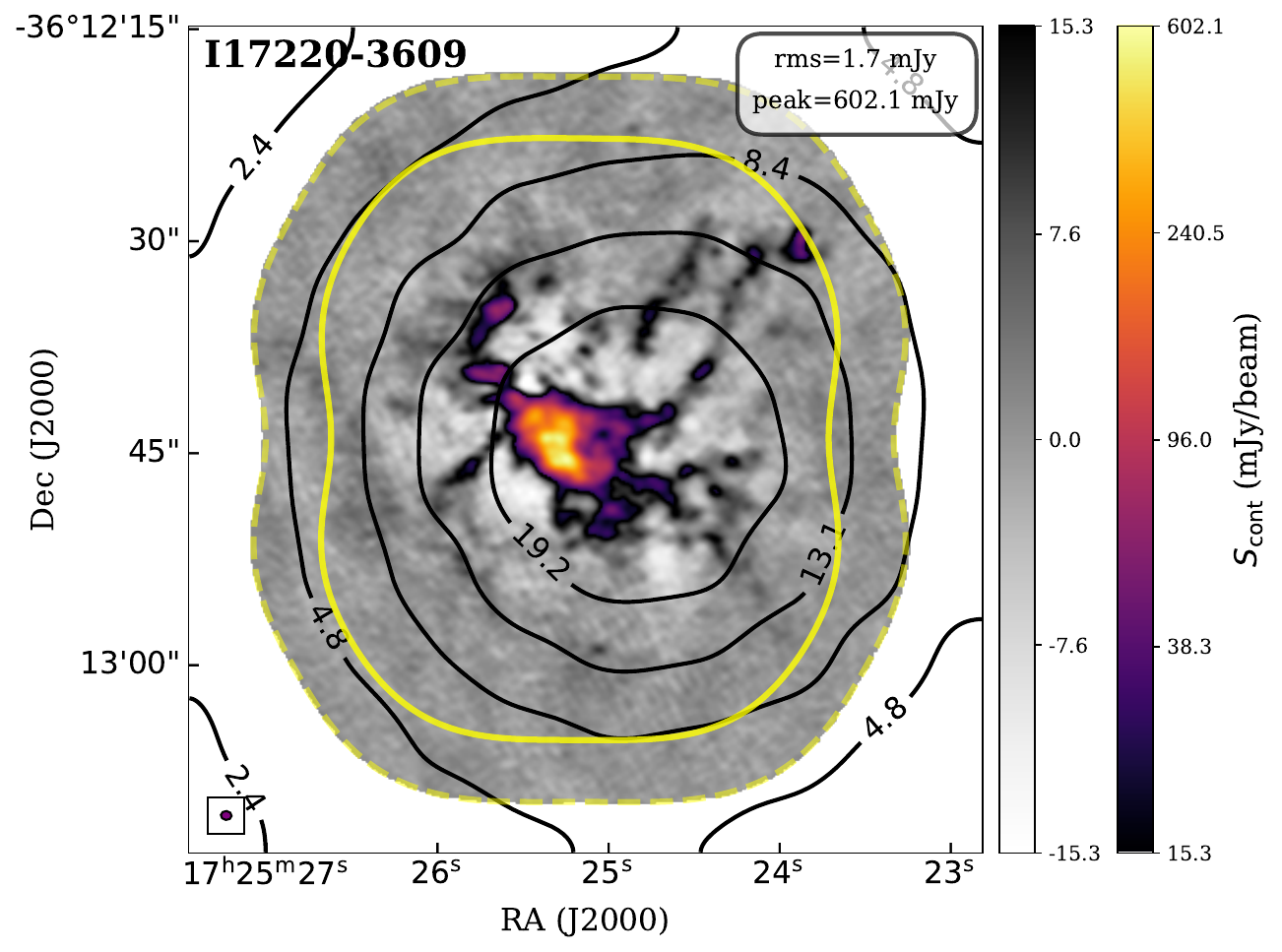}
\includegraphics[width=0.39\linewidth]{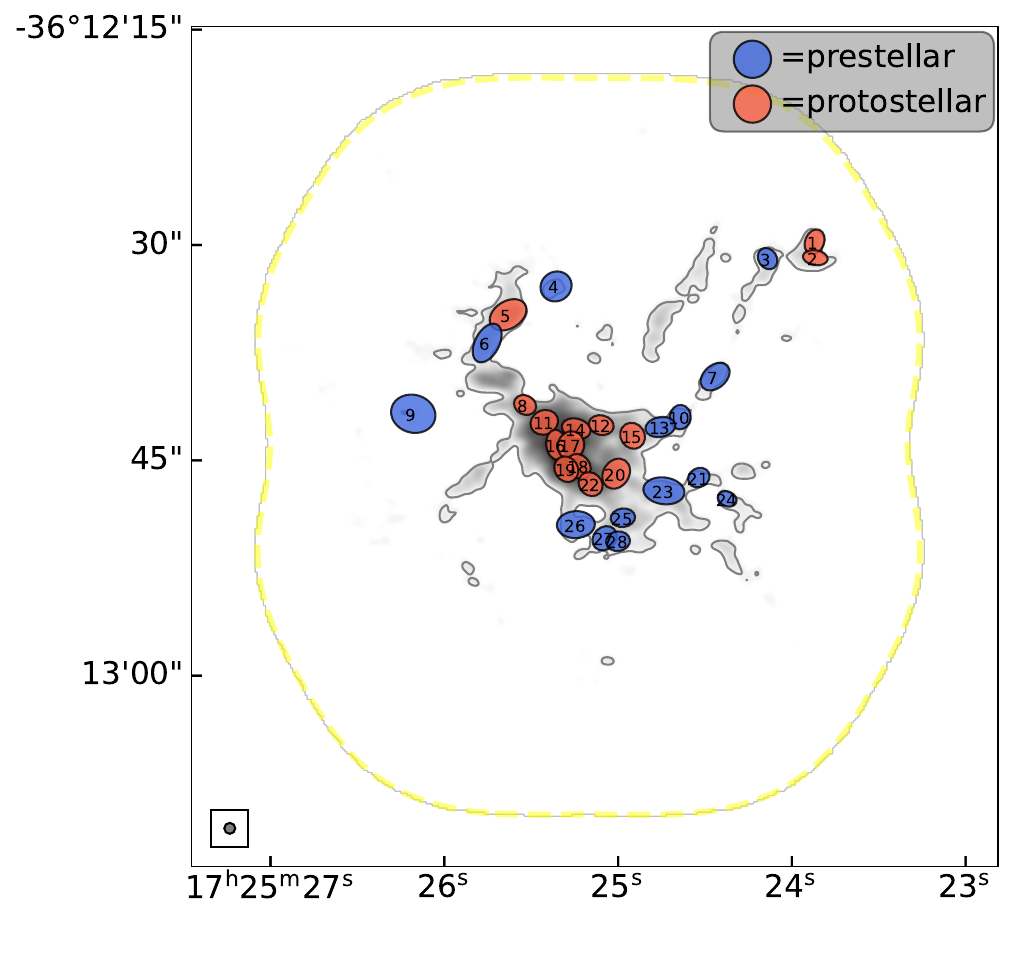}
\caption{Continued for three ASSEMBLE clumps (I6351-4722, I17204-3636, and I17220-3609).}
\end{figure*}

We here present the extraction of core-like structures (or cores) and the measurement of fundamental physical parameters including integrated flux, peak intensity, size, and position. The choice of core extraction algorithm should be carefully made based on actual physical scenarios and scientific expectations. In this work, we use the \getsf~extraction algorithm that spatially decomposes the observed images to separate relatively round sources from elongated filaments as well as their background emission \citep{Men2021getsf}. 
As suggested by \citet{Xu2023SDC335}, the \getsf~algorithm is a better choice than \texttt{astrodendro} in the case study of SDC335 (one of the ASSEMBLE sample), because it can: 1) deal with uneven background and rms noise; 2) can separate the blended sources/filaments; 3) extract extended emission features. 

We perform the \getsf~algorithm on the continuum emission maps without primary beam correction (\unpbcor). The \unpbcor~map is firstly smoothed into one with a circular beam whose size is equal to the major axis of the original beam. The \getsf~is set to extract sources whose sizes should be larger than the beam size but smaller than the MRS. As suggested by \citet{Men2021getsf}, significantly detected sources are defined as: 1) signal-to-noise ratio larger than unity; 2) peak intensity at least five times larger than the local intensity noise; 3) total flux density at least five times larger than the local flux noise; 4) ellipticity not larger than 2 to ensure a core-like structure; 5) footprint-to-major-axis ratio larger than 1.15 to rule out cores with abrupt boundary emission. After core extraction as well as fundamental measurement by \getsf, two flux-related parameters (integrated flux and peak intensity) are corrected by the primary beam response, depending on the core location in the continuum emission maps with primary beam correction (\pbcor). Fundamental measurements of the core parameters are listed in Table\,\ref{tab:fitcore}. 

To evaluate how much flux is recovered by the ALMA observations, we integrate the ATLASGAL 870\,$\mu$m flux over the field of view of the ASSEMBLE clumps. If all the sources and filaments extracted by \getsf~are included, then the recovered flux by ALMA ranges from 10\% to 25\%. Although the flux recovery can be further improved by including short-baseline observations (e.g., the Atacama Compact Array), some SMA/ALMA observations show a typical flux recovery between 10\% to 30\% \citep[e.g.,][]{Wang2014Snake,Sanhueza2017G28,Liu2018CMF,Sanhueza2019ASHES}. In our case, the maximum recoverable scale is $\sim9$\arcsec. Therefore, most of the mass in the massive clump is not confined in dense structures (cores). 

\begin{figure*}[!ht]
\centering
\includegraphics[width=0.48\linewidth]{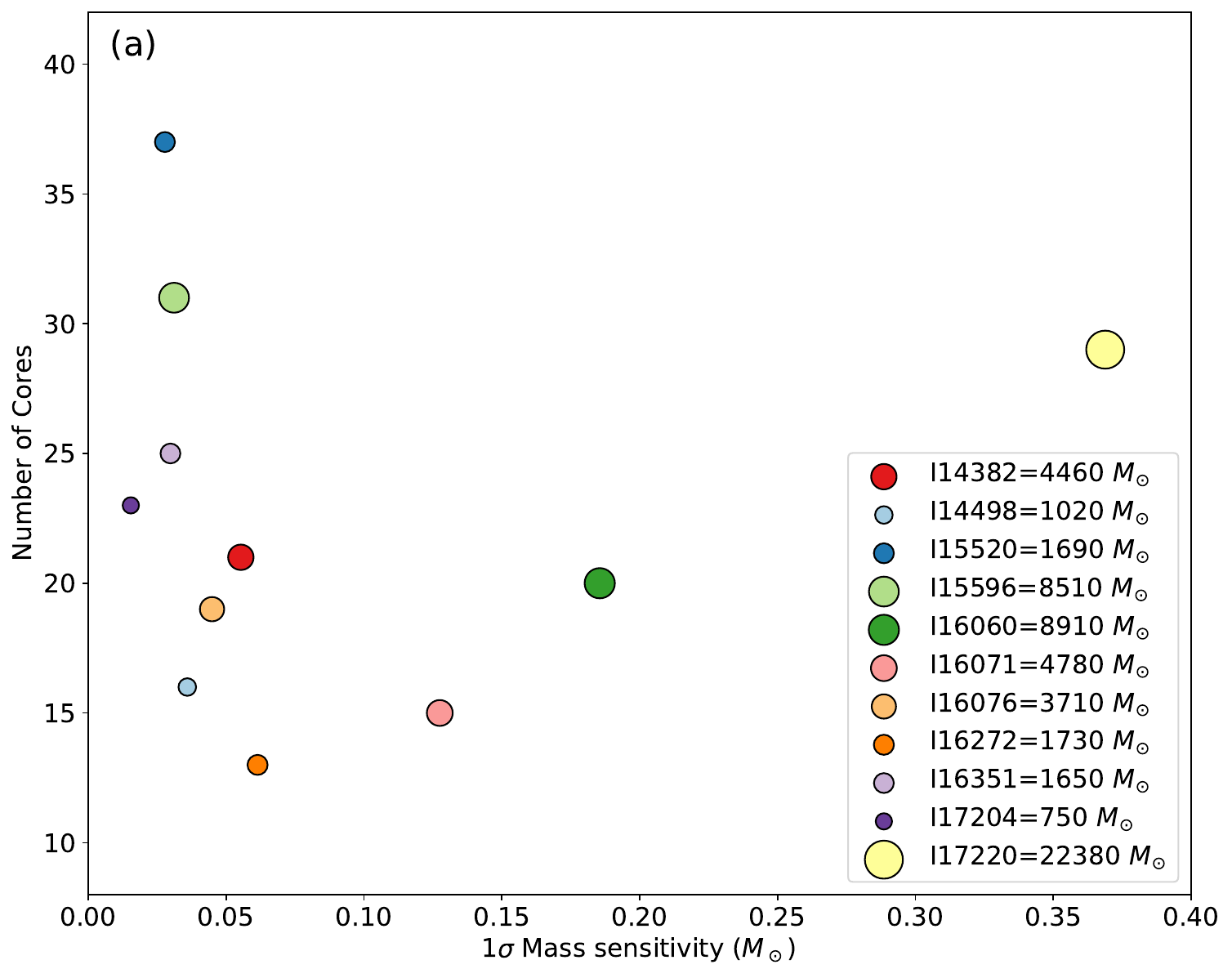}
\includegraphics[width=0.48\linewidth]{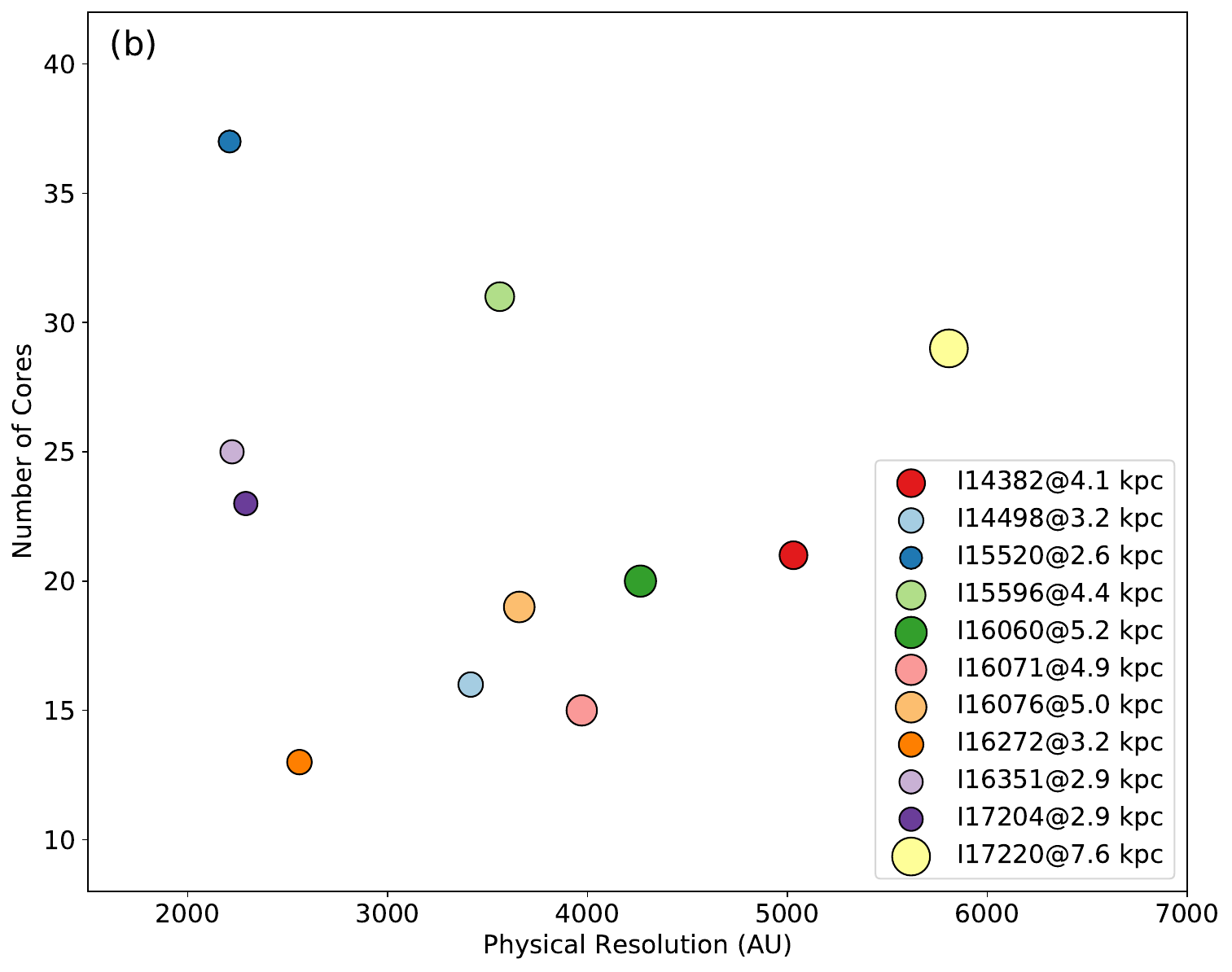}
\caption{(a) Number of cores detected against the $1\sigma$ mass sensitivity. The mass of clump is coded as the size of circle. (b) Number of cores detected against the physical resolution. The distance of clump is coded as the size of circle. \label{fig:det}}
\end{figure*}

As shown in panel (a) of Figure\,\ref{fig:det}, we found no overall correlation between the number of detected cores and the mass sensitivities with a Pearson correlation coefficient of 0.08. Likewise, panel (b) reveals no overall correlation between the number of detected cores and the physical resolution with a Pearson correlation coefficient of -0.04. Therefore, the number of detected cores is basically independent of the mass sensitivity and spatial resolution provided by the observations.


\begin{deluxetable*}{cccccccccc}
\tabletypesize{\small}
\tablewidth{4pt}
\tablecaption{Fundamental Measurements of Core Parameters from \getsf \label{tab:fitcore}}
\tablehead{
\multirow{2}{*}{\makecell[c]{ASSEMBLE \\ Clump}} & \multirow{2}{*}{\makecell[c]{Core \\ Name}} & \multicolumn{2}{c}{Position} & \colhead{Peak Intensity} & \colhead{Integrated Flux} & \colhead{$\theta_{\rm maj}\times\theta_{\rm min}$} & \colhead{PA} & \colhead{$\theta_{\rm deconv}$} & \multirow{2}{*}{\makecell[c]{Core \\ Classification\tablenotemark{a}}} \\
\cline{3-4}
 & & \colhead{$\alpha$(J2000)} & \colhead{$\delta$(J2000)} & \colhead{(\mjybeam)} & \colhead{(mJy)} & \colhead{($\arcsec\times\arcsec$)} & \colhead{(\degree)} & \colhead{(\arcsec)} & 
}
\colnumbers
\startdata
I14382 & 1 & 14:42:01.91 & -60:30:09.7 & 25.1(2.2) & 29.8(1.7) & 1.33$\times$1.13 & 100.7 & 1.23 & 0 \\
I14382 & 2 & 14:42:02.50 & -60:30:10.3 & 38.9(5.6) & 72.3(5.7) & 2.24$\times$1.40 & 93.7 & 1.3 & 0 \\
I14382 & 3 & 14:42:03.63 & -60:30:10.4 & 51.2(7.2) & 51.2(5.6) & 1.35$\times$1.11 & 58.7 & 1.23 & 0 \\
I14382 & 4 & 14:42:02.95 & -60:30:13.9 & 12.2(3.8) & 13.2(2.9) & 1.47$\times$1.18 & 142.7 & 1.23 & 0 \\
I14382 & 5 & 14:42:02.15 & -60:30:17.8 & 10.3(1.6) & 10.8(1.2) & 1.58$\times$1.01 & 72.7 & 1.23 & 0 \\
\enddata
\tablecomments{ASSEMBLE clump and extracted core ID are listed in (1) and (2). The core IDs are in order from the north to the south. The equatorial coordinate centers of the cores are listed in (3)--(4). The peak intensity and integrated flux are listed in (5)--(6). The fitted FWHM of the major and minor axes convolved with the beam and the position angle (anticlockwise from the north) are listed in (7)--(8). The deconvolved FWHM of the core size is shown in (9). The core classification in (10) is based on Section\,\ref{result:coreclass}. This table is available in its entirety in machine-readable form. }
\tablenotetext{a}{Core classification: 0 =  prestellar candidate, 1 = only molecular outflow is detected, 2 = only warm-core line is detected, 3 = both outflow and warm-core line are detected, and 4 = both outflow and hot-core line are detected, 5 = only hot-core line is detected.}
\end{deluxetable*}

\subsection{Core Classification and Evolutionary Stages} \label{result:coreclass}

\begin{figure*}
    \includegraphics[width=\linewidth]{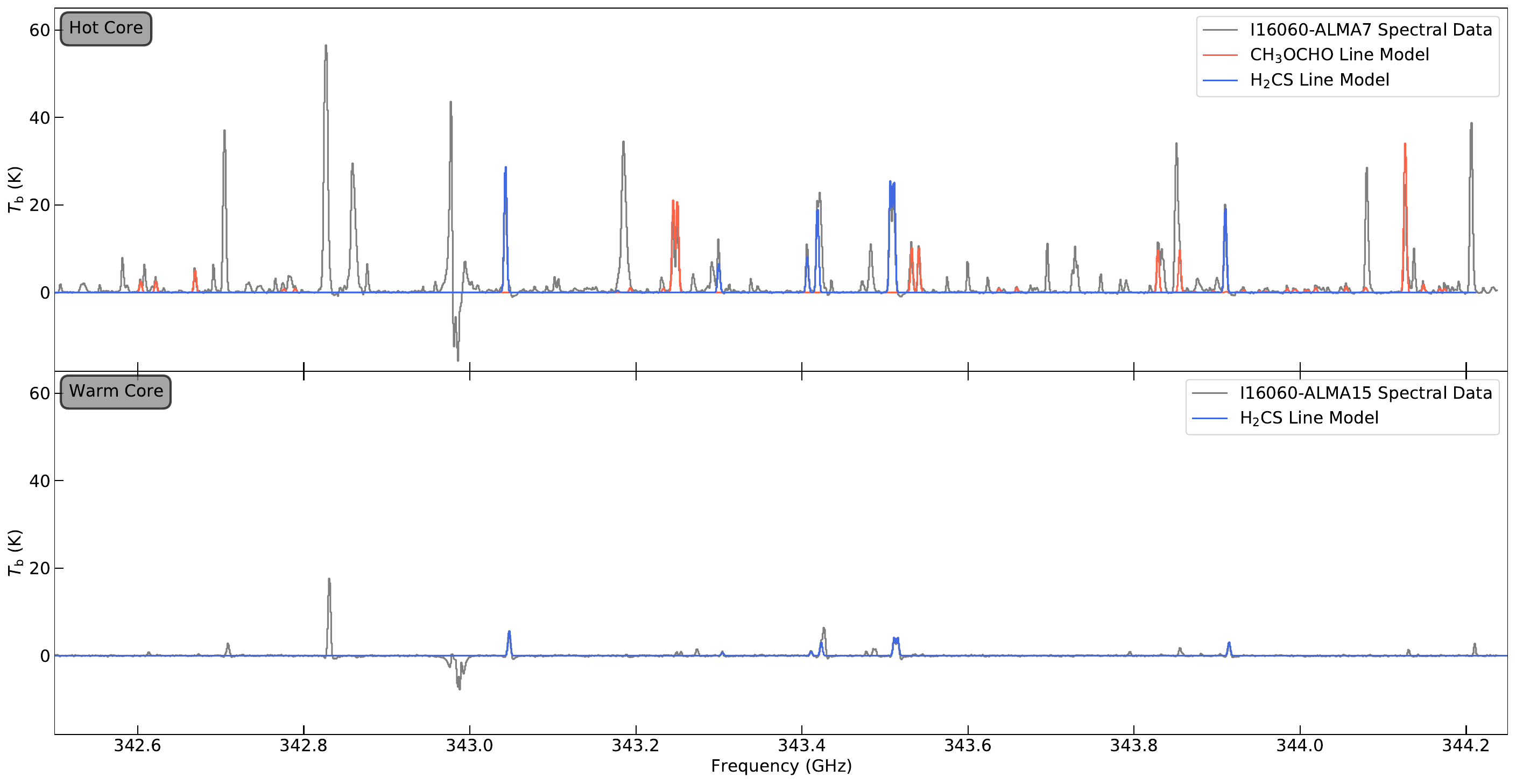}
    \caption{Examples of ``hot core'' and ``warm core'' spectra. The gray lines are the real spectral data extracted from I16060-ALMA7 and I16060-ALMA15 dense cores. The best-fit line models of \chtocho~and \htcs~are shown in red and blue, respectively. The core temperatures are assumed to be 112\,K and 89\,K for the hot and warm core, respectively. \label{fig:spectra}}
\end{figure*}

All the ASSEMBLE clumps have infrared bright signatures. As an example of a relatively early stage, I16272-4837 has extended 4.5\,$\mu$m emission, which is a common feature of outflows \citep{Cyganowski2008EGO}. A more evolved example of I14382-6017 is totally immersed in a cometary \hii~region traced by the PAH emission in the 8\,$\mu$m emission. Therefore, at least some cores in each clump are in an active star formation stage. 

The classification of the evolutionary stages of the 248 cores is based on the identification of star-formation indicators, including molecular outflows, \htcs~multiple transition lines, and \chtocho~multiple transition lines. 

For molecular outflows, \citet{Baug2020Outflow} used CO\,(3-2), HCN\,(4-3), and SiO\,(2-1) emission lines to confirm the presence of 32 bipolar and 41 unipolar outflows in the 11 ASSEMBLE clumps, and then a total of 42 continuum cores are associated with outflows. In this study, we updated the outflow catalogs by a channel-by-channel analysis of the outflow lobes to determine their association with the extracted cores, and subsequently assigned the outflows accordingly. A total of 39 ($\sim16$\%) cores are assigned bipolar or unipolar outflows. If a core is assigned outflows, then it is classified as protostellar \citep{Nony2023IMF-V}. Some cores even show multi-polar outflows \citep[e.g., I16272-4837 ALMA8;][]{Olguin2021DIHCA-I}, indicating either precession of accreting and outflowing protostars or presence of multiple outflows from multiple system. However, we should acknowledge that the method will miss those weak outflows associated with the lowest mass objects, especially for the more distant regions, naturally yielding a lower limit in the number of protostellar objects. 

Owing to its comparatively abundant nature, the emission of \htcs~is observed extensively in the core population (Chen et al., in preparation). However, the relatively low abundance of \chtocho~species restricts its detection to hot molecular cores with line-rich features. In this paper, we first classify those cores with robust ($>3\sigma$) detections of both \chtocho~and \htcs~multiple transition lines as ``hot cores'', especially those that have robust rotation temperature estimation by both \chtocho~and \htcs~molecules. Since the ``hot cores'' are believed to be the result of warm-up processes by central protostar(s) to 100--300\,K \citep{Gieser2019}, there should be a stage of dense cores with temperature of $<100$\,K and without line-rich features, which are called ``warm cores'' \citep{Sanhueza2019ASHES}. Then we define cores with only robust detection of \htcs~but without detection of \chtocho~lines as warm cores. We present examples of both ``hot cores'' and ``warm cores'' in Figure\,\ref{fig:spectra}, where I16060-ALMA7 is a typical hot core with line-rich feature, and evident detection of multiple transitions of \chtocho, as well as \htcs. However, I16060-ALMA15 has a paucity of hot molecular lines, including \chtocho, but with only \htcs. 

Among the 248 ASSEMBLE cores, \htcs~line emissions have been identified in 92 cores, of which 35 display ``line-rich'' features and are further categorized as hot cores, while the other 57 cores are classified as ``warm cores'' based on the detection of enough \htcs~lines. Among these warm cores, 22 have insufficient \htcs~transitions available for the calculation of temperature. 142 core without the star-forming indicators mentioned above (outflows, \htcs, or \chtocho~lines) are then classified as a prestellar core candidate, implying a stage preceding the protostellar phase. Based on the classification above, we mark the core in the column (10) of Table\,\ref{tab:fitcore}: 0 = prestellar candidate, 1 = only molecular outflow is detected, 2 = only \htcs~line is detected, 3 = both outflow and \htcs~line are detected, and 4 = both \chtocho~line and outflow are detected, 5 = only \chtocho~line is detected. 

Caveats of the core classification results: 1) external heating by hot cores in the vicinity can also excite \htcs~lines in some prestellar cores, so some warm cores can have no stars form inside; 2) prestellar core candidates may include both pre-protostellar cores that are gravitational bound, and cores that are not bound and unable to form star. To keep consistent, we don't distinguish the two and refer to them as prestellar core candidates in the following part of the paper. We note that spectral analyses of these cores can further constrain their dynamic states.

It is noteworthy that outflows have been observed in all of the ASSEMBLE clumps, providing evidence of star-forming activities with a 100\% occurrence rate in our clump sample. However, two massive star-forming clumps, namely I14382-6017 and I17204-3636, do not exhibit any detection of hot cores. This absence of hot cores has been confirmed by cross-matching with the ALMA Band-3 dataset, ensuring their non-existence \citep{Qin2022ATOMS-VIII}. In the case of the protocluster I14382-6017, the extended spherical morphology of H40$\alpha$ line emission is spatially consistent with the MeerKAT Galactic Plane Survey 1.28\,GHz data \citep[Goedhart et al. in prep.]{Padmanabh2023MeerKAT}. As identified by \citet{Zhang2023HII}, it represents an UC\hii~region with an electron density of 0.15--0.16$\times10^{4}$ cm$^{-3}$. The protocluster I14382-6017 is situated on the outskirts of the UC\hii~region, suggesting the possibility of a second generation of cores (refer to Figure\,\ref{fig:radio}). As a result, the absence of hot cores in this particular region can be attributed to the relatively young age of the newly formed protocluster. The absence of hot cores in I17204-3636 can be a different issue, as the H40$\alpha$ and the 1.28\,GHz emission are spatially correlated with dense cores (see Figure\,\ref{fig:radio}). But we note that I17204-3636 has the lowest mass of 760\,\msun, with the maximum core mass of 2.9\,\msun~(refer to Section\,\ref{result:coreproperty}). Furthermore, the temperature of the only warm core I17204-ALMA16 is 88($\pm$7)\,K (Section\,\ref{result:coreproperty}), which is not so high as $\gtrsim100$\,K to be a hot core. Therefore, in the case of I17204-3636, the cores may not be massive and hot enough to excite hot molecular lines or initiate hot core chemistry. 

\subsection{Core Physical Properties} \label{result:coreproperty}


\begin{deluxetable*}{ccccccccc}
\tabletypesize{\small}
\tablewidth{4pt}
\tablecaption{Calculated Properties for the Core Sample \label{tab:calcore}}
\tablehead{
\multirow{2}{*}{\makecell[c]{ASSEMBLE \\ Clump}} & \multirow{2}{*}{\makecell[c]{Core \\ Name}} & \multicolumn{2}{c}{$T_{\rm dust}$} & \colhead{$M_{\rm core}$} & \colhead{$R_{\rm core}$} & \colhead{$n$\,(H$_2$)} & \colhead{$\Sigma$} & \colhead{$N_{\rm peak}$\,(H$_2$)} \\
\cline{3-4}
 & & \colhead{(K)} & \colhead{Method\tablenotemark{a}} & \colhead{(\msun)} & \colhead{(au)} & \colhead{($\times10^6$\,cm$^{-3}$)} & \colhead{(g\,cm$^{-2}$)} & \colhead{($\times10^{23}$\,cm$^{-2}$)}
 }
\colnumbers
\startdata
I14382 & 1 & 28(5) & G & 1.3(0.9) & 2200 & 3.71(2.43) & 0.44(0.16) & 0.94(0.34) \\
I14382 & 2 & 28(5) & G & 3.2(2.1) & 4700 & 0.93(0.62) & 0.69(0.28) & 1.46(0.60) \\
I14382 & 3 & 28(5) & G & 2.3(1.6) & 2200 & 6.38(4.37) & 0.90(0.37) & 1.92(0.78) \\
I14382 & 4 & 28(5) & G & 0.6(0.4) & 2200 & 1.64(1.24) & 0.21(0.12) & 0.46(0.25) \\
I14382 & 5 & 28(5) & G & 0.5(0.3) & 2200 & 1.35(0.92) & 0.18(0.08) & 0.39(0.16) \\
\enddata
\tablecomments{ASSEMBLE clump and extracted core ID are listed in (1) and (2). Dust temperature and its estimation methods are listed in (3) and (4). The mass, radius, volume density, surface density, and peak column density are listed in (5)--(9).  This table is available in its entirety in machine readable form. 
\tablenotetext{a}Temperature estimation method: G = global clump-averaged temperature in column (9) of Table\,\ref{tab:sample}; H = \htcs~rotation temperature; C = \chtocho~rotation temperature.}
\end{deluxetable*}


\begin{deluxetable*}{ccccccccccc}
\tabletypesize{\small}
\tablewidth{0pt} 
\linespread{1.1} 
\tablecaption{Statistics of the ALMA Cores in Each Clump
\label{tab:corestats}}
\tablehead{
\multirow{2}{*}{\makecell[c]{ASSEMBLE \\ Clump}} & \multirow{2}{*}{\makecell[c]{$1\sigma$ Mass \\ Sensitivity}} & \multirow{2}{*}{\makecell[c]{Number \\ of Cores}} & \multicolumn{2}{c}{Core Mass} & \multicolumn{5}{c}{Mean Value of} & \multirow{2}{*}{\makecell[c]{Number of \\ Pre-/Proto-stellar \\ Cores}} \\
\cmidrule(r){4-5} \cmidrule(r){6-10}
\colhead{} & \colhead{} & \colhead{} & \colhead{Min.} & \colhead{Max.} & \colhead{Mass} & \colhead{Radius} & \colhead{$n$\,(H$_2$)} & \colhead{$\Sigma$} & \colhead{$N_{\rm peak}$\,(H$_2$)} \\
\colhead{} & \colhead{($M_\odot$)} & \colhead{} & \colhead{($M_\odot$)} & \colhead{($M_\odot$)} & \colhead{($M_\odot$)} & (au) & ($\times10^6$\,cm$^{-3}$) & (g\,cm$^{-2}$) & ($\times10^{23}$\,cm$^{-2}$) & \colhead{} 
}
\colnumbers
\startdata
I14382 & 0.044 & 21 & 0.39 & 4.65 & 1.73 & 2760 & 4.1 & 0.5 & 1.1 & 15/6 \\
I14498 & 0.030 & 16 & 0.37 & 9.70 & 2.25 & 3210 & 3.9 & 0.9 & 1.9 & 12/4 \\
I15520 & 0.026 & 37 & 0.12 & 3.75 & 0.99 & 2250 & 8.4 & 0.8 & 1.7 & 11/26 \\
I15596 & 0.029 & 31 & 0.46 & 8.14 & 2.14 & 3210 & 7.1 & 0.7 & 1.5 & 24/7 \\
I16060 & 0.182 & 20 & 1.67 & 52.61 & 14.34 & 4800 & 22.2 & 3.2 & 6.9 & 7/13 \\
I16071 & 0.128 & 15 & 1.57 & 49.03 & 8.82 & 3420 & 17.8 & 2.5 & 5.4 & 5/10 \\
I16076 & 0.046 & 19 & 0.56 & 12.07 & 2.41 & 3910 & 3.6 & 0.7 & 1.4 & 12/7 \\
I16272 & 0.054 & 13 & 0.17 & 19.28 & 4.58 & 2730 & 40.6 & 2.7 & 5.7 & 7/6 \\
I16351 & 0.029 & 25 & 0.18 & 2.76 & 1.28 & 2880 & 6.9 & 0.9 & 2.0 & 13/12 \\
I17204 & 0.014 & 23 & 0.22 & 2.89 & 1.05 & 2900 & 7.4 & 0.7 & 1.5 & 22/1 \\
I17220 & 0.372 & 28 & 5.37 & 52.27 & 19.57 & 6190 & 7.9 & 2.3 & 4.8 & 14/14 \\
\enddata
\tablecomments{ASSEMBLE clump is listed in (1). The $1\sigma$ mass sensitivity and the number of extracted cores are listed in (2) and (3). The minimum, maximum, and mean values of the mass are listed in (4)--(6). The mean values of the core radius, volume density, surface density, and peak column density are listed in (7)--(10). The numbers of prestellar and protostellar cores are listed in (11).}
\end{deluxetable*}

Temperature estimation utilizes three hybrid methods (clump-averaged temperature, \htcs~line, and \chtocho~line) based on the core properties. \htcs~lines are chosen due to their strong spatial correlation with dust as demonstrated in \citet{Xu2023SDC335}, and their widespread distribution (Chen et al. submitted). 
The ASSEMBLE spectral window encompasses multiple hyperfine components of the $J=10-9$ transitions, with upper energy levels from 90 to 420\,K \citep[see Table C1 in][]{Xu2023SDC335}. However, \htcs~lines could be optically thick towards massive hot cores, therefore only tracing the core envelope. To trace the dust temperature of hot cores, \chtocho~molecule with upper energy up to $\sim589$\,K is employed instead. Temperatures obtained from \chtocho~(mean value of 110\,K) are consistently higher than those derived from \htcs~(mean value of 95\,K), indicating that \chtocho~is a suitable tracer of the inner and denser gas. In cases where neither \htcs~nor \chtocho~lines are detected, it is assumed that the core either lacks sufficient column density or is too cold to excite the lines. This suggests that the core has not developed its own temperature gradient and thus is assumed to share the same temperature as the clump from the SED fitting. The temperature as well as the method to obtain it are listed in the column (3)--(4) of Table\,\ref{tab:calcore}. 

Assuming that all the emission comes from dust in a single $T_{\rm dust}$ and that the dust emission is optically thin, the core masses are then calculated using
\begin{equation}\label{eq:coremass}
	M_{\mathrm{core}} = \mathcal{R}\frac{F^\mathrm{int}_\nu D^2}{\kappa_\nu B_\nu (T_\mathrm{dust})},
\end{equation}
where $F^\mathrm{int}_\nu$ is the measured integrated dust emission flux of the core, $\mathcal{R}$ is the gas-to-dust mass ratio (assumed to be 100), $D$ is the distance, $\kappa_{\nu}$ is the dust opacity per gram of dust, and $B_\nu (T_\mathrm{dust})$ is the Planck function at a given dust temperature $T_\mathrm{dust}$. In our case, $\kappa_{\nu}$ is assumed to be 1.89\,cm$^2$\,g$^{-1}$ at $\nu\sim350$\,GHz \citep{Xu2023SDC335}, which is interpolated from the given table in \citet{O&H1994dust}, assuming grains with thin ice mantles and the MRN \citep{MRN1997} size distribution and a gas density of $10^6$\,cm$^{-3}$. Substituting the temperature in Equation\,\ref{eq:coremass}, the core masses are then calculated and listed in the column (5). 

Cores are characterized by 2D Gaussian-like ellipses with the FWHM of the major and minor axes ($\theta_{\rm maj}$ and $\theta_{\rm min}$), and position angle (PA) listed in the column (7)--(8) of Table\,\ref{tab:fitcore}. Following \citet{Rosolowsky2010BGPSII} and \citet{Contreras2013ATLASGAL}, the angular radius can be calculated as the geometric mean of the deconvolved major and minor axes:
\begin{equation}\label{eq:theta_core}
    \theta_{\rm core} = \eta\left[\left(\sigma^2_{\rm maj}-\sigma^2_{\rm bm}\right)\left(\sigma^2_{\rm min}-\sigma^2_{\rm bm}\right)\right]^{1/4},
\end{equation}
where $\sigma_{\rm maj}$ and $\sigma_{\rm min}$ are calculated from $\theta_{\rm maj}/\sqrt{8\ln2}$ and $\theta_{\rm min}/\sqrt{8\ln2}$ respectively. 
The $\sigma_{\rm bm}$ is the averaged dispersion size of the beam (i.e., $\sqrt{\theta_{\rm bmaj}\theta_{\rm bmin}/(8\ln2)}$ where $\theta_{\rm bmj}$ and $\theta_{\rm bmin}$ are the FWHM of the major and minor axis of the beam). $\eta$ is a factor that relates the dispersion size of the emission distribution to the angular radius of the object determined. 
We have elected to use a value of $\eta=2.4$, which is the median value derived for a range of models consisting of a spherical, emissivity distribution \citep{Rosolowsky2010BGPSII}. Therefore, the core physical radius can be directly calculated by $R_{\rm core} = \theta_{\rm core}\times D$, as shown in the column (6) of Table\,\ref{tab:calcore}.

The number density, $n$, is then calculated by assuming a spherical core,
\begin{equation} \label{eq:volume_density}
    n = \frac{M_{\rm core}}{(4/3) \pi \mu_{\ssstyle \rm H_2} m_{\ssstyle \rm H}R_{\rm core}^3},
\end{equation}
where $\mu_{\ssstyle \rm H_2}$ is the molecular weight per hydrogen molecule and $m_{\rm H}$ is the mass of a hydrogen atom. Throughout the paper, we adopt the molecular weight per hydrogen molecule $\mu_{\ssstyle \rm H_2} = 2.81$ \citep{Evans2022SlowSF}, and derive the number density of hydrogen molecule $n$(H$_2$). 

The core-averaged surface density can be calculated by $\Sigma=M_{\rm core}/(\pi R^2_{\rm core})$. The peak column density is estimated from
\begin{equation}
    N_{\rm peak}\,(\mathrm{H}_2) = \mathcal{R} \frac{F^{\rm peak}_{\nu}}{\Omega\mu_{\ssstyle \rm H_2} m_{\ssstyle \rm H}\kappa_{\nu}B_{\nu}(T_{\rm dust})},
\end{equation}
where $F^{\rm peak}_{\nu}$ is the measured peak flux of core within the beam solid angle $\Omega$\footnote{beam solid angle: $\Omega=\frac{\pi\theta_{\rm maj}\theta_{\rm min}}{4\ln(2)}$}. The calculated volume, surface and peak column densities are shown in (7)--(9) of Table\,\ref{tab:calcore}.

The major sources of uncertainty in the mass calculation come from the gas-to-dust ratio and the dust opacity. We adopt the uncertainties derived by \citet{Sanhueza2017G28} of 28\% for the gas-to-dust ratio and of 23\% for the dust opacity, contributing to the $\sim36$\% uncertainty of the specific dust opacity. The uncertainties of the core flux ($\sim$14\%), temperature ($\sim$20\%), and distance (assumed to be 10\%) are included. 
Monte Carlo methods are adopted for uncertainty estimation and $1\sigma$ confidence intervals are given for core mass, volume density, surface density and peak column density in (5), (7)--(9) of Table\,\ref{tab:calcore}. 

We also summarize the statistics of the core physical parameters in Table\,\ref{tab:corestats}. The number of cores in each clump is listed in column (3). The minimum, maximum, and mean core mass are listed in columns (4--6). The mean values of core radius, volume density, surface density, and column density are listed in columns (7--10). The numbers of prestellar and protostellar cores are listed in column (11).

\section{Discussion} \label{sec:discuss}

\subsection{Coevolution of Clump and Most Massive Core} \label{discuss:coevolve}

\begin{figure*}[!ht]
\centering
\includegraphics[width=0.98\linewidth]{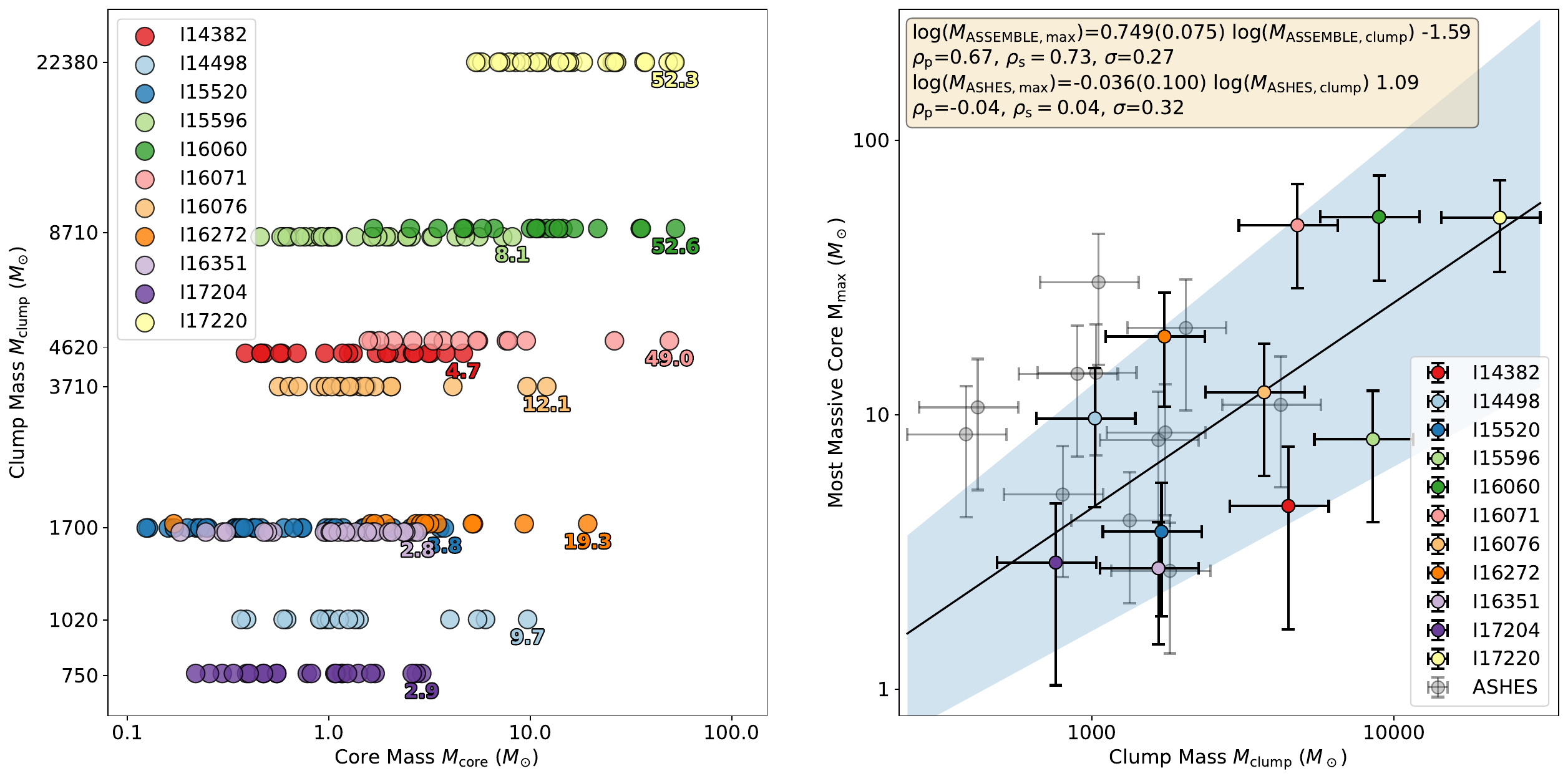}
\caption{Left: Core masses $M_{\rm core}$ versus the clump masses $M_{\rm clump}$. The masses of the most massive cores $M_{\rm max}$ in each clump are labeled with the corresponding colors. Right: The scaling relation between $M_{\rm max}$ and $M_{\rm clump}$, with the result of linear regression shown on the top left. The shaded area shows the $2\sigma$ uncertainty of the fitting result of the ASSEMBLE sample. The gray points from the ASHES Pilots show no correlation. \label{fig:coremass}}
\end{figure*}

The ability of a clump to form massive stars is directly linked to the amount of material within the natal clump \citep{Beuther2013IRDC18310}. Therefore, it is essential and straightforward to study the relation between clump and its most massive core (MMC), which is most likely to form massive stars inside the clump. The left panel of Figure\,\ref{fig:coremass} shows the core masses ($M_{\rm core}$) versus the mass of the clump ($M_{\rm clump}$) of the ASSEMBLE clumps, with the maximum value, that is, the mass of MMC ($M_{\rm max}$) labeled. As demonstrated in the right panel, a positive sublinear correlation is observed between $M_{\rm ASSEMBLE, max}$ and $M_{\rm ASSEMBLE, clump}$, with a power law index of 0.75(0.08). The Pearson and Spearman correlation coefficients are calculated to be 0.67 and 0.73, respectively. Significantly, both correlation coefficients exhibit p-values below 0.05, indicating a high level of statistical significance for the observed correlation. This positive correlation indicates a coevolution between the clump and MMC, i.e., a more massive clump contains a more massive core, which is consistent with what has been found in \citet{Anderson2021HFS}. 

Furthermore, the coevolution of the massive clump and its most massive core can be connected to gas kinematics in a dynamic picture. In massive star-forming regions, filamentary gas accretion flows frequently connect clump and core scales in both observations \citep{Peretto2013SDC335,Peretto2014SDC13,Liu2016Wide,Lu2018Filament,Yuan2018G22,Dewangan2020HFS,Sanhueza2021MagMAR,Li2022NGC6334S,Xu2023SDC335,Yang2023G310} and simulations \citep{Schneider2010DR21,NR2022Simulation}, which can play a crucial role in regulating mass reservoirs at different scales. Notably, \citet{Xu2023SDC335} found four spiral-like gas streams conveying gas from the natal clump directly to the most massive core, with a continuous and steady gas accretion rate across three magnitude. Therefore, we suggest that such a ``conveyor belt'' \citep{Longmore2014YMC} should be the main reason for coevolution. If all the massive clumps are undergoing a quick mass assembly, the sublinearity of the mass scaling relation also suggests that the clump-to-cores efficiency should vary among different clumps (Xu et al., in preparation). To more directly understand the dynamic picture of coevolution of clump and core, detailed gas kinematics analyses should be systematically performed in a sample with a wide range of evolutionary stages. 

\subsection{High-mass Prestellar Cores in Protoclusters?} \label{discuss:prestellar}
High-mass prestellar cores, defined as cores with masses greater than 30\,\msun\ (following the definition of \citet{Sanhueza2019ASHES}), are crucial in discriminating between different models of high-mass star formation. Specifically, they provide a key discriminator for the turbulent core accretion model \citep{Mckee2003TurbulentCore,Tan2013TurbulentCore,Tan2014PPVI} versus the competitive accretion model \citep{Bonnell2001Simulation,Bonnell2004Simulation} or the global hierarchical collapse model \citep{Heitsch2008GHC,VS2009GHC,VS2017GHC,BP2011GHC,BP2018GHC}. Despite numerous observational searches for high-mass prestellar cores \citep[e.g.,][]{Zhang2009Fragmentation,Zhang2011IRDC30,Wang2012Heating,Wang2014Snake,Cyganowski2014G11,Kong2017Hunter,Sanhueza2017G28,Louvet2018Review,Molet2019W43MM1,Svoboda2019HMSC,Sanhueza2019ASHES,Li2019IRDC,Morii2021G23}, only a few promising candidates have been identified to date: including G11P6-SMA1\citep{Wang2014Snake} and G28-C2c1a \citep{Barnes2023Dragon}. The rarity of high-mass prestellar cores suggests either that the initial fragmentation of massive clumps does not produce such massive starless cores or that these objects have short lifetimes.

It is worth noting that most of the efforts in the search for massive starless cores have been focused on IRDCs. However, several numerical simulations suggest that thermal feedback from OB protostars and strong magnetic field proto-stellar clusters can play a crucial role in reducing the level of further fragmentation and producing more massive dense cores \citep{Offner2009Simulation,Krumholz2007Simulation,Krumholz2011Simulation,Myers2013Simulation},
and hints for such a reduction of fragmentation for strong magnetic fields have actually been suggested observationally \citep{Palau2021Fragmentation}. Observations also suggest that a 5\,\msun~zero-age main sequence (ZAMS) star can produce radiation feedback to support high-mass fragments \citep{Longmore2011G8}. In particular, massive starless core candidates such as G9.62+0.19MM9 \citep{Liu2017G962} and W43-MM1\#6 \citep{Nony2018W43MM1} have been found in evolved protostellar clusters. 
Moreover, \citet{Contreras2018Infall} reported a relatively massive but highly subvirial collapsing prestellar core with mass 17.6\,\msun, that is heavily accreting from its natal cloud at a rate of $1.96\times10^{-3}$\,\massrate. If the accretion rate persists during the lifetime of the massive starless clump ($\lesssim1-3\times10^4$\,yr), then the mass of the prestellar core can be doubled at the beginning of the protostellar stage. Therefore, it would be even more promising to search for high-mass prestellar cores in protostellar clusters than in prestellar clusters.

Within the ASSEMBLE protoclusters, the most massive prestellar core I17220-ALMA9 has a mass of 18.3\,\msun~within 0.065\,pc, which is about two times larger than the ones found in the ASHES IRDCs. The second massive prestellar core I16060-ALMA17 has a mass of 16.5\,\msun~within 0.045\,pc. However, we should note that 1) the ASSEMBLE data only have the ALMA 12-m array configuration, so the core flux can be underestimated with extended flux filtered out; 2) we adopt the clump-averaged temperature as the temperature of the prestellar core, which can be overestimated, resulting in an underestimated core mass. Complementary short-baseline configuration and a better estimation of temperature should give a better estimate of the prestellar core mass. At any rate, the available evidence strongly suggests that: 1) prestellar cores are becoming more massive, which can be due to the continued mass accumulation along with the natal clump (see Section\,\ref{evolution:coregrow}); 2) high-mass prestellar cores can survive in protostellar clusters. However, to demonstrate the causality between the survival of high-mass prestellar cores and the protocluster environment, both a systematic search for high-mass prestellar cores in massive protoclusters and determination of environmental effects are needed. 

\subsection{Core Separation} \label{discuss:separation}

To study the spatial distribution of cores, we first build the minimum spanning tree (MST) for each ASSEMBLE core cluster; and the details can be found in Appendix\,\ref{app:mst}. 

\begin{figure}[!ht]
\centering
\includegraphics[width=1.0\linewidth]{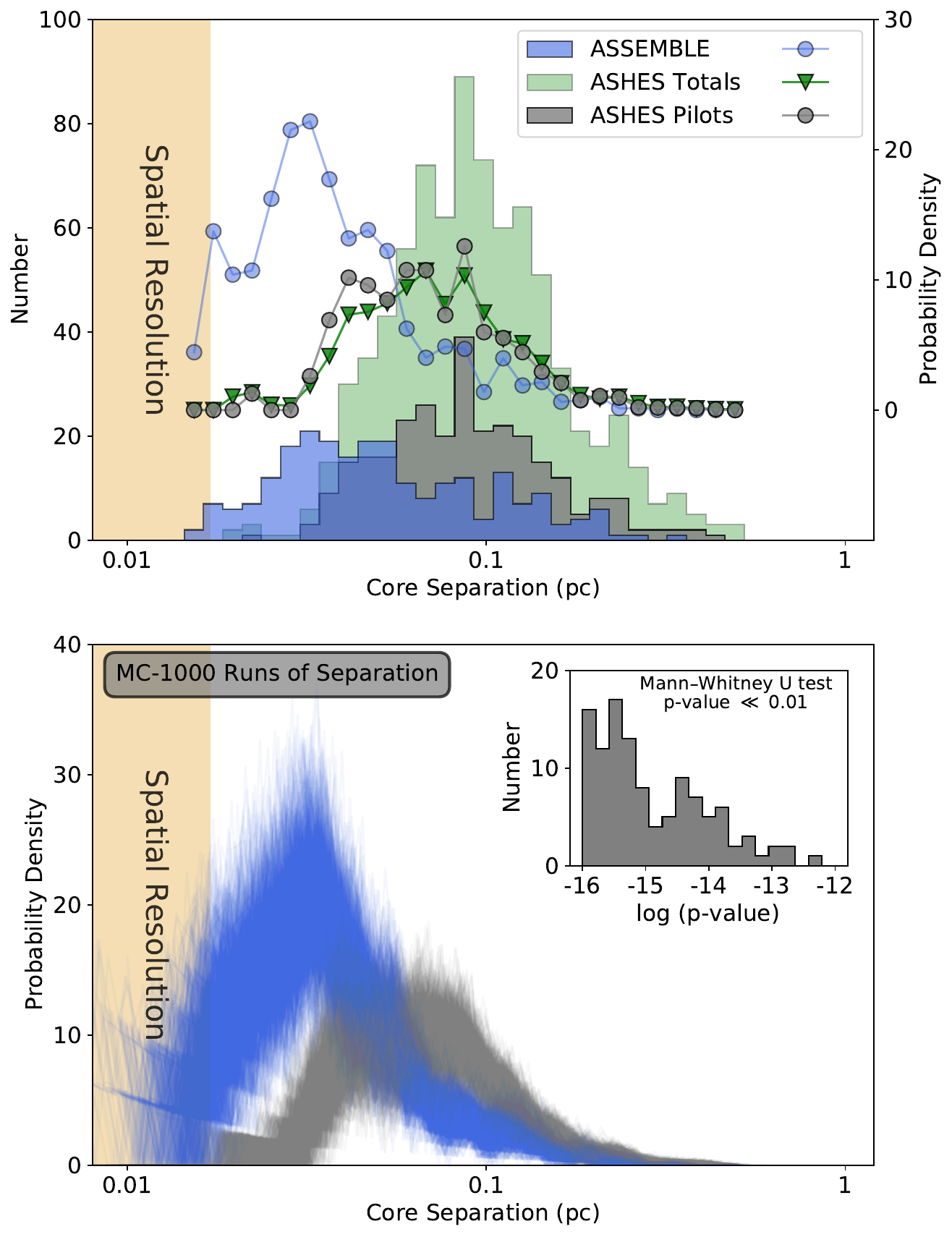}
\caption{Upper: the number distribution (indicated by stacked histogram) and probability density distribution (indicated by line-connected points) of core separation are presented in a logarithmic scale, where the ASSEMBLE, the ASHES Total \citep{Morii2023ASHES}, and the ASHES Pilot \citep{Sanhueza2019ASHES} samples are presented in blue, green, and gray colors, respectively. The mean spatial resolution of both the ASSEMBLE and the ASHES surveys are $\sim0.02$\,pc, shown with orange shadow. Lower: the 1000 Monte Carlo runs of the probability density distribution of core separation for the ASSEMBLE (blue lines) and the ASHES (gray lines), respectively, considering the Gaussian-like uncertainty of clump distance. The Mann-Whitney U test is performed on each of the sets of core separation distributions, and the distribution of the p-value is shown in the top right. The p-values are much lower than 0.01, showing that two samples share a significantly different distribution of core separation. \label{fig:separation}}
\end{figure}

Following the convention of \citet{Wang2016Filament,Sanhueza2019ASHES}, we take the ``edge'' of MST as the separation between the cores. A total of $N-1$ separation lengths are defined in each clump where $N$ is the core number. The upper panel of Figure\,\ref{fig:separation} shows the distributions of core separation of the ASSEMBLE sample in blue, the ASHES total sample \citep[ASHES Totals;][]{Morii2023ASHES} in green color, and the ASHES pilot sample \citep[ASHES Pilots][]{Sanhueza2019ASHES} in gray color, respectively. When normalized into probability density as shown with line-connected scatter plot, the Kolmogorov–Smirnov test between the distributions of the ASHES Totals and Pilots give a p-value of 0.57 $\gg0.1$, indicating that the ASHES Pilots share the same distribution with that of the ASHES Totals. Therefore, the ASHES Pilots are good enough to represent the ASHES Totals in the case of studying core separation. Since the sample size of the ASHES Pilots is comparable to that of the ASSEMBLE, we only compare core separations from the ASSEMBLE with those from the ASHES Pilots in the following analyses. 

The bias of the mass sensitivity and spatial distribution should be excluded. For example, if the ASSEMBLE mass sensitivity is higher than the ASHES one, we are about to detect more low-mass cores, reducing the separation. Thanks to comparable sensitivities of the two samples, we have detected the core population with the same truncation limited by the mass sensitivity. In addition, the ASSEMBLE and ASHES surveys share similar spatial resolutions, as indicated by the orange shadows, and therefore we can directly compare their core separations.

The Mann–Whitney U test \footnote{The Mann-Whitney U Test is a null hypothesis test, used to detect differences between two independent data sets. The test is specifically for non-parametric distributions, which do not assume a specific distribution for a set of data \citep{Mann-Whiteney-U-test}. Because of this, the Mann-Whitney U Test can be applied to any distribution, whether it is Gaussian or not.} between two groups of core separations gives a p-value $\ll0.01$, significantly excluding the null hypothesis that two distributions are the same. To further test the effects of the uncertainty of the clump distance, 1000 Monte Carlo runs are adopted to simulate the 1-$\sigma$ distribution dispersion, as shown in the blue and gray extent in the lower panel of Figure\,\ref{fig:separation}. The distribution of the p-value derived from the Mann-Whitney U test is shown with the subpanel on the upper right corner in the lower panel. Even perturbed by 1-$\sigma$ uncertainty from distance ($\sim$10--20\%), the majority of p-values are significantly lower than 0.01, suggesting that two distributions are truly different. In other words, the core separations in the ASSEMBLE protoclusters are systematically smaller than those in the ASHES protoclusters, suggesting that the cluster becomes tighter with closer separations during the clump evolution indicated by $L/M$. 

It should be noted that the ASSEMBLE core separation exhibits a significant peak at $\sim0.035$\,pc. The value is twice the spatial resolution (mean value of $\sim0.018$\,pc), suggesting it is not a result of resolution effects. Furthermore, both \citet{Tang2022W51} and \citet{Palau2018Fragmentation} have also observed two peaks in the separation histogram in W51 North and OMC-1S. One of these peaks falls within the range of 0.032 to 0.035\,pc, which aligns with the results we have obtained in our study. Such a consistency between three independent observations (with different spatial resolutions) might suggest a typical level of hierarchical fragmentation at this scale.

\subsection{The \texorpdfstring{$\mathcal{Q}$}{qvalue} Parameter} \label{discuss:Q}

To quantify the spatial distribution of cores, we follow the approach of \citet{Cartwright2004Cluster} and define the $\mathcal{Q}$ parameter as,
\begin{equation}
    \mathcal{Q} = \frac{\bar m}{\bar s},
\end{equation}
where $\bar m$ is the normalized mean edge length of the MST given by,
\begin{equation}
    \bar m = \sum^{N_c-1}_{i=1} \frac{L_i}{(N_c A)^{1/2}},
\end{equation}
where $N_c$ is the number of cores, $L_i$ is the length of each edge, and $A$ is the area of protocluster as $A = \pi R_{\rm cluster}^2$, with $R_{\rm cluster}$ calculated as the distance from the mean position of cores to the farthest core. $\bar s$ is the normalized correlation length,
\begin{equation}
    \bar s = \frac{L_{\rm av}}{R_{\rm cluster}},
\end{equation}
where $L_{\rm av}$ is the the mean separation length between all cores and $R_{\rm cluster}$ is the cluster radius.

The $\mathcal{Q}$ value serves as a measure of the degree of subclustering and the large-scale radial density gradient in a given region. As indicated by Fig. 5 in \citet{Cartwright2004Cluster}, a value of $\mathcal{Q} \gtrsim 0.8$ indicates a centrally condensed spatial distribution characterized by a radial density profile of the form $n(r)\propto r^{-\alpha}$. On the other hand, when $\mathcal{Q} \lesssim 0.8$, the $\mathcal{Q}$ parameter decreases from approximately 0.80 to 0.45 with an increasing degree of subclustering, ranging from a fractal dimension of $D = 3.0$ (representing a uniform number-density distribution without subclustering) to $D = 1.5$ (indicating strong subclustering).

\begin{figure}
    \centering
    \includegraphics[width=1.0\linewidth]{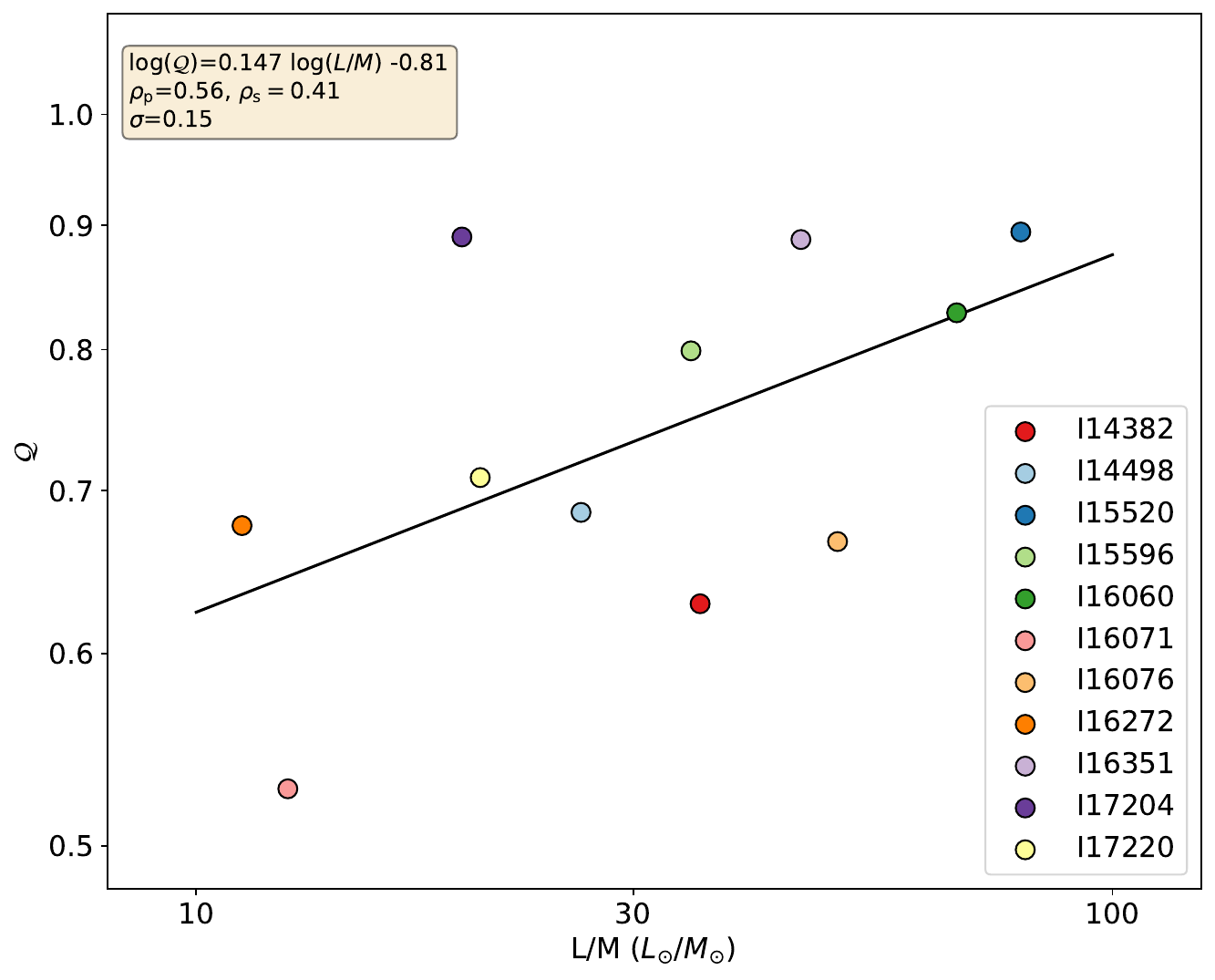}
    \caption{The $\mathcal{Q}$ values versus luminosity-to-mass ratio $L/M$ in the ASSEMBLE protoclusters. The linear regression results including the fitting model, Pearson correlation coefficient $\rho_{\rm p}$, Spearman rank correlation coefficient $\rho_{\rm s}$, and scatter $\sigma$ are shown on the upper left. \label{fig:Qpar_assemble}}
\end{figure}

From the MST results, the derived $\mathcal{Q}$ parameters for the ASSEMBLE clumps range from 0.53 to 0.89, with a median value\footnote{The uncertainty of median value $\sigma_{\rm med}$ is estimated from the median absolute deviation (MAD) as, $\sigma_{\rm med}\simeq1.4826\times$\,MAD, based on the assumption of normality.} of 0.71(0.13). We note that there are four protoclusters I15520, I16060, I16351, and I17204 that have $\mathcal{Q}$ greater than 0.8, indicative of a centrally condensed spatial distribution. As shown in Figure\,\ref{fig:Qpar_assemble}, the $\mathcal{Q}$ parameter shows a weak correlation with luminosity-to-mass ratio, with Pearson correlation coefficient $R_{\rm p}=0.56$. The positive correlation suggests that a protocluster is becoming more centrally condensed as it evolves. In Section\,\ref{evolution:Q}, a correlation among a sample of both ASSEMBLE and ASHES could be more instructive, since a wider dynamic range of $L/M$ is available.

\subsection{Mass Segregation} \label{discuss:segregation}

\begin{figure*}[!ht]
\centering
\includegraphics[width=0.98\linewidth]{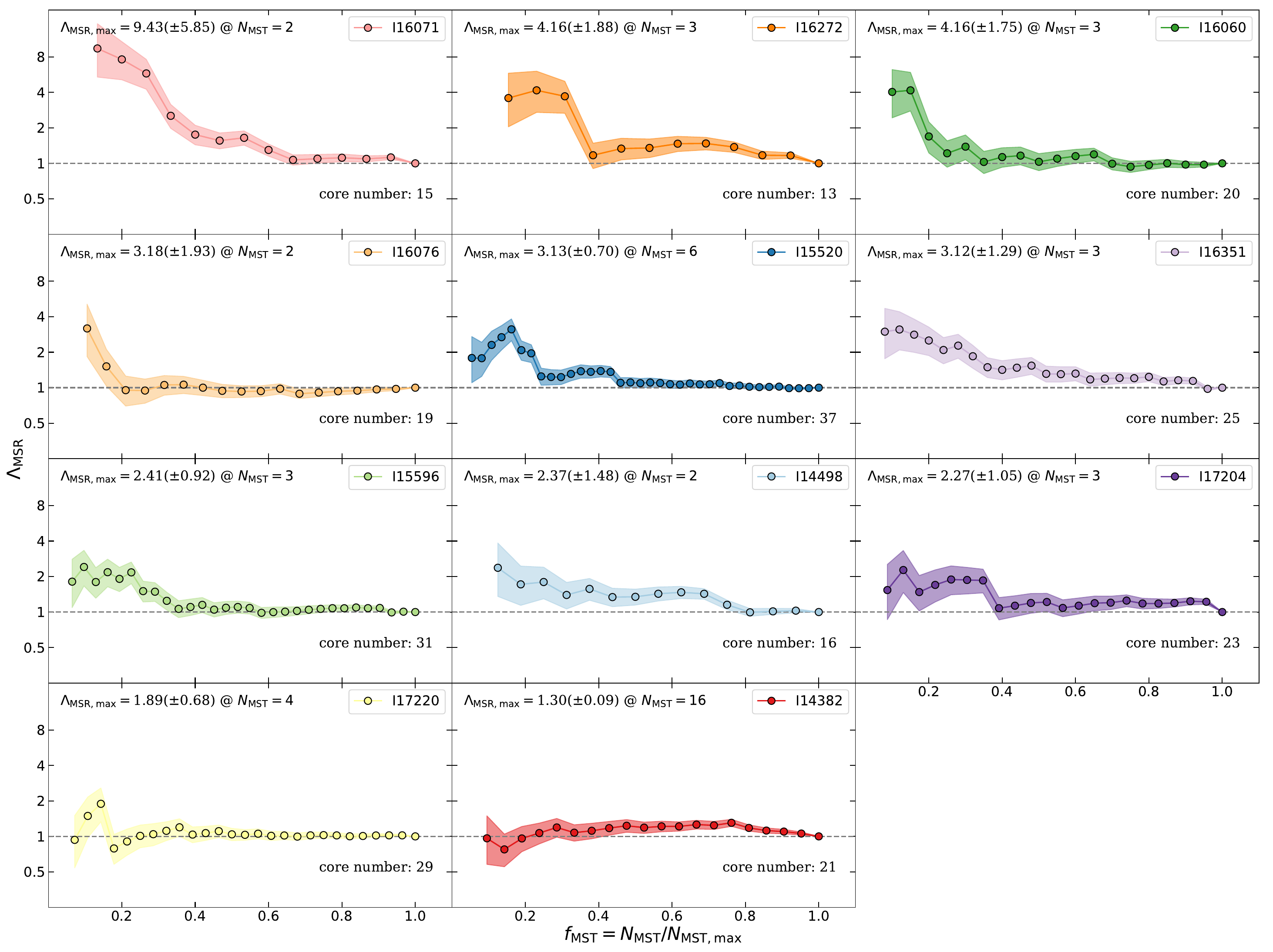}
\caption{The mass segregation ratio $\Lambda_{\rm MSR}$ plots: $\Lambda_{\rm MSR}$ is presented as a function of the fraction of core number $f_{\rm MST}=N_{\rm MST}/N_{\rm MST,max}$ for the 11 protoclusters. The shaded regions in each panel represent the local uncertainties. The upper left corner of each panel shows the maximum value of $\Lambda_{\rm MSR,max}$ and its corresponding $N_{\rm MST}$. The $\Lambda_{\rm MSR}$ versus $f_{\rm MST}$ panels are arranged in descending order of $\Lambda_{\rm MSR,max}$ for the corresponding protoclusters. The total core number is listed at the lower right corner. \label{fig:segregation}}
\end{figure*}

As defined in \citet{Allison2009Segregation,Parker2015Segregation}, mass segregation refers to a more concentrated distribution of more massive objects with respect to lower mass objects than that expected by random chance. For dynamically old bound systems (i.e. relaxed and virialized clusters), the process of two-body relaxation has redistributed energy between stars and they approach energy equipartition whereby all stars have the same mean kinetic energy. Therefore, more massive stars will have a lower velocity dispersion, and they will sink into the deeper gravitational potential, i.e., the center of the cluster \citep{Spitzer1969Equipartition}. 

Despite the observed mass segregation in old stellar clusters, it does not have to be from canonical two-body relaxation dynamical process. If we observe mass segregation in a region that is so young that two-body encounters cannot have mass segregated the stars, then the mass segregation must be set by some aspect of the star formation process, and is often called ``primordial mass segregation'' \citep{Parker2015Segregation}, which has been found in some simulations of star formation \citep[e.g.,][]{Moeckel2009Segregation,Myers2014Simulation}. Observationally, \citet{Sanhueza2019ASHES} have only found weak mass segregation in 4 out of 12 IRDCs and no mass segregation in the others. The overall conclusion is that there is no significant evidence of primordial mass segregation in IRDCs \citep{Sanhueza2019ASHES,Morii2023ASHES}. In contrast, at a similar physical resolution of 2400\,au and the same band (1.3\,mm) by ALMA, \citet{Dib2019Cluster} have found massive star-forming region W43 exhibits evident mass segregation with maximum mass segregation ratio $\Lambda^{\rm max}_{\rm MSR}=3.49$ (see definition in Equation\,\ref{eq:Lambda}). 

\subsubsection{\texorpdfstring{$\Lambda$}{Lambda} Plots: Characterisation of Mass Segregation} \label{segregation:Lambda}

To quantify the mass segregation in the protoclusters, we adopt the mass segregation ratio (MSR), $\Lambda_{\rm MSR}$, which is defined by \citet{Allison2009Segregation} and shown to perform best compared to three other methods by \citet{Parker2015Segregation}. The value of $\Lambda_{\rm MSR}$ at $N_{\rm MST}$ is given by
\begin{equation} \label{eq:Lambda}
\Lambda_{\rm MSR} (N_{\rm MST}) = \frac{\langle l_{\rm random}\rangle}{l_{\rm massive}} \pm \frac{\sigma_{l, \rm random}}{l_{\rm massive}},
\end{equation}
where $l_{\rm random}$ is the mean MST edge length of an ensemble of $N_{\rm MST}$ cores randomly chosen from the protocluster and $l_{\rm massive}$ is the mean MST length of the top-$N_{\rm MST}$ most massive cores. In our analyses, we performed 1000 Monte Carlo runs of choosing $N_{\rm MST}$ random cores to obtain a set of $l_{\rm random}$, calculating the mean value $\langle l_{\rm random} \rangle$ and its standard deviation $\sigma_{l, \rm random}=\sqrt{\langle (l_{\rm random}-\langle l_{\rm random} \rangle)^2 \rangle}$. For each $N_{\rm MST}$, $\Lambda_{\rm MSR}$ is meant to measure how much the MST length of the top-$N_{\rm MST}$ most massive cores deviate from the MST length of the entire protocluster. If the MST length of the top-$N_{\rm MST}$ ensemble is shorter than the MST length of the entire protocluster, it is suggested that massive cores have a more concentrated distribution.

By definition, $\Lambda\simeq1$ means that the massive cores were distributed in the same way as the other cores (i.e., no mass segregation); $\Lambda>1$ means that the massive cores were concentrated (i.e., mass segregation), and $\Lambda<1$ means that the more massive cores were spread out relative to the other cores (i.e., inverse-mass segregation).

Figure\,\ref{fig:segregation} presents mass segregation ratio $\Lambda_{\rm MSR}$ versus the fraction of the selected core number to the total core number $f_{\rm MST}=N_{\rm MST}/N_{\rm MST,max}$, which is called ``$\Lambda_{\rm MSR}$ plot'' hereafter. We arrange the protoclusters in descending order of the maximum value of the mass segregation ratio $\Lambda_{\rm MSR,max}$. For example, the protocluster I16071 in the first panel has the highest $\Lambda_{\rm MSR,max}$ of $8.72(\pm3.69)$, which implies strong mass segregation. In contrast, the protocluster I14382 has no mass segregation or even a weak inverse-mass segregation, as shown in the last panel.

There are three notable features that deserve additional explanations in the $\Lambda$ plots: 

$\bullet$ $\Lambda_{\rm MSR}$ peak at small $f_{\rm MST}$: protoclusters have a wide range of dense core mass while there are a small number of massive dense cores. When $f_{\rm MST}$ or $N_{\rm MST}$ are small, massive dense cores should account for a large proportion in the ensemble, so $l_{\rm massive}$ should be significantly smaller than $\langle l_{\rm random}\rangle$ if mass segregation exists. 

$\bullet$ $\Lambda_{\rm MSR}$ decrease with $f_{\rm MST}$: when $N_{\rm MST}$ increase, $l_{\rm massive}$ will involve more low-mass cores so that the mass segregation trend, if it exists, will be washed out; furthermore, when $f_{\rm MST}$ is larger, the ensembles of cores used to compute both $l_{\rm massive}$ and $\langle l_{\rm random} \rangle$ are more similar to the entire core sample so that both quantities theoretically approach the same value, the MST length of the entire core sample. 

$\bullet$ Diverse $\Lambda_{\rm MSR}$ profiles or diverse fractions of cores involved in mass segregation: $\Lambda_{\rm MSR}$ drops with $f_{\rm MST}$ with different rates. The clump with the strongest mass segregation, I16071, has its $\Lambda_{\rm MSR}$ dropping toward 1 around $f_{\rm MST}$ $\simeq0.6$, while the clump with the second largest mass segregation, I16060, has its $\Lambda_{\rm MSR}$ rapidly dropping toward 1 around $f_{\rm MST}$ $\simeq0.2$. Such diversity is also true among the clumps with lower degrees of mass segregation (e.g., I15596 vs. I14498). Therefore, it is of great interest to understand why the different protoclusters can show such different profiles of $\Lambda_{\rm MSR}$ plots in the future. 

\subsubsection{\texorpdfstring{$\mathcal{I}^{\rm MSR}_{\Lambda}$}{Integral_MRS}: Mass Segregation Integral (MSI)} \label{segregation:integral}

$\Lambda_{\rm MSR}$ plots are difficult to compare with each other, because $\Lambda_{\rm MSR}$ by definition depends on $N_{\rm MST}$ or $f_{\rm MST}$. In other words, to fully characterise the degree of mass segregation of a protocluster, two main factors need to be take into account: 1) $\Lambda_{\rm MSR,max}$, directly determines what the largest deviation from the random process is, according to the definition of Equation\,\ref{eq:Lambda}; 2) $N_{\rm MST,crit}$ or $f_{\rm MST,crit}$, which determines at what point the mass segregation ratio of cluster disappears for parameter $N_{\rm MST}$ or $f_{\rm MST}$.

Here we propose mass segregation integral (MSI) $\mathcal{I}^{\rm MSR}_{\Lambda}$ to describe how a cluster is segregated,
\begin{equation}
    \mathcal{I}^{\rm MSR}_{\Lambda} \equiv \frac{\sum\limits_{i=2}^{N_{\rm MST,max}} \Lambda_{\mathrm{MSR},i}}{N_{\rm MST,max}-1},
\end{equation}
where $\Lambda_{\mathrm{MST},i}$ is the mass segregation ratio at given $f_{\rm MST}$ or $N_{\rm MST}$. The MSI is meant to record every deviation from $\Lambda_{\mathrm{MSR},i}=1$ (when there is no mass segregation) at each $N_{\rm MST}$. 

In Section\,\ref{evolution:segregation}, we will examine the significant mass segregation observed in the ASSEMBLE protoclusters and explore its possible origins using the MSI. However, it is important to acknowledge limitations of using the MSI. First, the MSI collapses the $\Lambda_{\rm MSR}$ profile into a scalar value, thus disregarding the potentially various spatial distributions that can produce the same MSI value. Second, the physical and mathematical interpretations of the MSI are not yet fully understood, as it only records the deviations from the random distribution of cores within a protocluster. To enhance our understanding, future studies could establish a correlation between the MSI and the evolutionary timescale of a protocluster.

\section{Evolution of Massive Protoclusters} \label{sec:evolution}

The ASSEMBLE clumps have evolved to a late stage in the formation of massive protoclusters, with an $L/M$ ratio ranging from 10 to 100\,$L_{\odot}/M_{\odot}$. On the contrary, the ASHES clumps are in an early stage, with a $L/M$ ratio between 0.1 and 1\,$L_{\odot}/M_{\odot}$. Therefore, the ASSEMBLE and ASHES clumps can serve as mutual informative comparison groups, as the basis of our dynamic view of protocluster evolution. As introduced in Section\,\ref{result:coreproperty}, the statistics of the core parameters are summarized in Table\,\ref{tab:corestats}, which can be directly compared to Table 5 in \citet{Sanhueza2019ASHES}. To highlight the quiescent and active nature of ASHES and ASSEMBLE clumps, respectively, the samples also have the second names infrared dark clouds (IRDCs) and infrared bright clouds (IRBCs), respectively.

\subsection{Core Growth and Mass Concentration} \label{evolution:coregrow}

\begin{figure*}[!ht]
\centering
\includegraphics[width=0.98\linewidth]{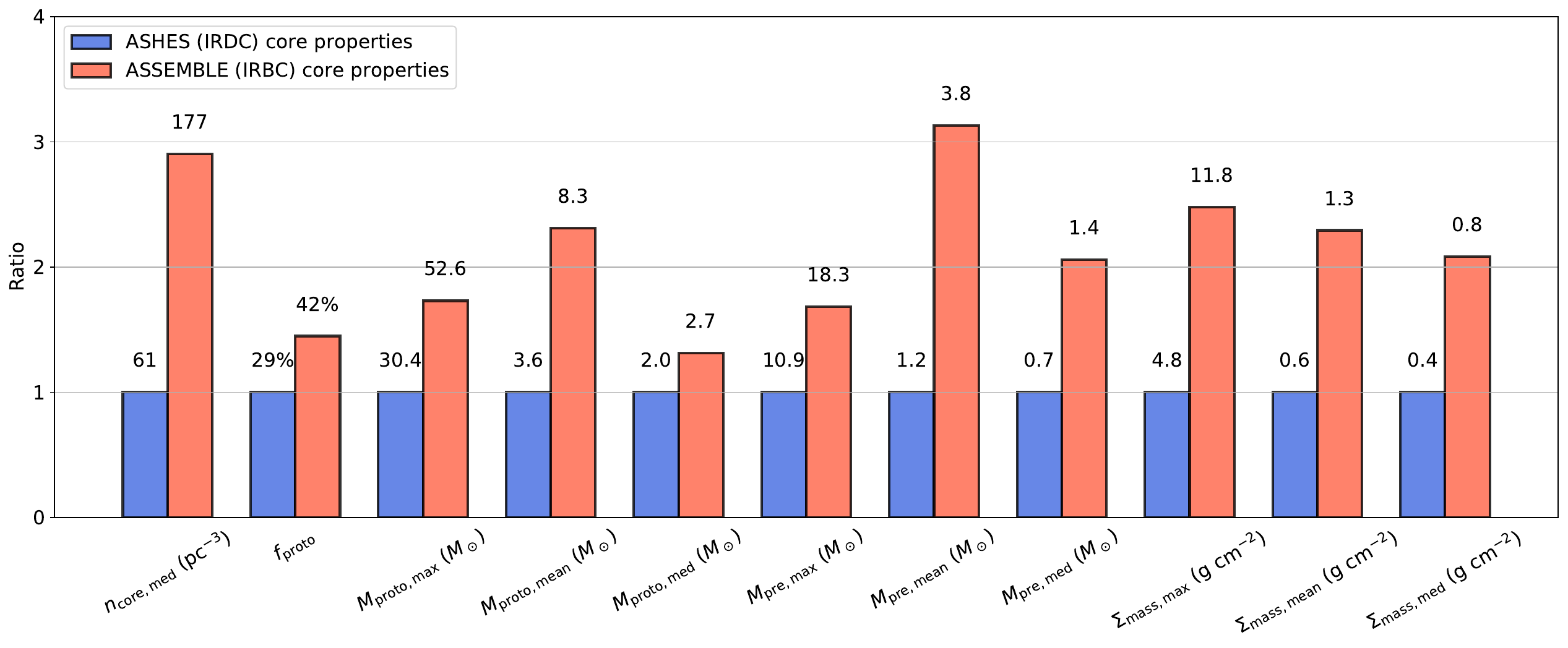}
\caption{Comparison of parameters including median volume density of the core number $n_{\rm core,med}$, number fraction of protostellar cores $f_{\rm proto}$, the maximum/mean/median value of mass of protostellar cores $M_{\rm proto,max}$/$M_{\rm proto,mean}$/$M_{\rm proto,med}$, mass of prestellar cores $M_{\rm pre,max}$/$M_{\rm pre,mean}$/$M_{\rm pre,med}$, and core surface mass density $\Sigma_{\rm mass,max}$/$\Sigma_{\rm mass,mean}$/$\Sigma_{\rm mass,med}$. The IRDCs and IRBCs are in blue and red, respectively. The y value is normalized by the ratio to the value of IRDCs. \label{fig:compare}}
\end{figure*}

As shown in Figure\,\ref{fig:compare}, the IRBCs exhibit a median volume density of the core number $n_{\rm core,med}$ of 177 per pc$^3$, which is approximately three times greater than the 61 per pc$^3$ observed in the IRDCs. We consider the potential effects from the different mass sensitivities between the two projects. The slightly worse sensitivity ($\bar\sigma=0.089$\,\msun) of the ASSEMBLE compared to the ASHES ($\bar\sigma=0.078$\,\msun) shows that correcting for sensitivity would only increase the core density in IRBCs. To exclude the effects of different source extraction algorithms, we also perform the source extraction using \getsf~in the 12m-alone data of ASHES as it was done for the ASSEMBLE in Appendix\,\ref{app:mst}, only obtaining a much lower core number of 66, mostly due to two factors: 1) \texttt{getsf} tends to extract spherical cores but miss those irregular ones; 2) the 12m-alone data filter out large-scale structures that are previously identified by \texttt{astrodendro} algorithm. Therefore, correcting the effects from the array configuration and source extraction algorithm will only result in even larger difference between the two sets of parameters mentioned above. In any circumstances, the core number densities in the ASSEMBLE clumps are considerably higher than those of IRDCs.

As demonstrated in simulations by \citet{Camacho2020Simulation}, massive clumps accrete mass and increase density as they evolve, resulting in a decrease in the free-fall timescale. Consequently, the dense cores formed by Jeans fragmentation collapse to form protostars more quickly, leading to a higher fraction of protostellar cores, $f_{\rm proto}$. As shown in the second column of Figure\,\ref{fig:compare}, $f_{\rm proto}$ increases significantly from 29\% in the early-stage IRDCs to 42\% in the late-stage IRBCs on average. The increasing trend of $f_{\rm proto}$ with respect to evolutionary stage has also been previously reported by \citet{Sanhueza2019ASHES} and is consistent with the fragmentation reported by \citet{Palau2014Fragmentation,Palau2015Fragmentation,Palau2021Fragmentation} in more evolved IRBCs, because in these works most of the cores are protostellar (given their higher masses and compactness compared to the ASSEMBLE sample).

Furthermore, we provide several pieces of evidence for the growth of the core mass from IRDC to IRBC in columns (3--11) in Figure\,\ref{fig:compare}. Parameters including the maximum, mean, and median mass of protostellar cores $M_{\rm proto,max}$, ${M}_{\rm proto,mean}$, and ${M}_{\rm proto,med}$, respectively; those of prestellar cores $M_{\rm pre,max}$, ${M}_{\rm pre,mean}$, and ${M}_{\rm pre,med}$, respectively; and those of surface mass density of total core population ${\Sigma}_{\rm mass,mean}$, $\Sigma_{\rm mass,max}$, and ${\Sigma}_{\rm mass,med}$, respectively, all exhibit systematic increments from IRDCs to IRBCs. These mass or surface density increments have also been observed in another comparative work between IRDCs and IRBCs with hub-filament structures \citep{Liu2023HFS}, where gas inflow is thought to be responsible for the hierarchical and multiscale mass accretion \citep{GM2010W33A,Liu2022ATOMS-V,Liu2022ATOMS-IX,Liu2023HFS,Xu2023SDC335,Yang2023G310}. Very recent statistical studies of dense cores in both the Dragon infrared dark cloud \citep{Kong2021Dragon}, the Orion Molecular Cloud \citep[OMC;][]{Takemura2023Orion}, the ASHES IRDC sample \citep{Li2023Dynamics}, and the ALMA-IMF protoclusters \citep{Nony2023IMF-V,Pouteau2023IMF-VI} also suggest that the protostellar cores are considerably more massive than the starless cores, suggesting that cores grow with time. If the missing flux in the ASSEMBLE sample were recovered, the effect would be to increase the core mass and surface density, strengthening the arguments above. 

\subsection{``Nurture'' but not ``Nature'': A Dynamic View of the \texorpdfstring{$M_{\rm max}$ versus $M_{\rm clump}$}{Mmax-Mclump} Relation} \label{evolution:Mmax_Mclump}

As discussed in Section\,\ref{discuss:coevolve}, a positive correlation between $M_{\rm max}$ and $M_{\rm clump}$ is observed within the ASSEMBLE protoclusters, suggesting a close relationship between the natal clump and the most massive core through multiscale gas accretion \citep{Xu2023SDC335}. On the contrary, the ASHES pilots, represented by the gray data points in the right panel of Figure\,\ref{fig:coremass}, exhibit no significant $M_{\rm max}$ versus $M_{\rm clump}$ correlation. The Spearman rank correlation coefficient is calculated to be 0.04, with a corresponding p-value of 0.90. This lack of correlation aligns with the concept of thermal Jeans fragmentation \citep[e.g.,][]{Palau2015Fragmentation,Palau2018Fragmentation,Sanhueza2019ASHES}, where the clump's thermal Jeans mass is primarily determined by its dust temperature within a narrow range of 10--20\,K \citep{Morii2023ASHES}, rather than the turbulence whose energy is governed by clump's gravity assuming energy equipartition \citep{Palau2015Fragmentation}. This finding supports the notion that early-stage cores are characterized by dominant initial fragmentation rather than gravitational accretion. In that case, no correlation is naturally expected between the mass of the natal clump and the mass of the core resulting from fragmentation. Therefore, we propose a dynamic picture of the clump-core connection. 

$\bullet$ At the beginning, initial Jeans fragmentation produces a set of dense cores whose mass is not associated with clump-scale gravitational potential (mass) and turbulence.

$\bullet$ As a massive clump evolves, multiscale continuous gas accretion help build up the connection between clump and core scales, for example the mass correlation that we've observed.

\subsection{Implications of the \texorpdfstring{$M_{\rm max}$ versus $M_{\rm cluster}$}{Mmax-Mcluster} Relation} \label{evolution:Mmax_Mcluster}

\begin{figure}[!ht]
\centering
\includegraphics[width=0.98\linewidth]{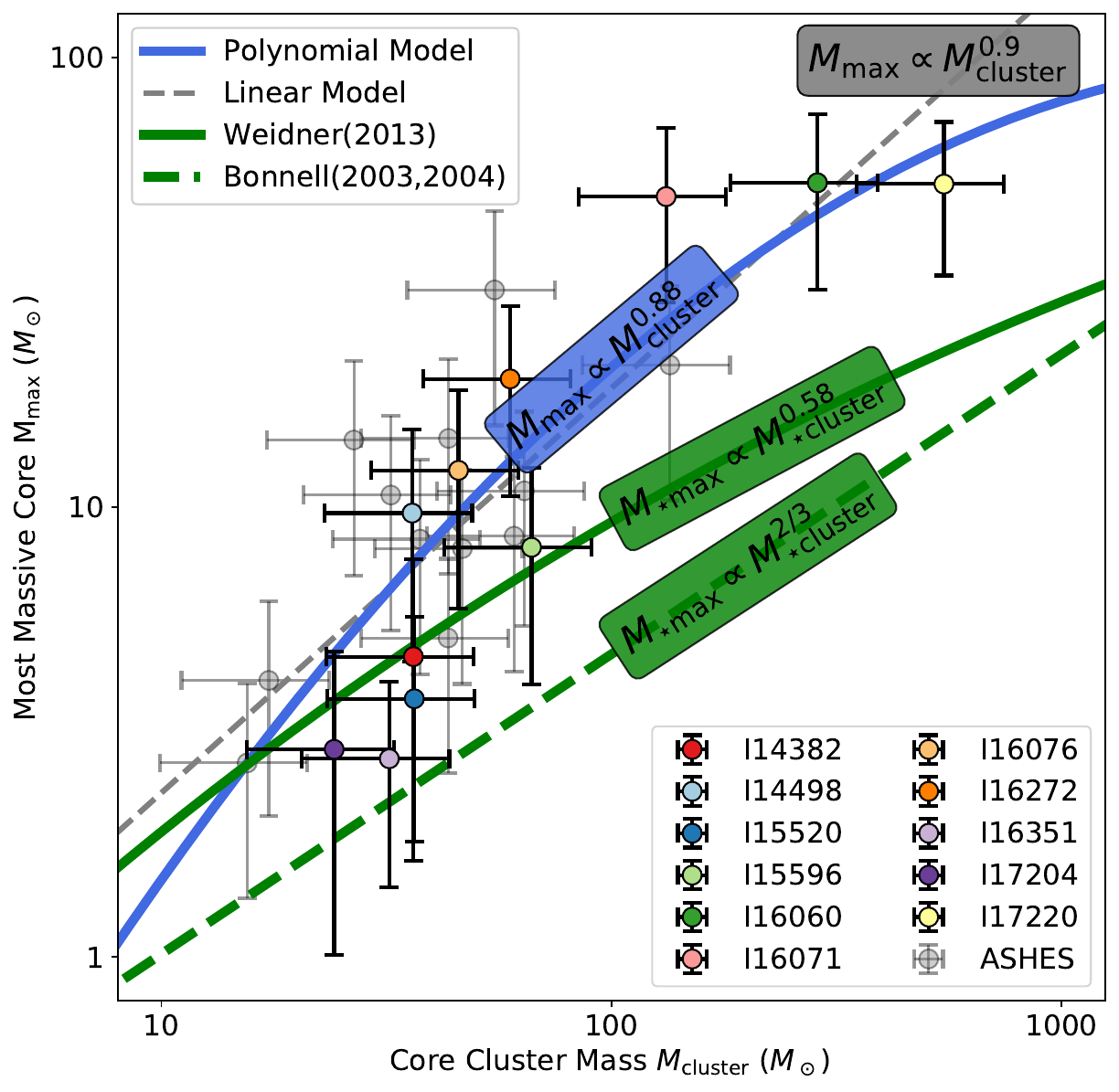}
\caption{The $M_{\rm max}$ versus $M_{\rm cluster}$ relation. The ASSEMBLE and the ASHES Pilots are plotted with errorbars. The blue solid line shows the second-order polynomial fitting of all the data points. $M_{\rm max}\propto M_{\rm cluster}^{0.88}$ in the blue box indicates the power-law index in the first-order approximation. The gray dashed line shows the linear regression of all the data points. The green solid line shows the polynomial fitting of the data from 139 star clusters \citep{Weidner2013Cluster}, with first-order approximation of $M_{\rm \star max} \propto M_{\rm \star cluster}^{0.58}$ in the upper green box. The dashed green line as well as the lower green box show the relation from smoothed particle hydrodynamics simulation \citep{Bonnell2003Simulation,Bonnell2004Simulation}. \label{fig:max_cluster}}
\end{figure}

The relation $M_{\rm \star max}$ versus $M_{\rm \star cluster}$, which describes the relationship between the mass of the most massive star ($M_{\rm \star max}$) and the total mass of the star cluster ($M_{\rm \star cluster}$), has been previously established both analytically \citep{Weidner2004Cluster} and observationally \citep[e.g.,][]{Testi1999Clusters,Weidner2006Cluster}. This relation highlights the systematic variation of the typical upper mass limit with the overall mass of the star cluster. It suggests that the formation of stars within cloud cores is primarily influenced by growth processes occurring in an environment with limited resources. This finding underscores the significance of resource availability in shaping the stellar population within star clusters. 

Protoclusters provide a retrospective glimpse towards the early version of star clusters. Throughout this paper, we refer $M_{\rm cluster}$ as the sum of all the core masses in a protocluster. Note that $M_{\rm cluster}$ is different than the total mass of a stellar cluster ($M_{\rm \star cluster}$). As shown in the left panel of Figure\,\ref{fig:coremass}, $M_{\rm cluster}$ are plotted with the masses of the most massive cores $M_{\rm max}$ in Figure\,\ref{fig:max_cluster}. Contrary to what has been found in the ``$M_{\rm max}$ versus $M_{\rm clump}$'' plane (shown in the right panel of Figure\,\ref{fig:coremass}), both the ASHES and the ASSEMBLE protoclusters have positive correlation between $M_{\rm max}$ and $M_{\rm cluster}$. A second-order polynomial model ``$y = -0.27 x^2 + 1.96 x - 1.51$'' fits the $M_{\rm max}$ versus $M_{\rm cluster}$ relation best, with a correlation coefficient of 0.68, which is shown in the blue solid line. Besides, we also present the relation in stellar clusters given by observations \citep{Weidner2013Cluster} in green solid line and smoothed particle hydrodynamics simulations \citep{Bonnell2003Simulation,Bonnell2004Simulation} in green dashed line.

In order to facilitate a direct comparison between the stellar cluster and protostellar cluster, we have also performed a first-order approximation of the polynomial models represented by the blue and green solid lines. By utilizing the mean value theorem, the first-order power-law index can be estimated by considering the average derivatives within the given value range. The estimation of the power-law index is indicated in the blue and green boxes, which are overlaid on the respective solid lines. As shown in the gray dashed line, we directly perform the linear regression to derive a power-law index of 0.9, validating our first-order approximation of 0.88. 

Despite uncertainties, the slope of the $\log M_{\rm max}$ versus $\log M_{\rm cluster}$ relation is notably steeper compared to that of the $\log M_{\rm \star max}$ versus $\log M_{\rm \star cluster}$ relation. To reconcile this disparity within the context of protocluster evolution, we take into account the influence of multiple star systems on massive star formation. As depicted in Figure 1 by \citet{Offner2022Multiple}, the probability of events involving multiplicity is nearly 100\%. In other words, it is highly likely that massive cores give rise to the formation of more than one massive star. Hence, it is natural to expect that the slope of the $\log M_{\rm max}$ versus $\log M_{\rm cluster}$ relation can evolve into a shallower version, akin to what is observed in the $\log M_{\rm \star max}$ versus $\log M_{\rm \star cluster}$ relation.

Another noteworthy finding is the similarity in the total cluster mass distribution between the ASHES and ASSEMBLE protoclusters, particularly within the range of 10--100 \msun, when excluding four outliers with $M_{\rm cluster}>100$\,\msun. Since these outliers also exhibit higher clump masses (refer to Figure\,\ref{fig:coremass}), a more fundamental question arises: Why do these protoclusters with different evolutionary stages consistently maintain a mass proportion of cluster to clump ($M_{\rm cluster}/M_{\rm clump}$) between 1--10\%? This proportion can be regarded as the dense gas fraction (DGF), which is often closely associated with star formation efficiency \citep{Ge2023Filament}. Consequently, it is of great significance to investigate the evolution of DGF in relation to massive star-forming clumps (Xu et al., in preparation).

\subsection{Gravitational Contraction: Protoclusters Evolve to Greater Compactness} \label{evolution:compactness}

As shown in Figure\,\ref{fig:separation}, the core separation distribution of the ASSEMBLE protoclusters have two prominent features. One is a peak at 0.035\,pc as discussed in Section\,\ref{discuss:separation}, systematically smaller than what has been found in the ASHES, meaning that the spatial distribution becomes tighter and more compact as the protocluster evolves. The other one is an extended tail at 0.06--0.3\,pc, numerically consistent with what has been found in the ASHES protoclusters of 0.06--0.24\,pc (refer to green or gray histograms of Figure\,\ref{fig:separation}), which are assumed to be the residuals of the initial fragmentation at the early stage. In this section, we discuss how gravity leads to the tightening process of protoclusters and complete the dynamic picture of fragmentation and gravitational contraction.

We can make a simple semi-quantitative calculation. Given that the thermal Jeans fragmentation is observed to dominate at the early stage of massive star formation \citep{Sanhueza2019ASHES,Li2022NGC6334S,Morii2023ASHES,Palau2014Fragmentation,Palau2015Fragmentation,Palau2021Fragmentation}, we assume the initial condition that the cores could have initially fragmented on Jeans length scales of $\sim0.14$\,pc \citep[mean Jeans lengths in the ASHES;][]{Sanhueza2019ASHES}. If dense cores are moving toward the center of the clump by gravity, then the velocity of the cores should be free-fall velocity as $v_{\rm ff}=\frac{R_{\rm cl}}{t_{\rm ff}}$. Adopting the typically observed massive clump size and density of $R_{\rm cl}=1$\,pc and $n_{\rm H_2}=10^4$\,cm$^{-3}$, the core separation will be tighter by,
\begin{equation} \label{eq:gravity}
\begin{split}
    \Delta l_{\rm gc} & = v_{\rm ff} t_{\rm life} = \frac{R_{\rm cl}}{t_{\rm ff}} \times t_{\rm life} \\ 
    & =0.088\left(\frac{R_{\rm cl}}{1\,\mathrm{pc}}\right)\left(\frac{n_{\ssstyle \rm H_2}}{10^4\rm cm^{-3}}\right)^{0.5}\left(\frac{t_{\rm life}}{5\times10^4\,\mathrm{yr}}\right)\,\mathrm{pc}
\end{split}
\end{equation}
where $t_{\rm life}\sim0.2-1\times10^5$\,yr \citep{Motte2018Review} are the free-fall time scale and the statistical lifetime of massive starless clumps, respectively. Therefore, the core separation should tighten by  $\Delta l_{\rm gc}\simeq0.04--0.18$\,pc in the protoclusters by gravitational contraction, numerically consistent with the shift from extended tail (0.06--0.3\,pc) to the observed separation (peaked at 0.035\,pc). 

The simple gravitational contraction model fits the observations well, indicating ongoing bulk motions from the global gravitational collapse of massive clumps \citep{Beuther2018Fragmentation,VS19GHC}. But note that we are still unable to rule out another possibility that the closer separation is due to hierarchical fragmentation to produce a series of condensations inside a massive core.

\subsection{Evolution of the \texorpdfstring{$\mathcal{Q}$}{qvalue} Parameter} \label{evolution:Q}

\begin{figure}[!ht]
\centering
\includegraphics[width=0.98\linewidth]{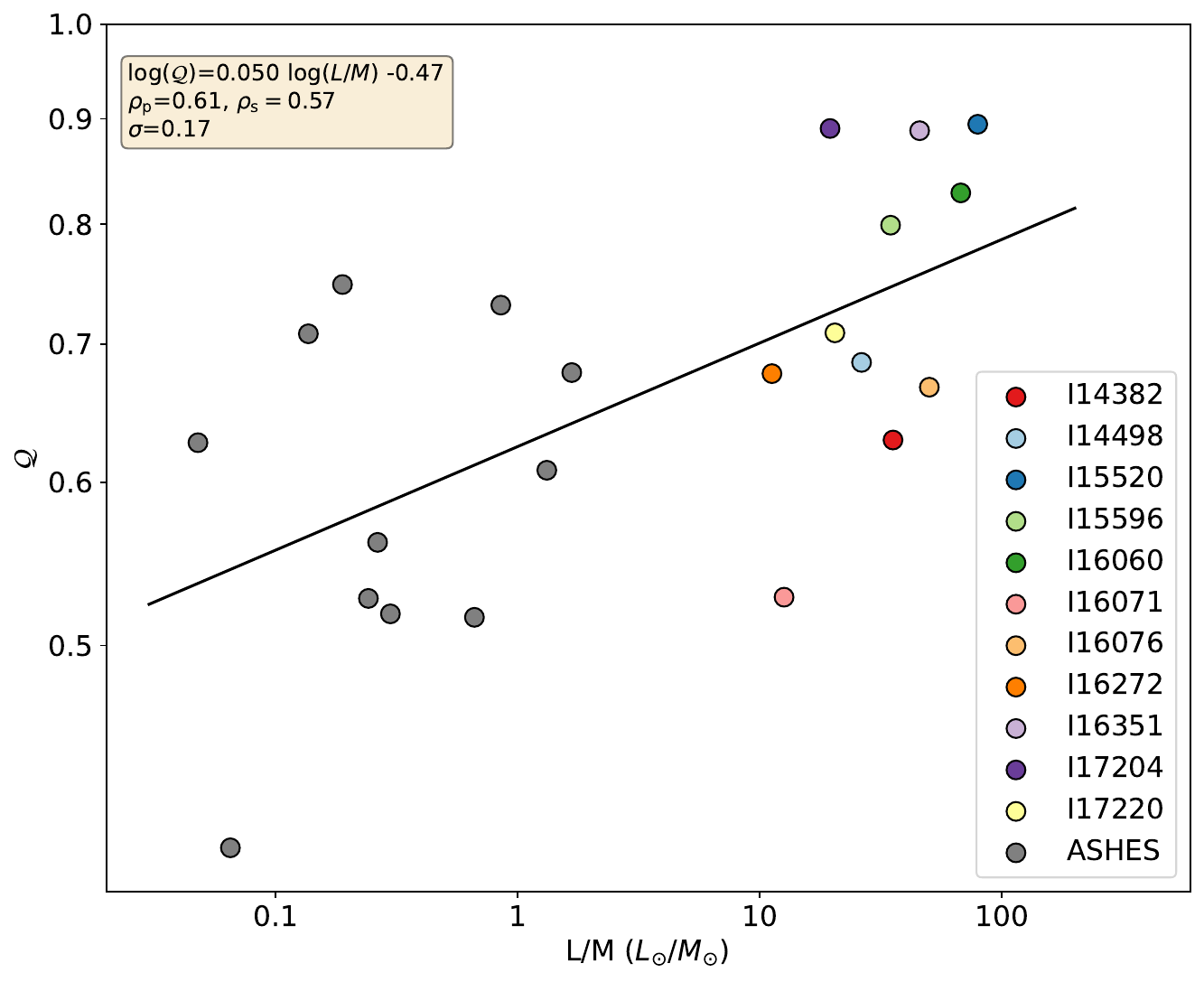}
\includegraphics[width=0.98\linewidth]{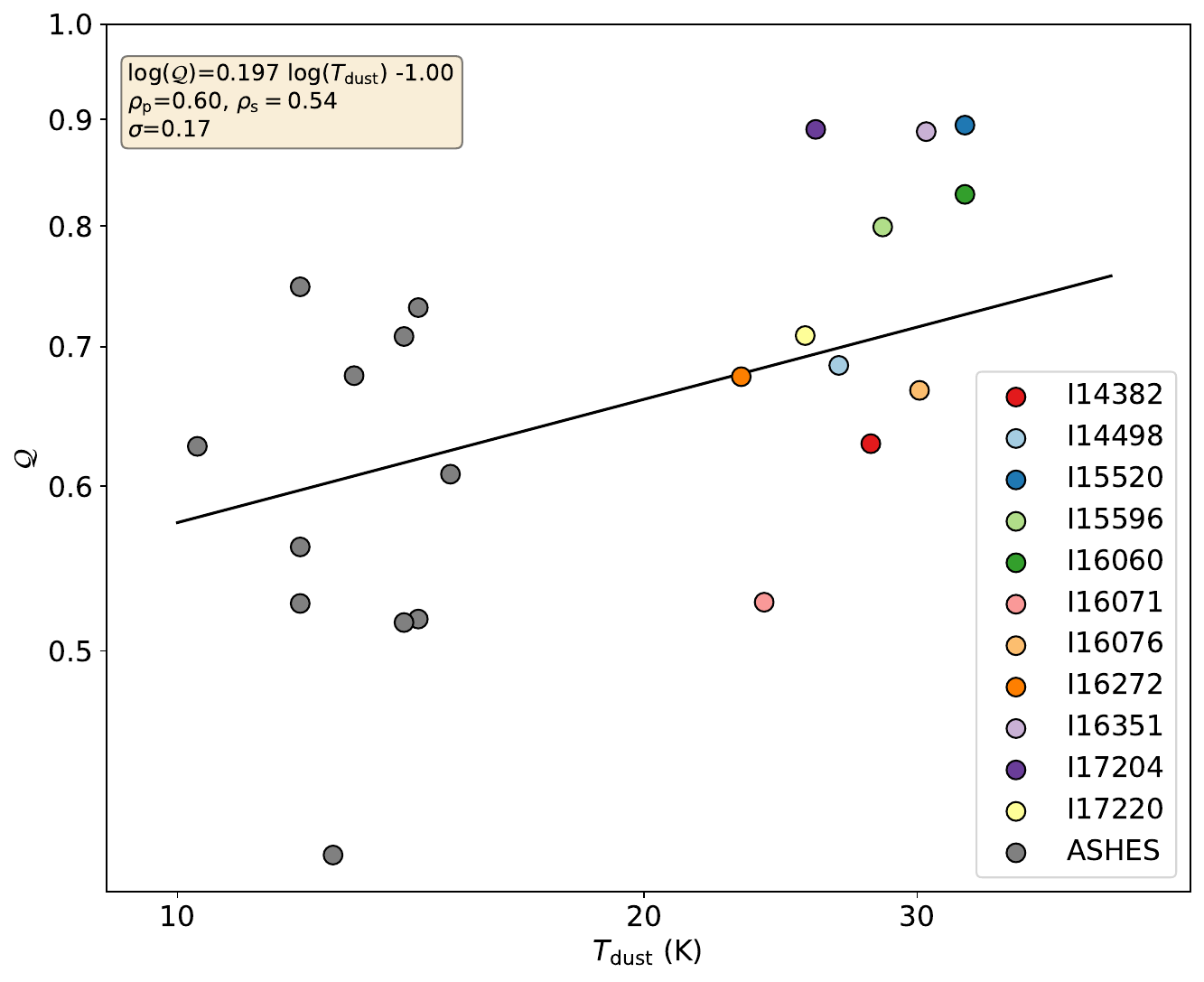}
\caption{The $\mathcal{Q}$ values versus luminosity-to-mass ratio $L/M$ (left) and temperature $T_{\rm dust}$ (right). The linear regression results including the fitting model, Pearson correlation coefficient $\rho_{\rm p}$, Spearman rank correlation coefficient $\rho_{\rm s}$, and scatter $\sigma$ are shown on the upper left. ASHES clumps are shown in the $\mathcal{Q}$ versus $T_{\rm dust}$ panel with gray points. \label{fig:Qpar}}
\end{figure}

As clumps evolve over time, the primordial distribution of cores dissolves due to dynamical relaxation, leading to a more radially concentrated structure as predicted by simulations \citep{Guszejnov2022Starforge}. Consequently, more-evolved clumps are predicted to have higher $\mathcal{Q}$ values. In the 12 ASHES Pilots, \citet{Sanhueza2019ASHES} used the fraction of protostellar cores $f_{\rm proto}$ to gauge the evolutionary stage. Due to the narrow parameter space of similar evolutionary properties such as dust temperature $T_{\rm dust}$ (10--15\,K) and luminosity-to-mass ratio $L/M$ (0.1--1\,$L_{\odot}/M_{\odot}$), only a weak correlation between $\mathcal{Q}$ and $f_{\rm proto}$ was found \citep{Sanhueza2019ASHES}. However, \citet{Dib2019Cluster} found that the most active star forming region W43 has a higher $\mathcal{Q}$ value compared to more quiescent regions (L1495 in the Taurus, Aquila, and Corona Australis). These studies inspire a larger sample with wide evolutionary stages to shed more light on the interplay between star formation in clouds and the spatial distribution of dense cores.

The combination of the ASSEMBLE and ASHES clumps provides a systematic sample with a wide dynamic range of evolutionary stage ($L/M$ of 0.1--100\,$L_{\odot}/M_{\odot}$ and $T_{\rm dust}$ of 10--35\,K). To make the comparison between two samples more directly, we have simulated the mock 0.87\,mm continuum data with only the 12-m array configuration (see details in Appendix\,\ref{app:mock}). Following the same procedure of core extraction, we have an updated ASHES core catalog used for the MST algorithm (see more details in Appendix\,\ref{app:mst}). The $\mathcal{Q}$ parameters for the ASHES sample range from 0.40 to 0.75, with a median value of 0.61(0.13). The Mann–Whitney U test between the $\mathcal{Q}$ parameters of the ASSEMBLE and ASHES samples has a p-value of 0.03 ($<$0.05), showing the two samples have significantly different $\mathcal{Q}$ parameters.

The linear regressions between the $\mathcal{Q}$ parameters and the evolutionary indicators ($L/M$ and $T_{\rm dust}$) are performed in the $\log$ versus $\log$ space and shown in Figure\,\ref{fig:Qpar}. The positive correlations between both $L/M$ and $T_{\rm dust}$, indicates that the $\mathcal{Q}$ parameters evolve with time. The correlations are confirmed to be statistically significant by the high Pearson correlation coefficients $\rho_{\rm p}$ of 0.61 and 0.60 with p-values of 0.0024 and 0.0034 for $L/M$ and $T_{\rm dust}$, respectively. Moreover, the Spearman correlation coefficients $\rho_{\rm s}$ are 0.57 and 0.53 with p-values of 0.0056 and 0.0088. Statistically, it's tentatively evident that the $\mathcal{Q}$ parameter of the protostellar clusters should increase in later evolutionary stages, indicating more sub-clustering distribution at an early stage but more centrally condensed structure when the cluster evolves, which agrees with the results and predictions in \citet{Sanhueza2019ASHES}. 

\subsection{Origin of Mass Segregation} \label{evolution:segregation}

\begin{figure}[!ht]
    \centering
    \includegraphics[width=0.98\linewidth]{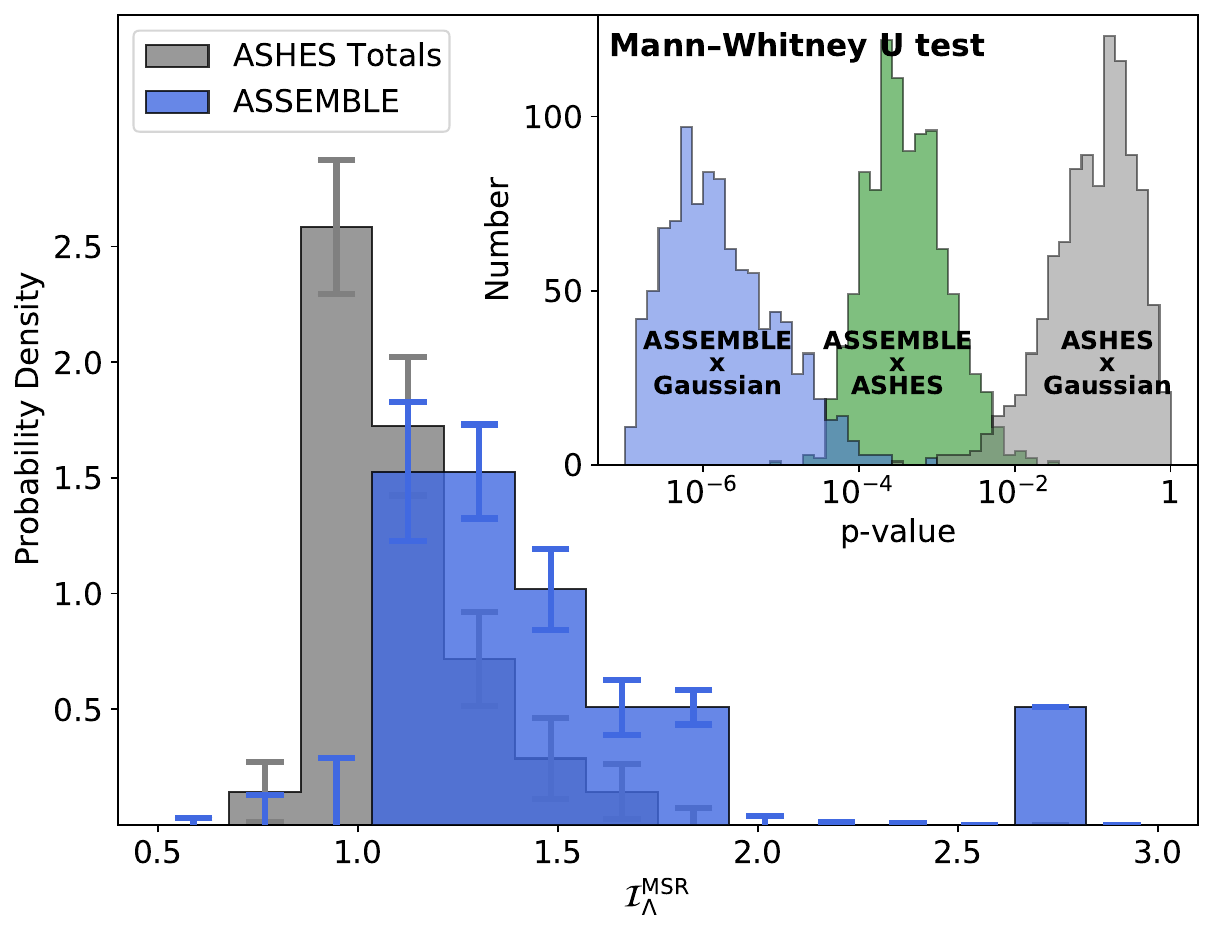}
    \caption{The mass segregation integrals $\mathcal{I}_{\Lambda}^{\rm MSR}$ of the ASSEMBLE and the ASHES are shown as blue and gray histograms with errorbars. The p-values of the Mann-Whitney U test between ASSEMBLE, ASHES, and Gaussian distributions ($\mu=1$,$\sigma=0.1$) are shown in three colors, which are attached in the upper right corner. \label{fig:IntMSR}}
\end{figure}

Compared to the early-stage clusters previously reported by \citet{Sanhueza2019ASHES,Morii2023ASHES}, we have identified three main differences in mass segregation according to the $\Lambda$-plots in Figure\,\ref{fig:segregation}. Firstly, evident mass segregation (with $\Lambda_{\rm MSR}\gtrsim3$) was found in 73\% (8 out of 11) ASSEMBLE protoclusters, which is $>5$ times more than it was identified in the ASHES sample \citep[13\%, 5 out of 39;][]{Morii2023ASHES}. Secondly, the mass segregation ratios we observed were significantly higher, with some clusters exhibiting values as large as $\sim9$. Finally, we observed $\Lambda_{\rm MSR}>1$ even for larger core numbers ($\gtrsim10$) in certain protoclusters such as I16351, I15520, and I15596. 

Using the mass segregation integral (MSI) introduced in Section\,\ref{segregation:integral}, we present a direct comparison between the ASSEMBLE and the ASHES Totals, as illustrated in Figure\,\ref{fig:IntMSR}. The Mann-Whitney U test reveals significant differences between the ASSEMBLE and the ASHES protoclusters, as indicated by the green histogram of p-values. To establish a reference sample for statistical analysis, we simulate 100 clusters with mean MSI of $\mu=0$ (no mass segregation), and with random perturbation of $\sigma=0.1$ (assumed to be the same as typical uncertainties when calculating the MSI) in MSI. The Mann-Whitney U test rejects the null hypothesis that the MSI of the ASSEMBLE protoclusters follows the random perturbation (p-value $\ll$ 0.01), highlighting the presence of evident mass segregation. In contrast, the null hypothesis cannot be confidently rejected for the ASHES protoclusters, with mean and median p-values of 0.18 and 0.12, respectively. Thus, the ASSEMBLE protoclusters exhibit robust evidences of mass segregation, whereas the mass segregation in the ASHES protoclusters is weak to moderate.

In the context of protocluster evolution, the degree of mass segregation increases unambiguously from the ASHES clusters to the ASSEMBLE clusters (this work). Therefore, the natural question is the origin of mass segregation. Here, we test whether mass segregation can result from the canonical dynamical relaxation by two-body relaxation.

To analyze the dynamics of the cluster, we adopt the formulation of \citet{Reinoso2020Cluster}, who extended the framework of \citet{Spitzer1987Book} to include the effect of a gas potential. The crossing time of the cluster is then given as,
\begin{equation}
t_{\rm cross} = \frac{R}{V_{\rm vir}},
\end{equation}
with velocity under the virial equilibrium $V_{\rm vir}$,
\begin{equation}
V_{\rm vir} = \sqrt{\frac{GM_{\rm core}}{R}} (1+q),
\end{equation}
and $q=M_{\rm gas}/M_{\rm core}$. Here, $M_{\rm gas}$ and $M_{\rm cluster}$ are ambient low-density gas mass and total dense core cluster mass, respectively. $R$ is radius of cluster. The relaxation time is then given as,
\begin{equation}
t_{\rm relax} = 0.138 \frac{N(1+q)^4}{\ln(\gamma N)}t_{\rm cross},
\end{equation}
where $N$ is the number of core in a cluster and $\gamma$ is a constant of proportionality in the term of virial velocity. The $\gamma$ value is between 0.42 and 0.38 for the polytropes of index of the cluster system between 3 and 5, and the $\gamma=0.4$ provides a reasonably good approximation for most systems \citep{Spitzer1969Equipartition}. So we use $\gamma=0.4$ here.

We take the ASHES sample as the initial condition for our protocluster analyses, assuming a typical value of $R=0.5$\,pc, $N=25$, and $M_{\rm cluster}=100$\,\msun~(i.e., the mass of the protoclusters). For the gas mass, we used two different methods. The first method is calculating the total gas mass based on an average volume density of $5\times10^4$\,cm$^{-3}$ and a radius of $R=0.5$\,pc \citep[from Table\,1 in][]{Morii2023ASHES}, resulting in a gas mass of approximately $\sim200$\,\msun. The second method considers that the ALMA recovered flux only comes from the dense cores that we identified and the missing flux should come from diffuse gas, both of which are covered by the ATLASGAL emission. As shown in Section\,\ref{result:extractcores}, the flux ratio of ALMA to ATLASGAL has a mean value of $\sim20$\%. Therefore, the total gas mass should be four times larger than the total mass of the core cluster, giving a value of 400\,\msun. Two independent methods yield $M_{\rm gas}$ within the range of 200--400\,\msun. By considering both methods, we derived the $q$ value of 2--4. Taking all factors into account, we found that the typical relaxation time of a protocluster was as long as $70-500$\,Myrs, which is much longer than the typical lifetime of massive star formation (several Myrs). Considering the short formation timescale of massive stars, the mass segregation is unlikely to be caused by dynamical relaxation \citep{Zhang2022G35}, as is the case for more evolved stellar clusters.
In the context of a stellar cluster, such mass segregation should be considered ``primordial'', although it has already evolved from its initial stage.

If the observed mass segregation is not induced by traditional dynamical processes by cores/stars themselves, what could be its origin? We propose that this could be naturally due to the gravitational concentration of the entire clump or gas accretion toward the center. The ALMA observations of IRDCs have already revealed a large number of sub-Jeans-mass cores during the initial fragmentation \citep{Sanhueza2019ASHES,Morii2023ASHES}. In the late stage, the most massive cores are always located at the centers of the clumps or of their gravitational potentials. Our work supports the predictions of numerical simulations where members near the center of the gravitational potential will become the most massive cores during the evolution due to their privileged location in the forming cluster \citep{Bonnell1998Segregation,Bonnell2006Simulation}. 

\section{Conclusion} \label{sec:conclusion}

The ALMA Survey of Star formation and Evolution in Massive protoclusters with Blue Profiles (ASSEMBLE) is aimed at a comprehensive examination of the mass assembly process of massive star formation in a dynamic view, including fragmentation and accretion, and their relevance to theories. To this end, the survey employed ALMA 12\,m mosaicked observations to capture both continuum and spectral line emissions in 11 massive ($M_{\rm clump}\gtrsim10^3$\,\msun) and luminous ($L_{\rm bol}\gtrsim10^4$\,\lsun) clumps protoclusters with blue profiles. This paper releases the continuum data, characterizes the core physical properties, and presents the analyses of the evolution of the protostellar clusters, while outlining the conclusions drawn from the analysis as follows:

\begin{figure*}[!ht]
	\includegraphics[width=0.98\linewidth]{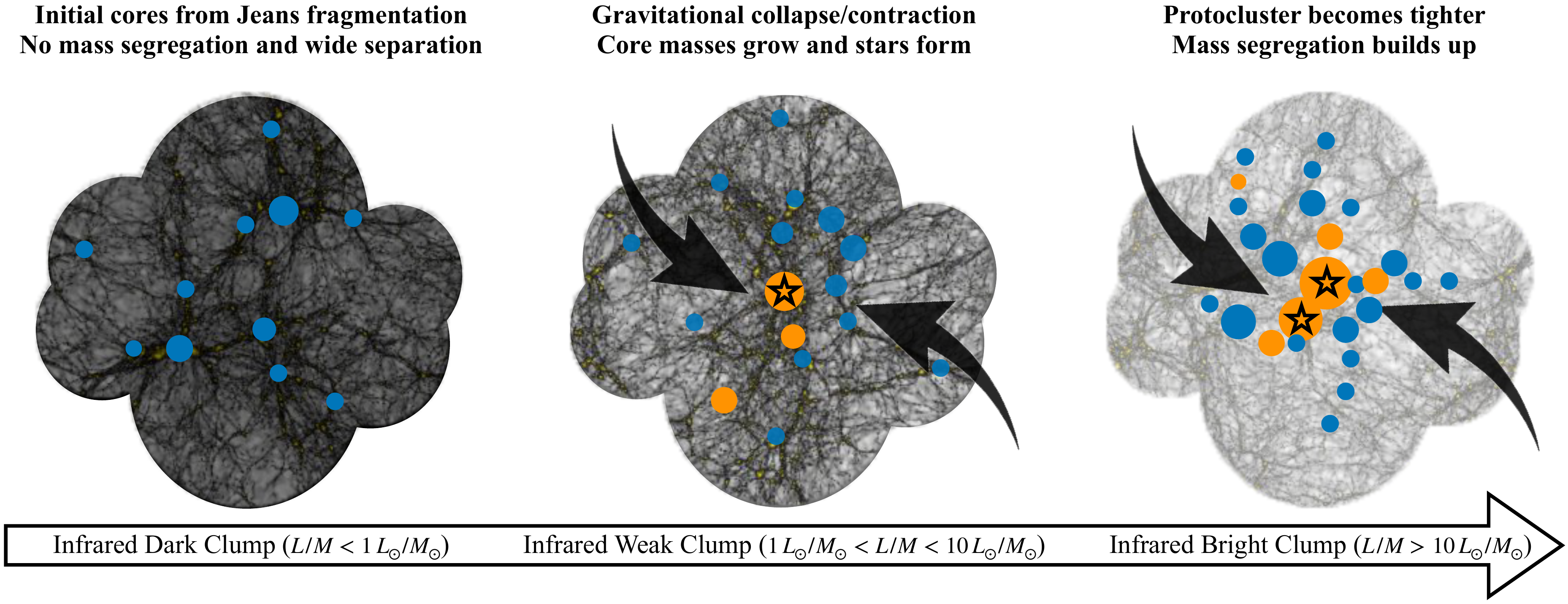}
	\caption{The cartoon of protocluster evolution from Infrared dark, to infrared weak, and to infrared bright. The black filamentary structures connect the dense cores at the early stage \citep{Morii2023ASHES} and transfer gas inwards \citep{Xu2023SDC335}, and then fade away as the protocluster evolves \citep{Zhou2022HFS}. The black arrows indicate inflow gas streams. \includegraphics[scale=0.08,align=c]{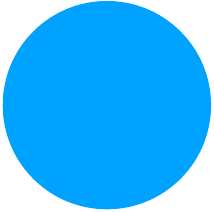} = prestellar cores; \includegraphics[scale=0.08,align=c]{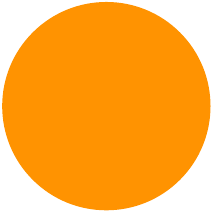} = protostellar cores; \includegraphics[scale=0.12,align=c]{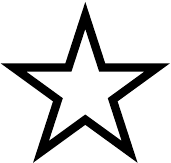} = OB stars. \label{fig:cartoon}}
\end{figure*}

\begin{enumerate}
    \item With a high angular resolution of $\sim$0.8--1.2\arcsec, the 870\,$\mu$m dust continuum emission reveals fragmentation with diverse morphologies. Applying the \getsf~algorithm to the continuum data, we identified a total of 248 cores across the 11 massive protoclusters, with the number of cores per clump ranging from 15 to 37. 
    \item We classified the cores on the basis of molecular outflows and line identification. Of the 248 cores, 142 were classified as prestellar core candidates, while 106 were identified as protostellar cores. To estimate the temperature, we used the rotational temperature derived from the multi-transition lines of \htcs~and \chtocho. If neither of the two lines are detected, we used the clump-averaged temperature for the prestellar core candidates. The properties of \htcs~lines in the ASSEMBLE sample will be discussed in a forthcoming article.
    \item Compared to early-stage ASHES protoclusters, the more evolved ASSEMBLE protoclusters show systematic increases in the average and maximum mass as well as in the surface density of both protostellar and prestellar cores. These increases indicate ongoing mass accretion onto these dense cores, which aligns with the gas accretion process observed in these massive clumps with blue profiles.
    \item The mass of the most massive core (MMC) $M_{\rm max}$ correlates with the mass of the clump $M_{\rm clump}$ as $M_{\rm max}\propto M_{\rm clump}^{0.75}$, with a Spearman correlation coefficient of 0.73. The sublinear correlation indicates a coevolution between clump and MMC potentially by multiscale gas accretion. In contrast, the correlation is not observed in the early-stage ASHES protoclusters, consistent with the idea that early-stage cores are characterized by dominant initial fragmentation rather than clump-scale gravitational accretion.
    \item The correlation between the mass of MMC $M_{\rm max}$ and the mass of protoclusters $M_{\rm cluster}$ is almost linear with a power index of $\sim$0.9 in the first-order approximation. Despite uncertainties, the slope of the $\log M_{\rm max}$ versus $\log M_{\rm cluster}$ relation is steeper compared to that of the $\log M_{\rm \star max}$ versus $\log M_{\rm \star cluster}$ relation found in star clusters, which can be reconciled by an increasing trend of stellar multiplicity with mass.
    \item The most massive prestellar cores found in our study have an average mass of 18.6\,\msun, which is approximately two times larger than that found in the ASHES Pilots. Furthermore, the median and mean masses of the prestellar cores in the protoclusters are $\sim$ 2--3 times higher than in the IRDCs. This suggests that prestellar cores are becoming more massive as a result of the continued mass accumulation within the natal clump and that high-mass prestellar cores can potential survive in protostellar clusters. Thus, we recommend a systematic search for high-mass prestellar cores in massive protoclusters.
    \item Using the Minimum Spanning Tree (MST) algorithm, the cores within each cluster are connected by edges. The core separations in the ASSEMBLE sample are systematically smaller than those in the ASHES sample, indicating that the cluster becomes tighter with closer separations during its evolution. The $\mathcal{Q}$ parameters are observed to be positively correlated with both luminosity to mass ratio $L/M$ and dust temperature $T_{\rm dust}$, indicating a more sub-clustered distribution at an early stage, but a more centrally condensed structure as the cluster evolves. 
    \item According to the mass segregation ratio ($\Lambda$) plots and the mass segregation integral (MSI) that we defined in this paper, mass segregation is commonly found (8 out of 11) and clearly evident in the ASSEMBLE protoclusters. The MSI of the ASHES sample shows an insignificant difference from the random spatial distribution without mass segregation, indicating a weak or no mass segregation in the initial stage. It was further proposed that the mass segregation should arise from gas accretion processes and gravitational concentration, as opposed to arising from dynamical interactions between point masses when the gas has already gone from the systems. 
\end{enumerate}

Leveraging the results and discussions presented above, we are proposing a comprehensive dynamic perspective on protocluster evolution as shown in Figure\,\ref{fig:cartoon}. At the initial stage, the protocluster originates from thermal Jeans fragmentation in infrared dark ($L/M<1 L_\odot/M_\odot$) clumps, with wide separation and no mass segregation. Subsequently, filamentary structures, especially hub-filament system \citep{Morii2023ASHES}, act as ``conveying belts'' and facilitate mass transfer toward the cores, by which the connection between the clump and the cores is gradually established \citep{Xu2023SDC335}. Concurrently, protostars form from dense cores, leading to the heating of gas and dust within the clump, transitioning it into an infrared weak state ($1 L_\odot/M_\odot<L/M<10 L_\odot/M_\odot$). Due to the effects of persistent global gravitational collapse and contraction, the protocluster becomes even tighter with narrower core separations and the mass segregation builds up in the late stage ($L/M>10 L_\odot/M_\odot$).

The ASSEMBLE project not only provides valuable insights into the mass segregation and clustering properties of massive protoclusters but also can be used to investigate outflows \citep{Baug2021Outflow}, chemistry, and core-scale infall motion. When combined with Band-3/6 data from the ATOMS project \citep[PI: Tie Liu, see the survey description in][]{Liu2020ATOMS-I}, the ASSEMBLE project's data can facilitate more kinematic analyses, further illuminating how gas is transferred inward and how efficient accretion is at the clump scale and in a dynamic view. In this paper, our analyses and their statistical significance are mainly limited by sample size. As the ASSEMBLE project aims to expand its sample to include a wider range of parameters such as evolutionary stage ($L/M$) and clump mass ($M_{\rm clump}$), even more statistically significant results are expected. 


\section*{Acknowledgment}

We thank the anonymous referee for the constructive comments. 
This work has been supported by the National Science Foundation of China (11973013, 12033005, 11721303, 12073061, 12122307, 12203011, 12103045) and the National Key R\&D Program of China (No. 2022YFA1603100, 2019YFA0405100).
FWX and KW acknowledge support from the China Manned Space Project (CMS-CSST-2021-A09, CMS-CSST-2021-B06), and the High-Performance Computing Platform of Peking University. 
TL acknowledges the support by the international partnership program of Chinese academy of sciences through grant No.114231KYSB20200009, and Shanghai Pujiang Program 20PJ1415500. 
MYT acknowledges the support by Yunnan provincial Department of Science and Technology through grant No.202101BA070001-261. 
K.M is financially supported by Grants-in-Aid for the Japan Society for the Promotion of Science (JSPS) Fellows (KAKENHI Number 22J21529). 
PS was partially supported by a Grant-in-Aid for Scientific Research (KAKENHI Number JP22H01271 and JP23H01221) of JSPS.
HLL is supported by Yunnan Fundamental Research Project (grant No.\,202301AT070118).
AP, GCG and EVS acknowledge financial support from the UNAM-PAPIIT IG100223 grant. AP acknowledges financial support from the Sistema Nacional de Investigadores of CONAHCyT, and from the CONAHCyT project number 86372 of the `Ciencia de Frontera 2019' program, entitled `Citlalc\'oatl: A multiscale study at the new frontier of the formation and early evolution of stars and planetary systems', M\'exico. GCG acknowledges support from UNAM-PAPIIT IN10382 grant.
A.S., G.G., and L.B.\ gratefully acknowledge support of ANID through the BASAL project FB210003. A.S.\ also gratefully acknowledges support from the Fondecyt Regular (project code 1220610). This work is sponsored (in part) by the Chinese Academy of Sciences (CAS), through a grant to the CAS South America Center for Astronomy (CASSACA) in Santiago, Chile. 
C.W.L. is supported by the Basic Science Research Program through the National Research Foundation of Korea (NRF) funded by the Ministry of Education, Science and Technology (NRF- 2019R1A2C1010851), and by the Korea Astronomy and Space Science Institute grant funded by the Korea government (MSIT; Project No. 2023-1-84000).
K.T. was supported by JSPS KAKENHI (Grant Number JP20H05645).
This paper uses the following ALMA data: ADS/JAO.ALMA\#2017.1.00545.S and 2019.1.00685.S, and ADS/JAO.ALMA\#2015.1.01539.S, 2017.1.00716.S, and 2018.1.00192.S. ALMA is a partnership of ESO (representing its member states), NSF (USA) and NINS (Japan), together with NRC (Canada), MOST and ASIAA (Taiwan), and KASI (Republic of Korea), in cooperation with the Republic of Chile. The Joint ALMA Observatory is operated by ESO, AUI/NRAO, and NAOJ. Data analysis was in part carried out on the High-Performance Computing Platform of Peking University, and the open-use data analysis computer system at the Astronomy Data Center (ADC) of the National Astronomical Observatory of Japan.
The MeerKAT telescope is operated by the South African Radio Astronomy Observatory, which is a facility of the National Research Foundation, an agency of the Department of Science and Innovation.

%

\vspace{5mm}
\facility{ALMA}


\software{Astropy \citep{Astropy2013,Astropy2018}, CASA \citep{McMullin2007CASA}, getsf \citep{Men2021getsf}
}



\clearpage

\appendix

\section{Radio Counterpart and Environment} \label{app:radio}

\begin{figure*}[!ht]
\centering
\includegraphics[width=0.32\linewidth]{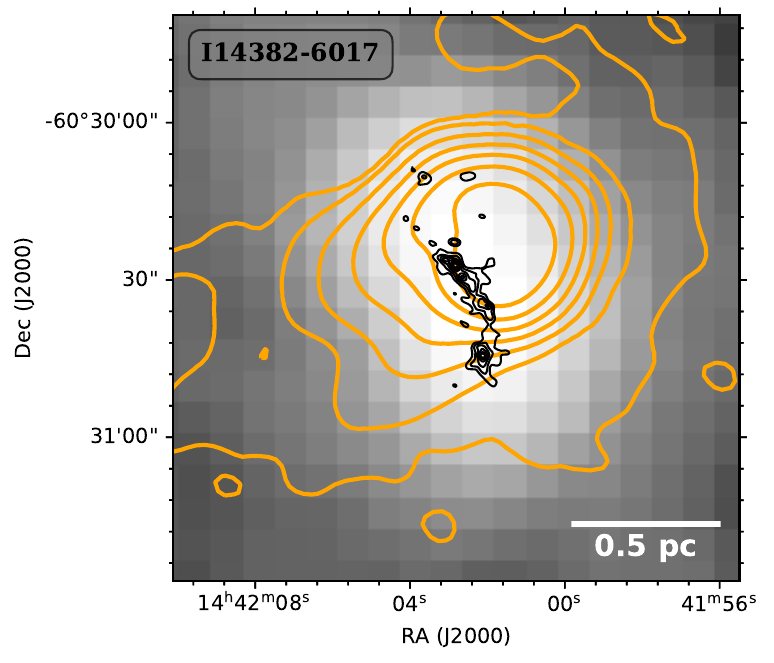}
\includegraphics[width=0.32\linewidth]{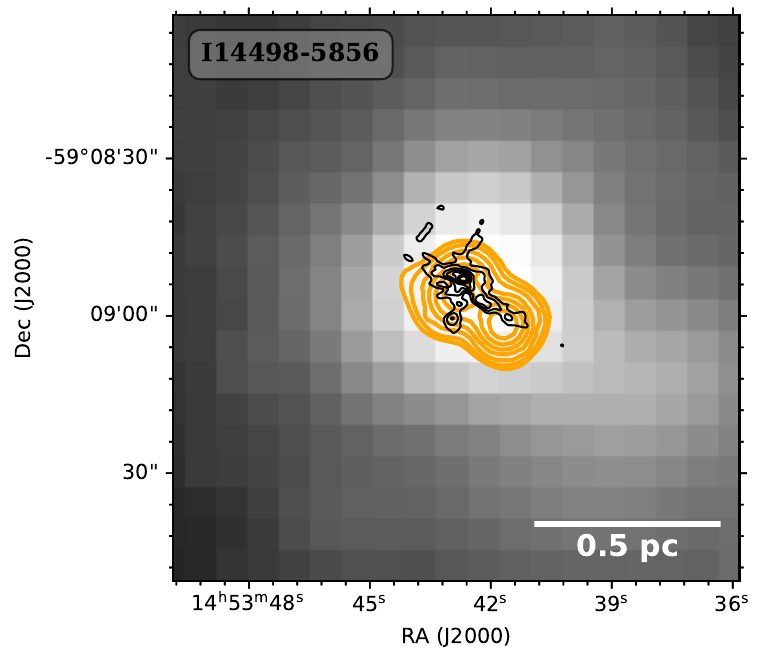}
\includegraphics[width=0.32\linewidth]{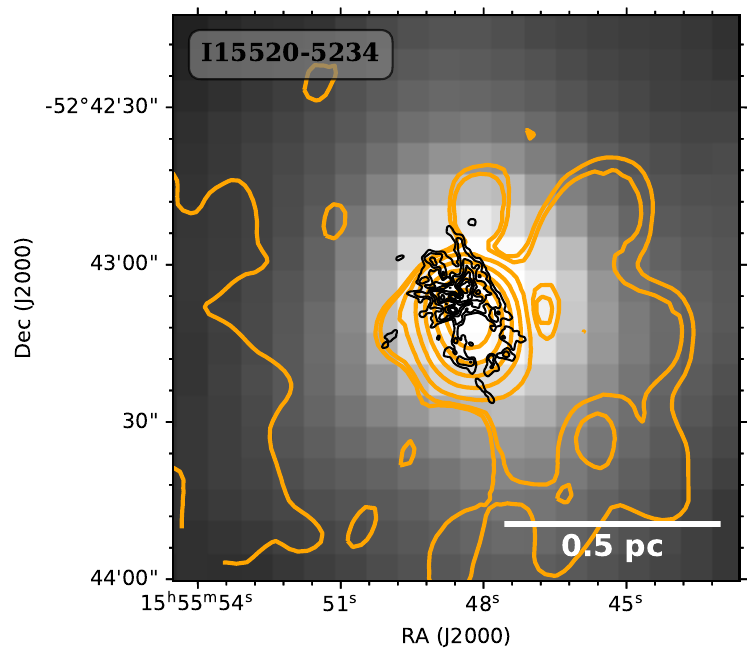}
\includegraphics[width=0.32\linewidth]{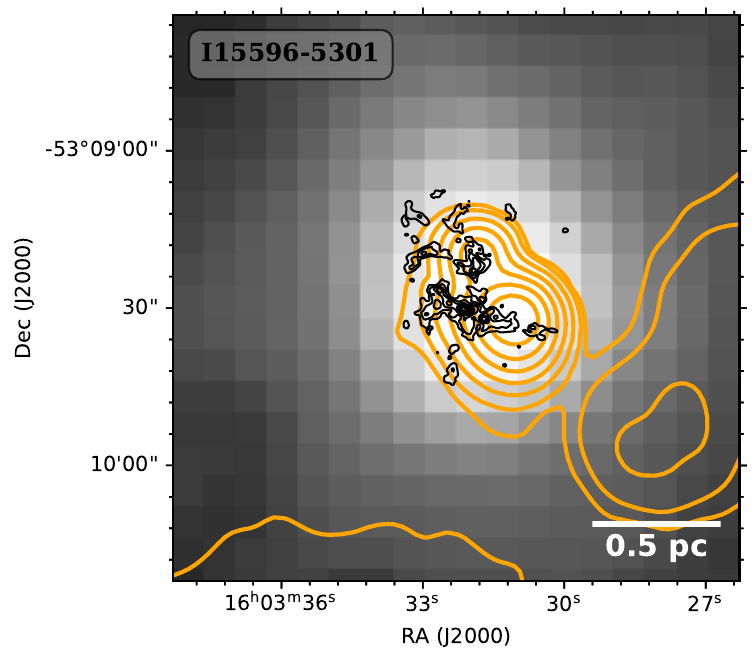}
\includegraphics[width=0.32\linewidth]{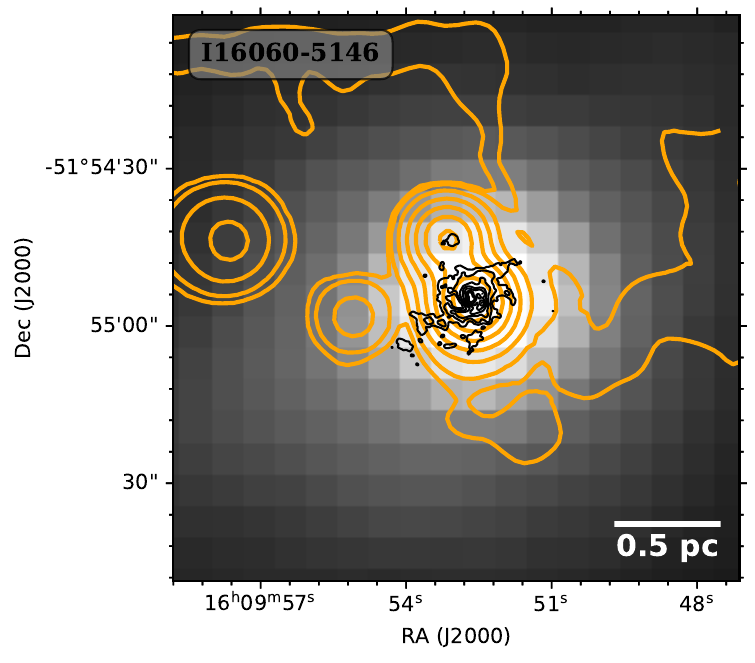}
\includegraphics[width=0.32\linewidth]{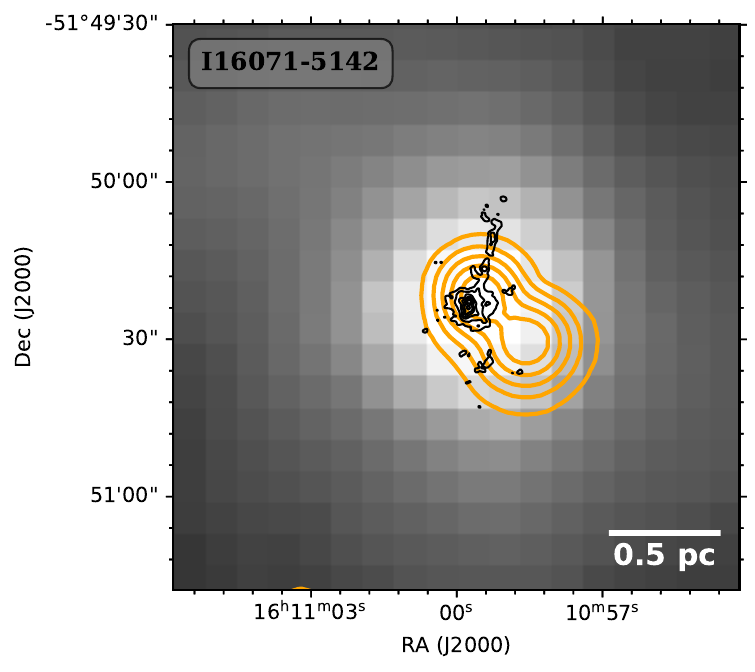}
\includegraphics[width=0.32\linewidth]{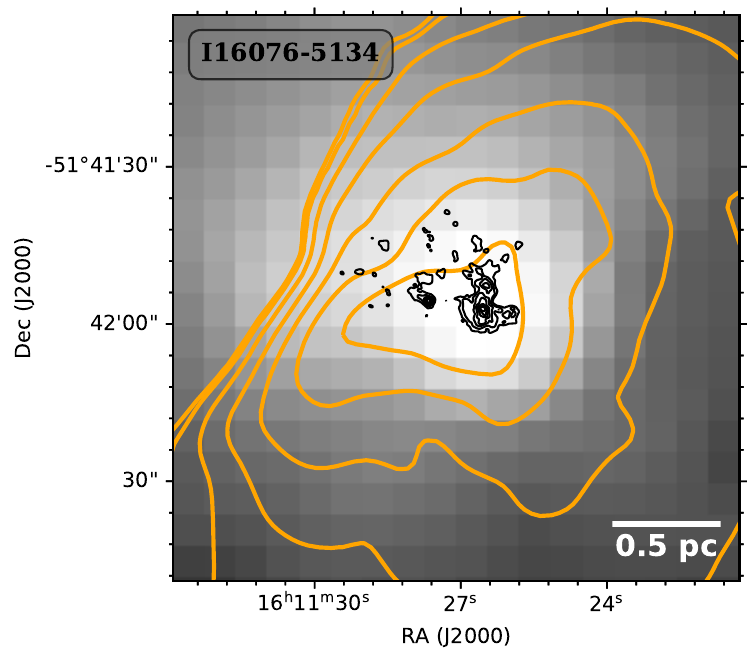}
\includegraphics[width=0.32\linewidth]{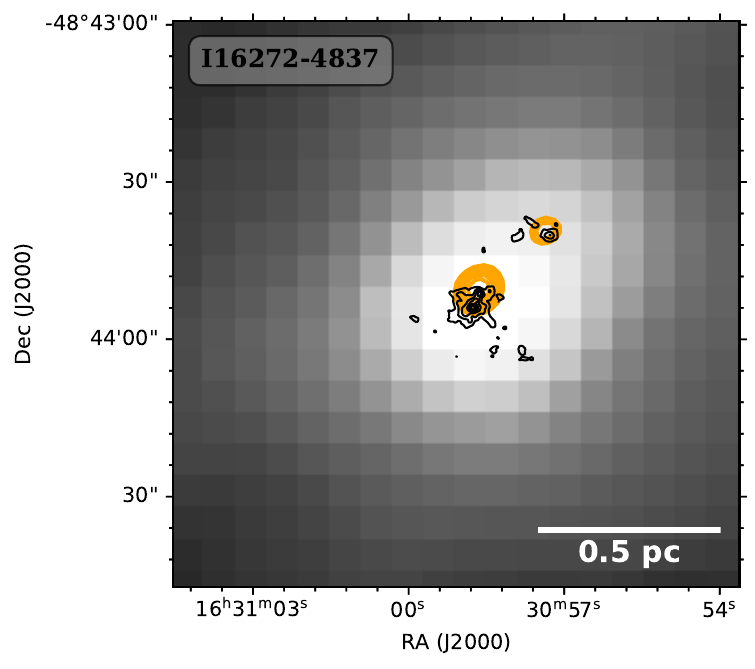}
\includegraphics[width=0.32\linewidth]{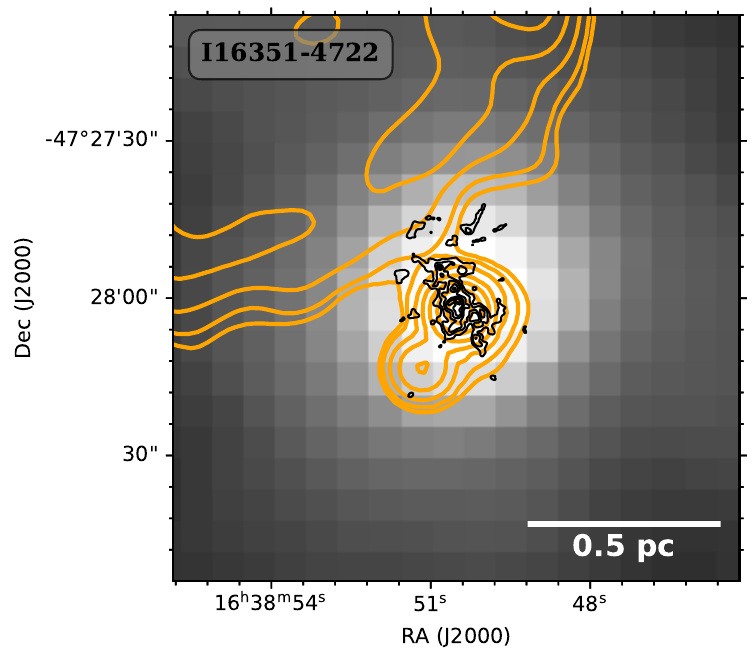}
\includegraphics[width=0.32\linewidth]{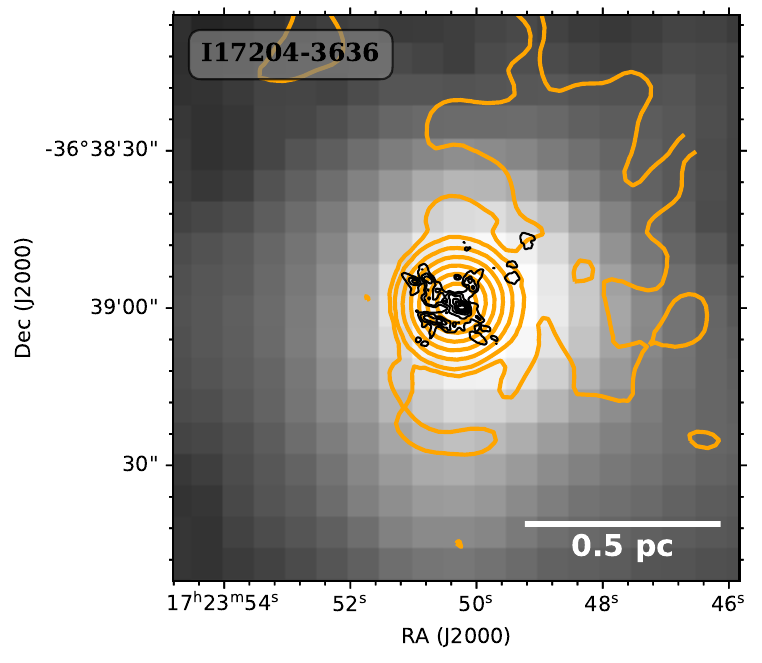}
\includegraphics[width=0.32\linewidth]{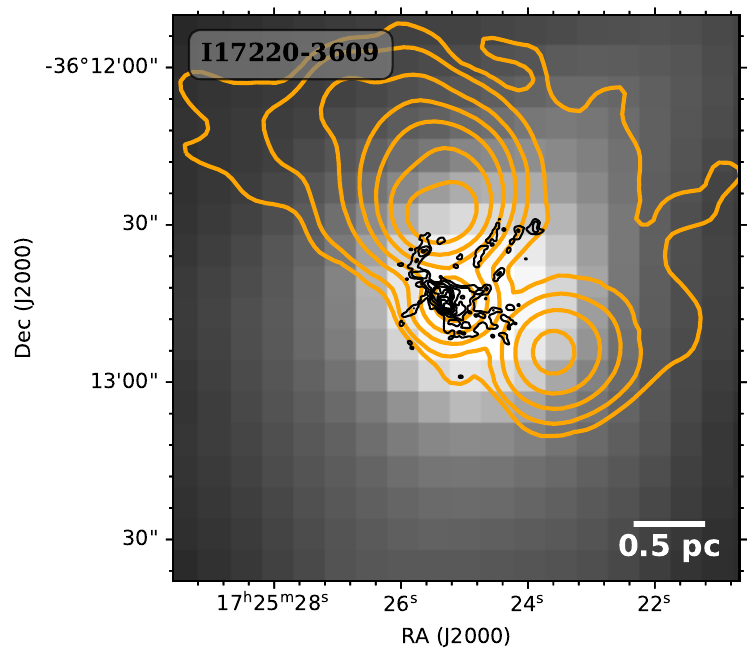}
\caption{The 1.28\,GHz radio emissions of the ASSEMBLE protoclusters are shown with yellow contours, with logarithmically spaced levels starting from $5\sigma$ to the peak flux. The background gray color maps show the ATLASGAL 870\,$\mu$m continuum emission. The overlaid black contours show the ALMA 870\,$\mu$m continuum emission as the right panels in Figure\,\ref{fig:continuum}. \label{fig:radio}}
\end{figure*}

The radio counterpart and the environment atlas of the ASSEMBLE protoclusters are shown in Figure\,\ref{fig:radio}. The 1.28\,GHz MeerKAT images \citep[Goedhart et al. in prep.]{Padmanabh2023MeerKAT} of the ASSEMBLE protoclusters are shown with yellow contours, with logarithmically spaced levels starting from $5\sigma$ to the peak flux. The background gray color maps show the ATLASGAL 870\,$\mu$m continuum emission. The overlaid black contours show the ALMA 870\,$\mu$m continuum emission as the right panels in Figure\,\ref{fig:continuum}.

\section{Minimum Spanning Tree Methods} \label{app:mst}

\begin{figure*}[!ht]
\centering
\includegraphics[width=0.32\linewidth]{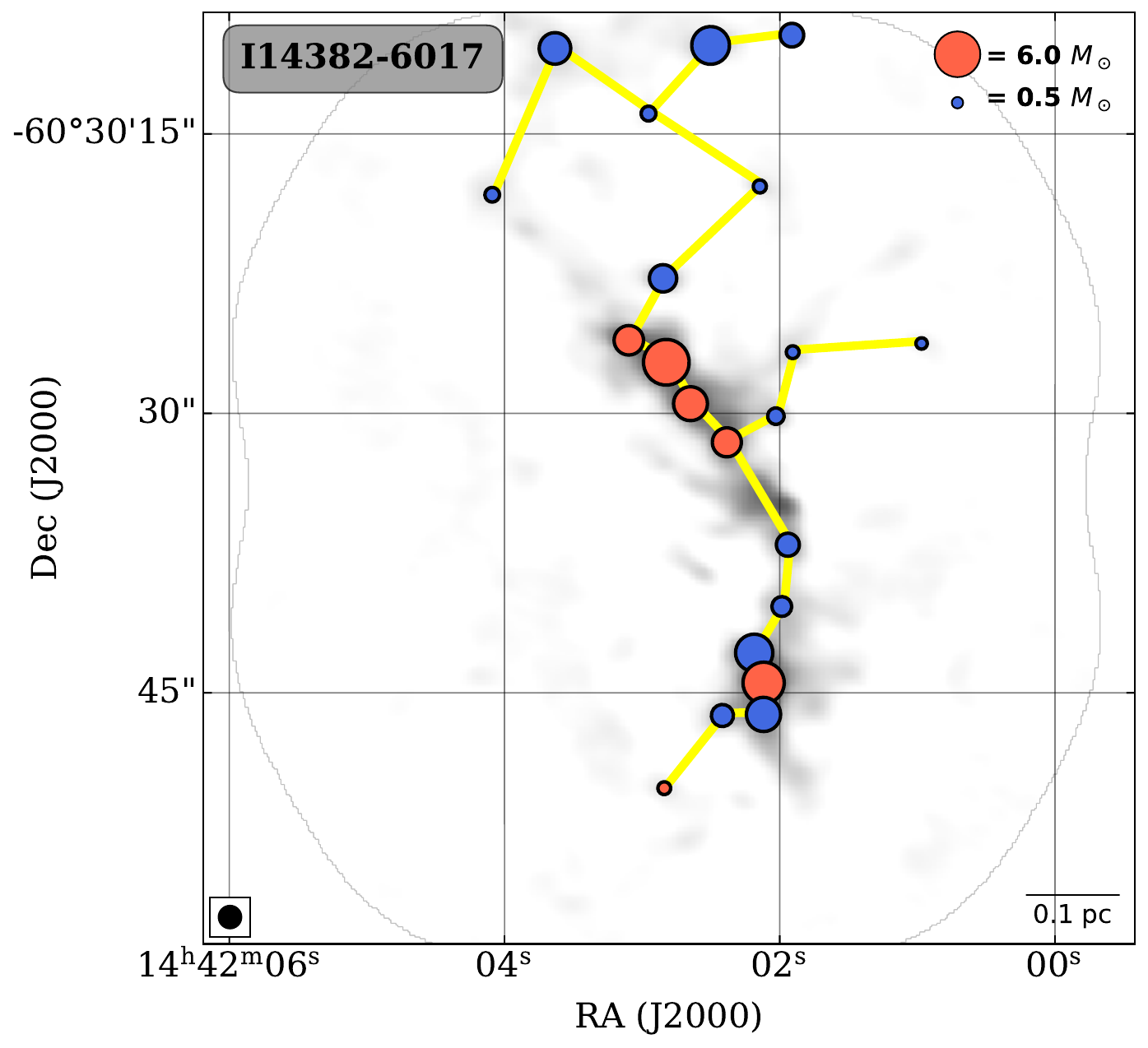}
\includegraphics[width=0.32\linewidth]{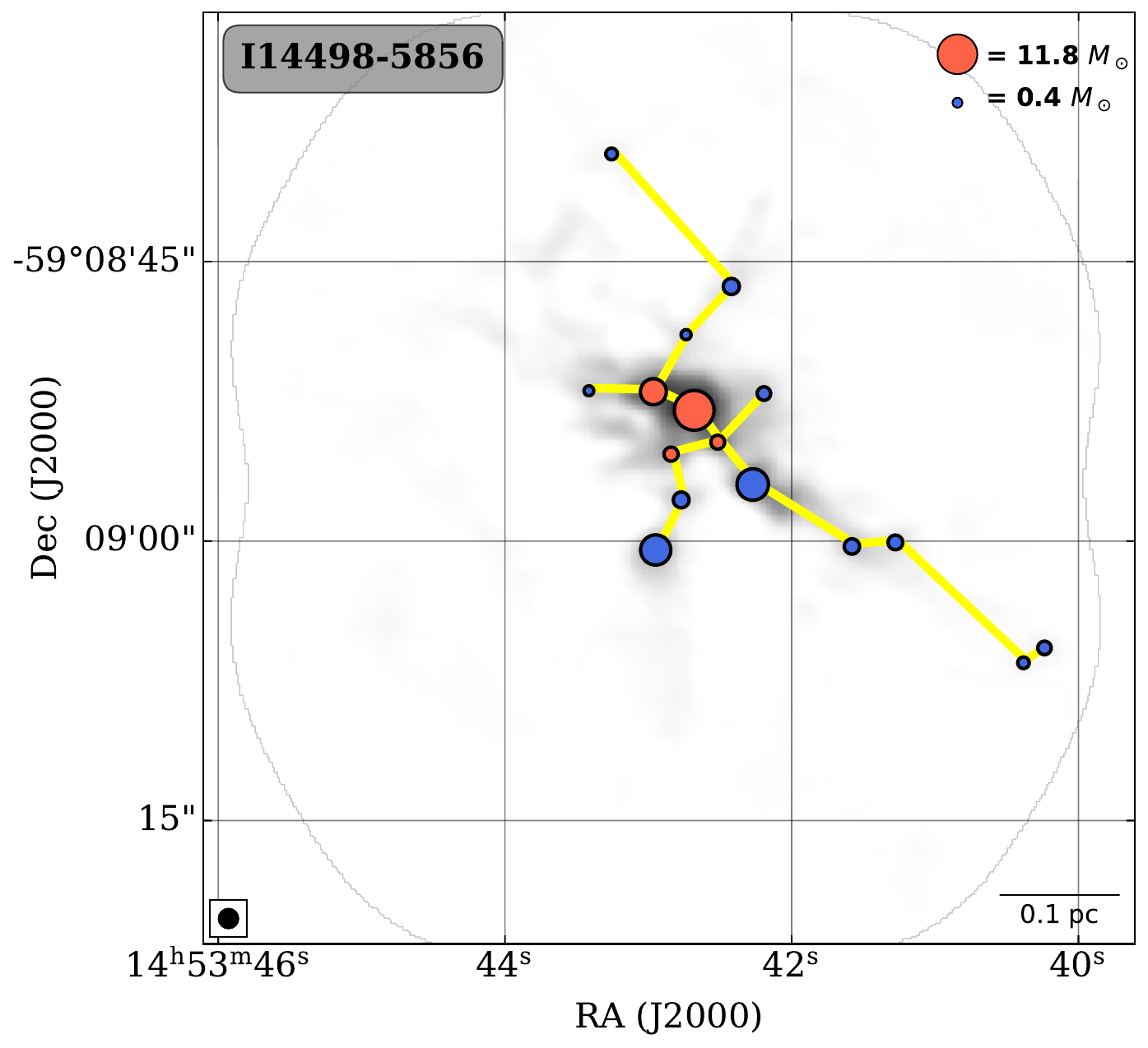}
\includegraphics[width=0.32\linewidth]{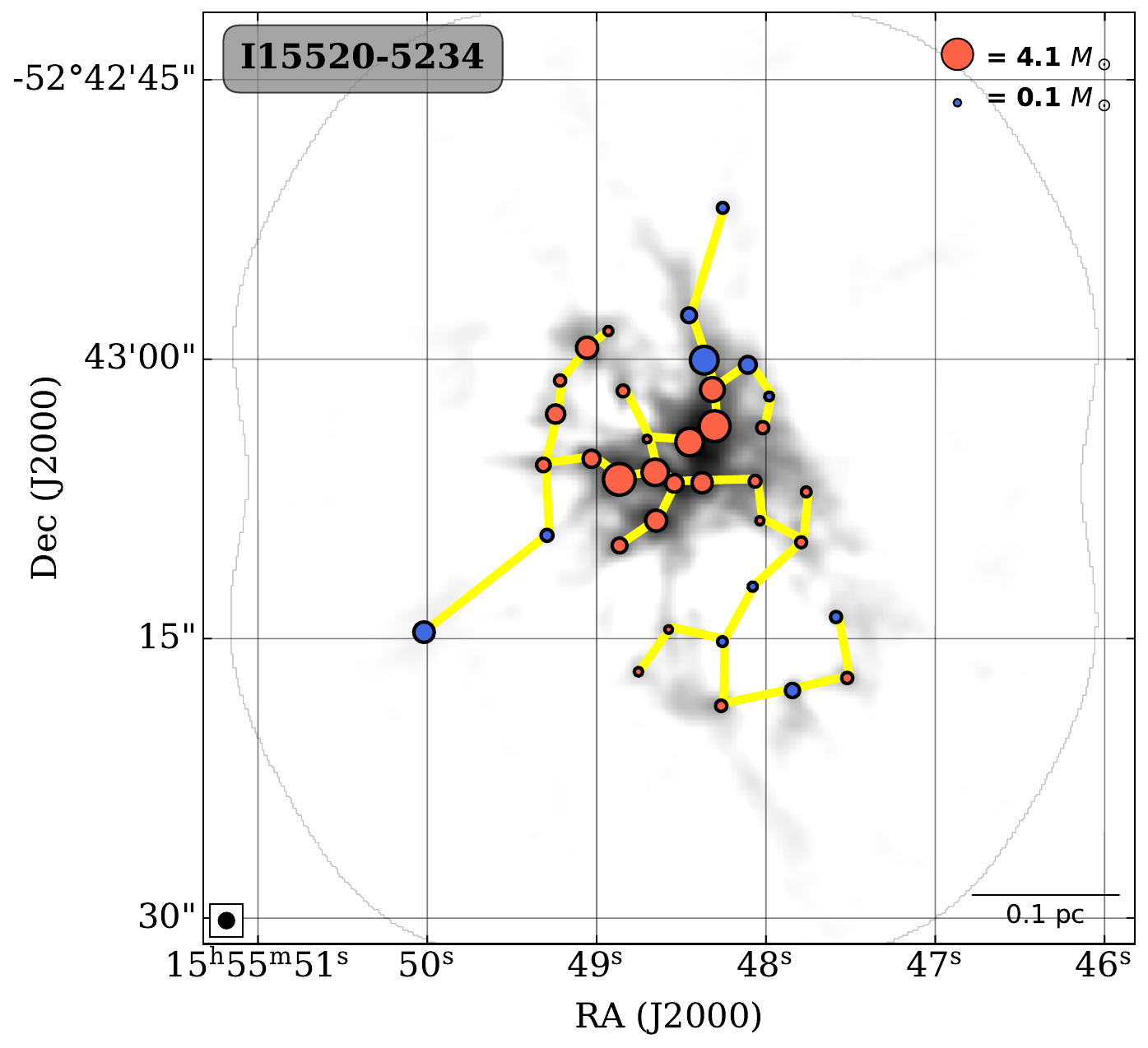}
\includegraphics[width=0.32\linewidth]{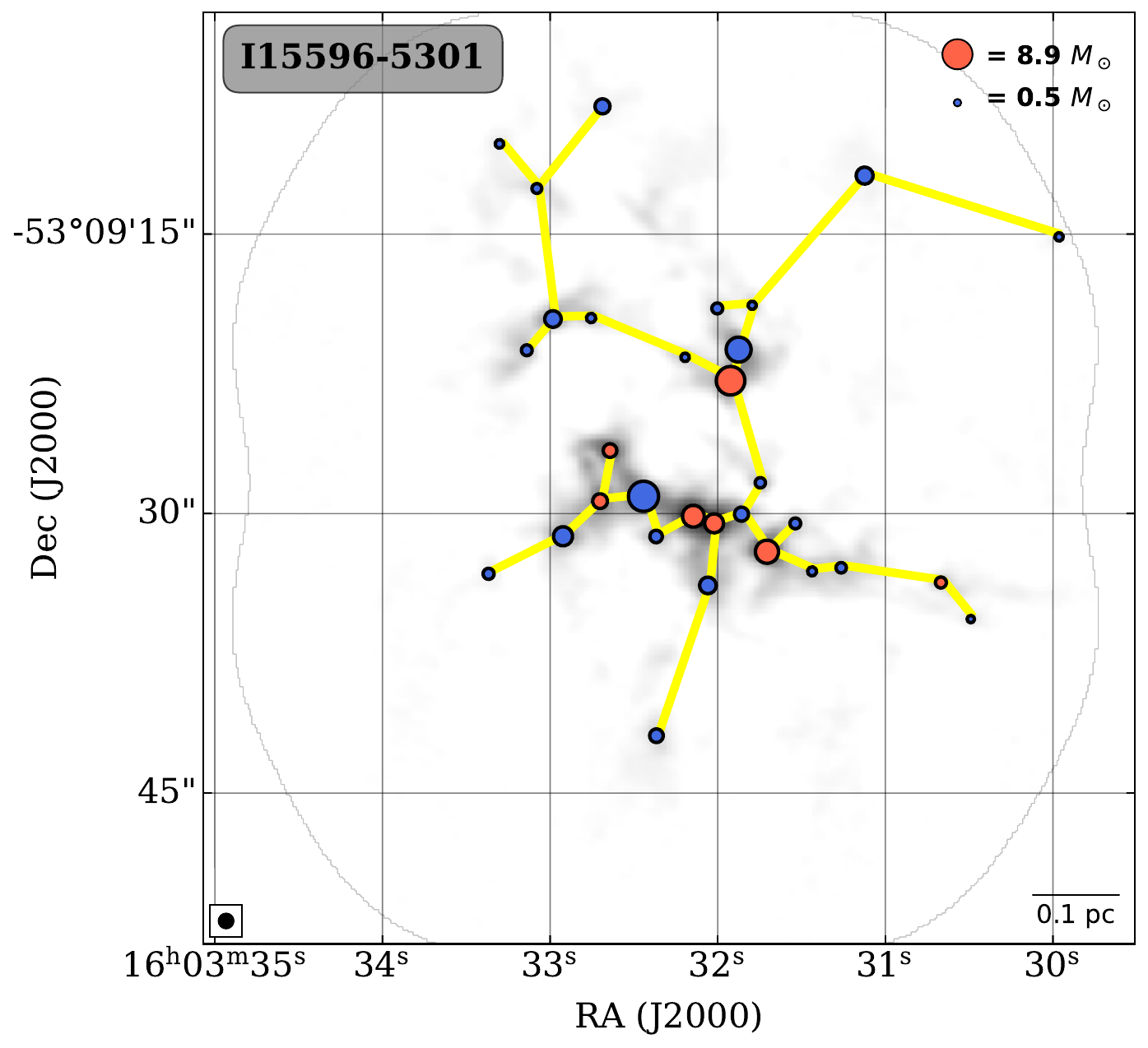}
\includegraphics[width=0.32\linewidth]{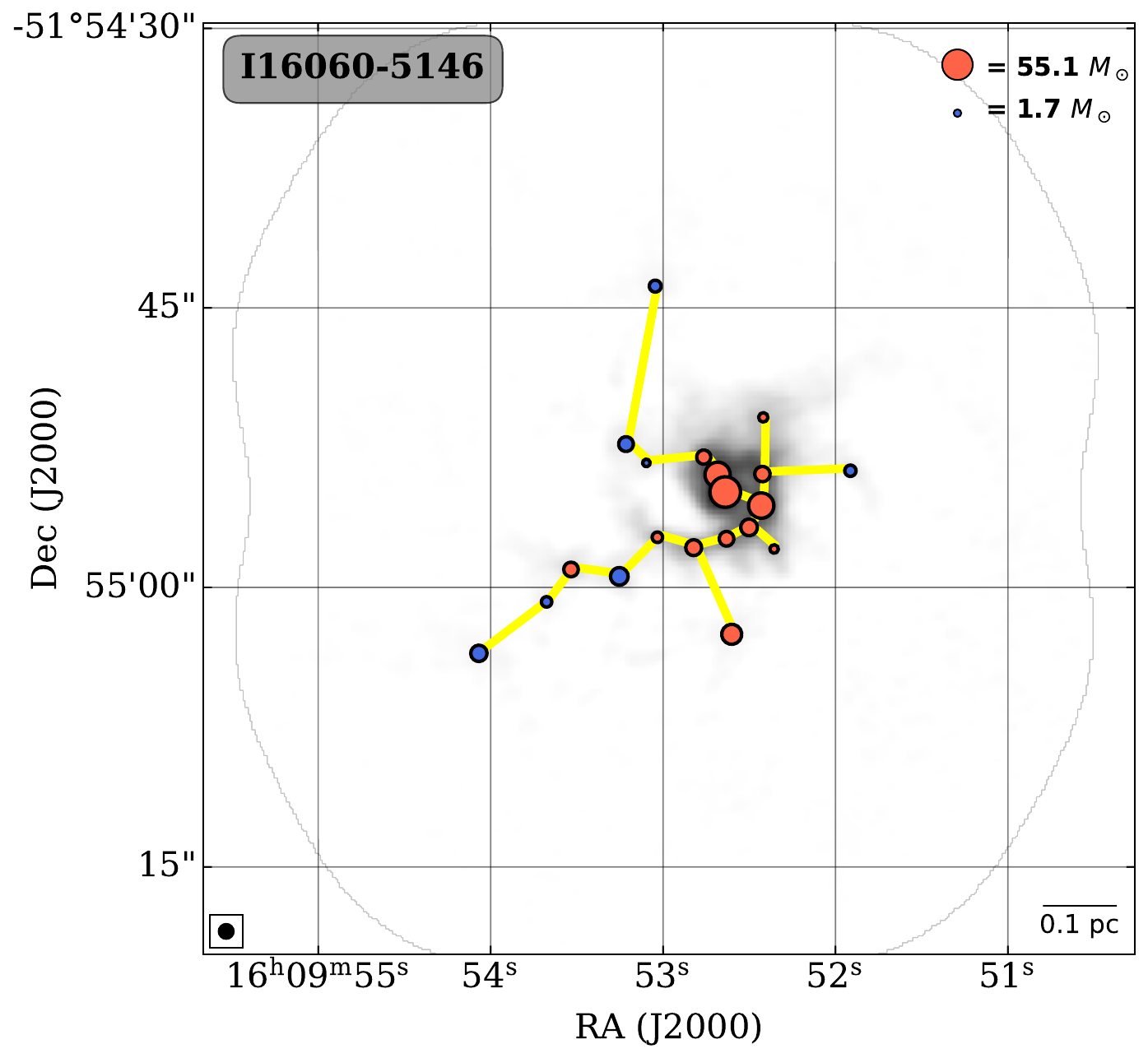}
\includegraphics[width=0.32\linewidth]{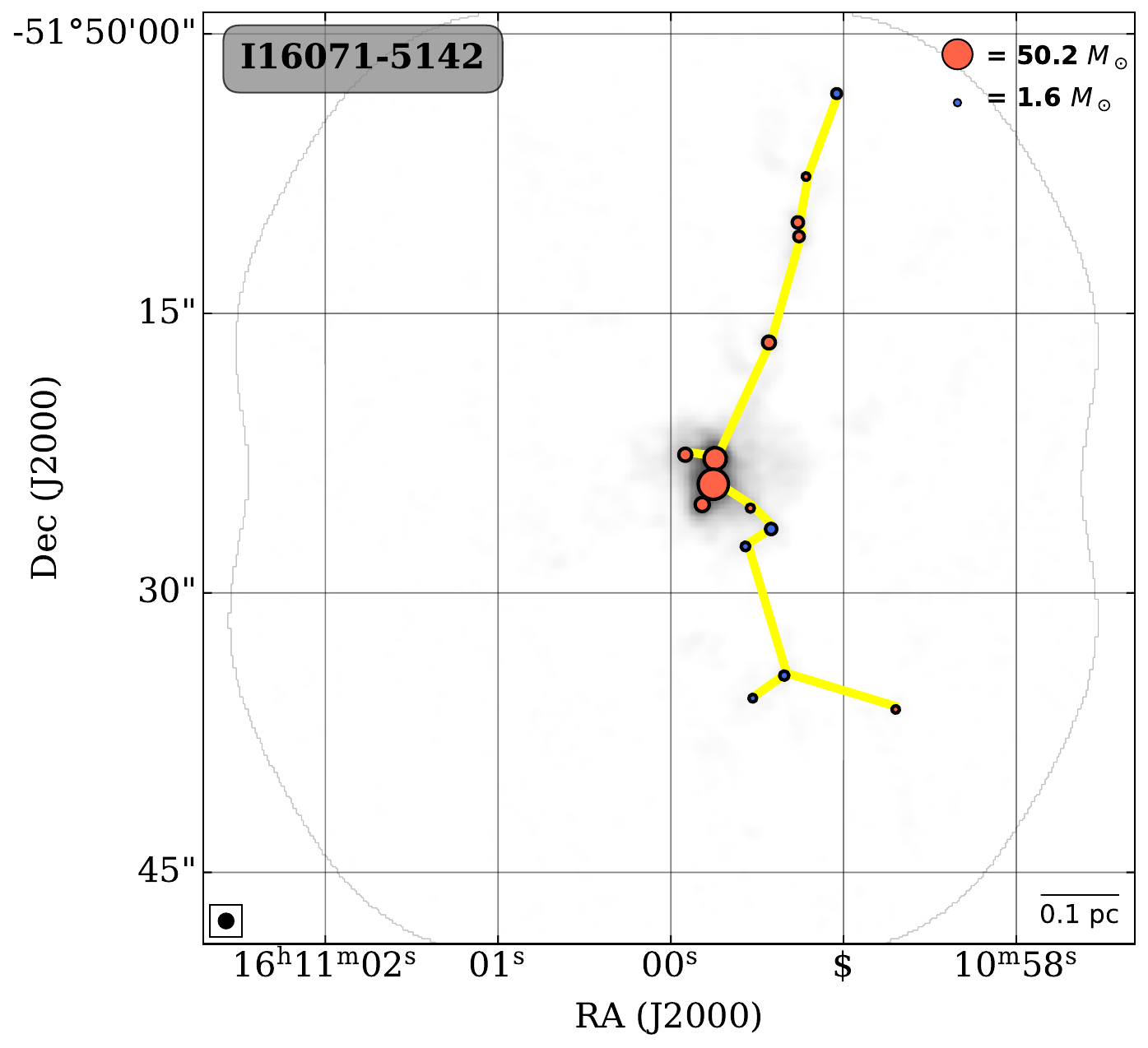}
\includegraphics[width=0.32\linewidth]{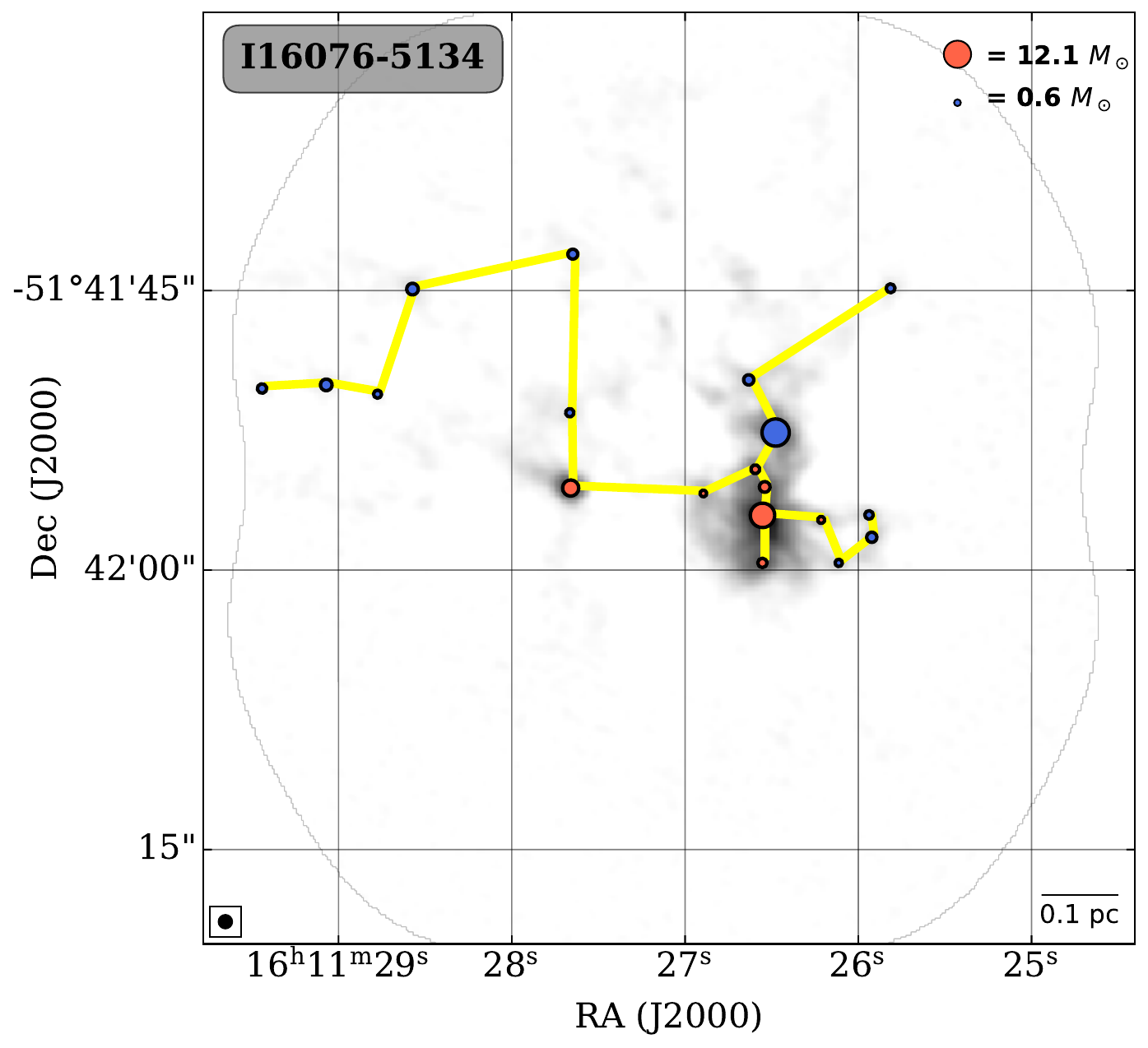}
\includegraphics[width=0.32\linewidth]{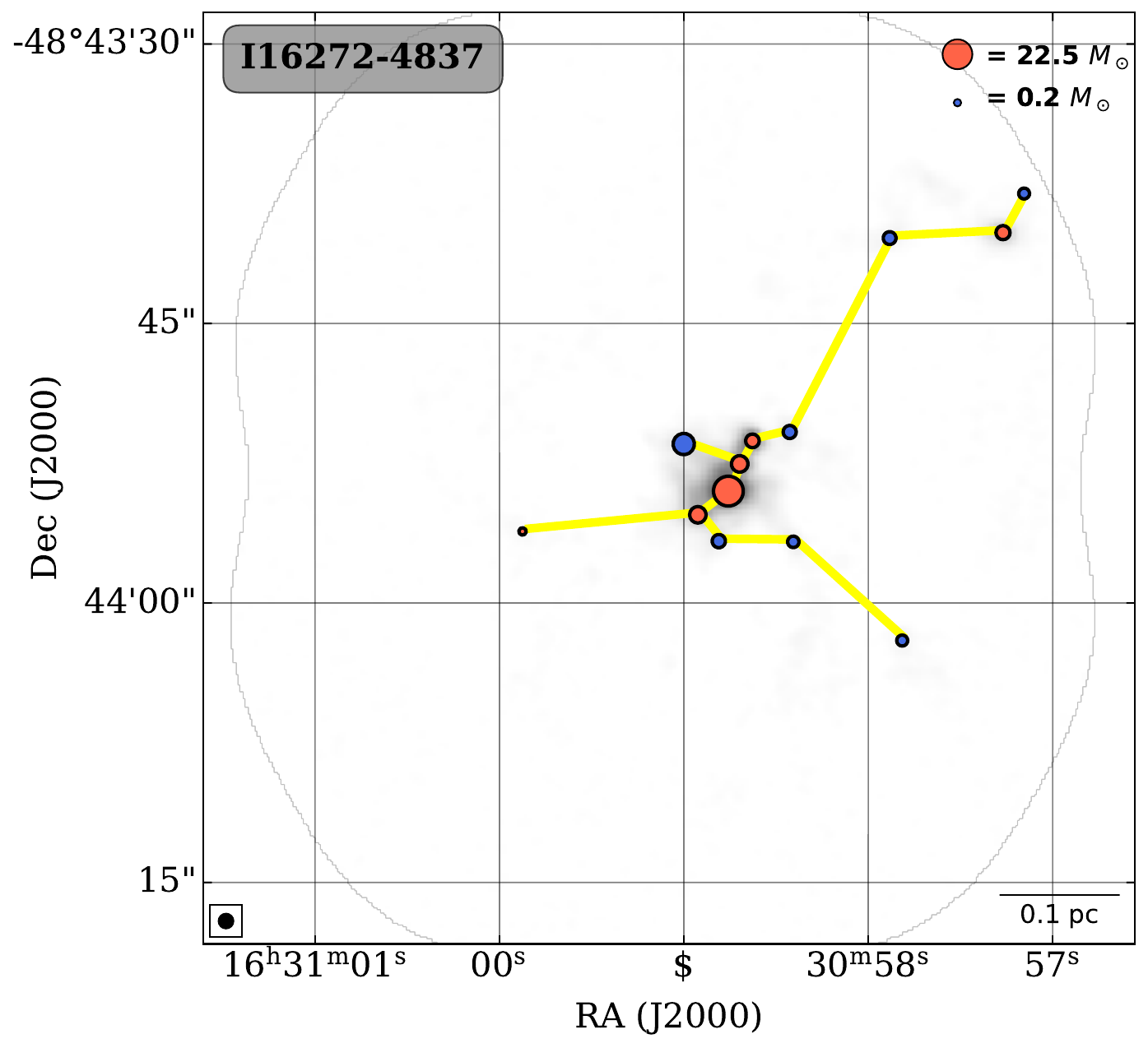}
\includegraphics[width=0.32\linewidth]{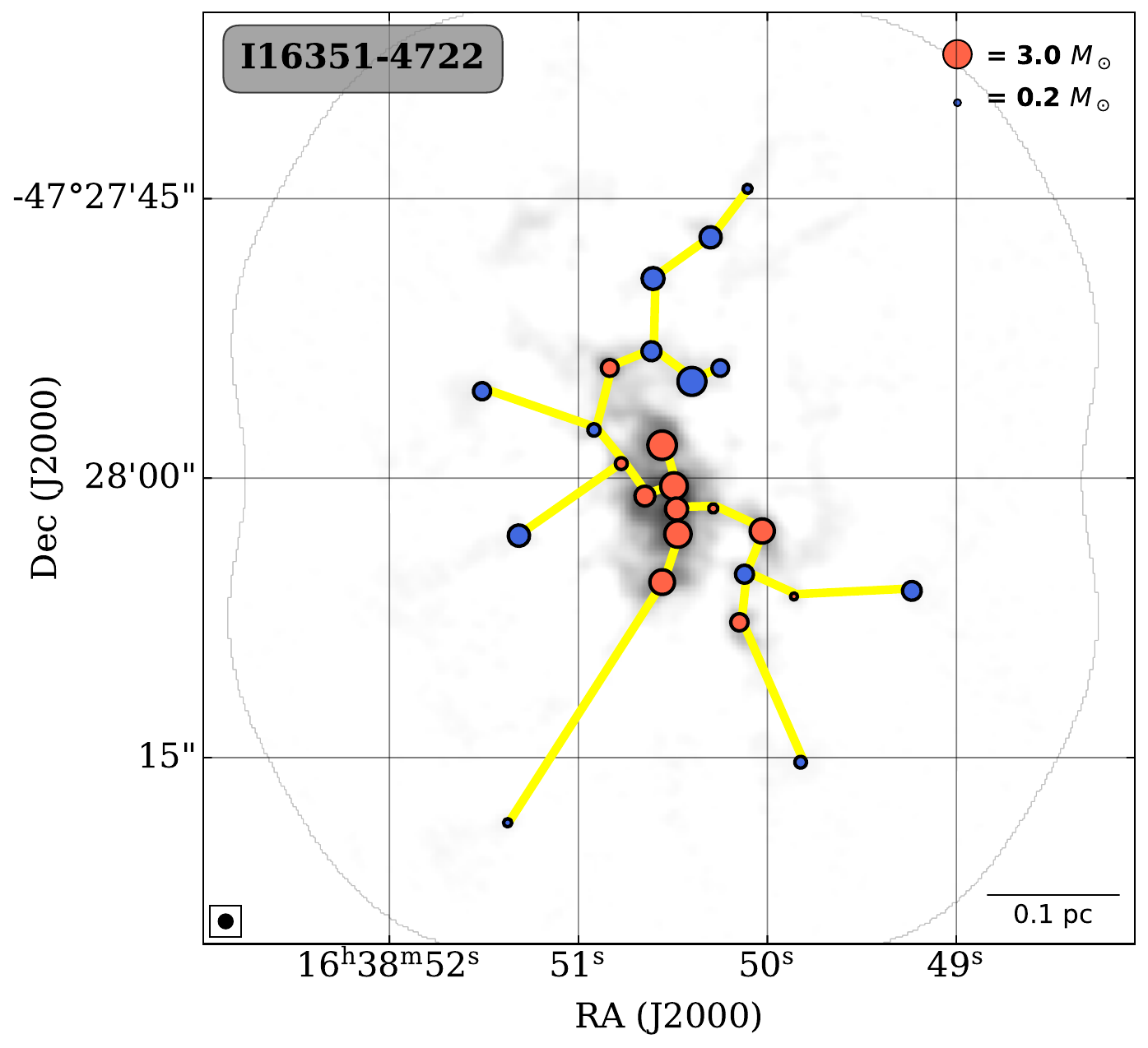}
\includegraphics[width=0.32\linewidth]{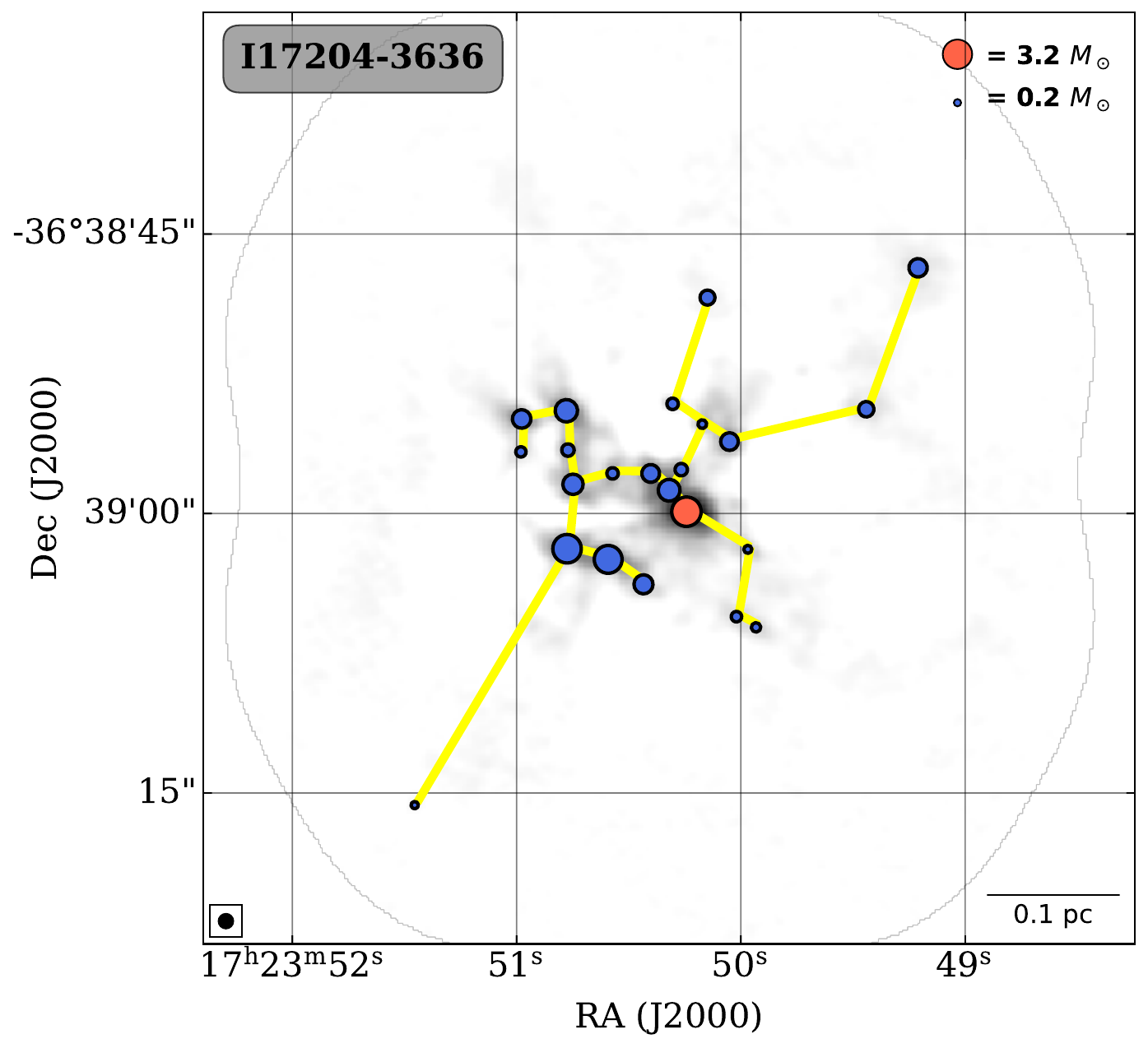}
\includegraphics[width=0.32\linewidth]{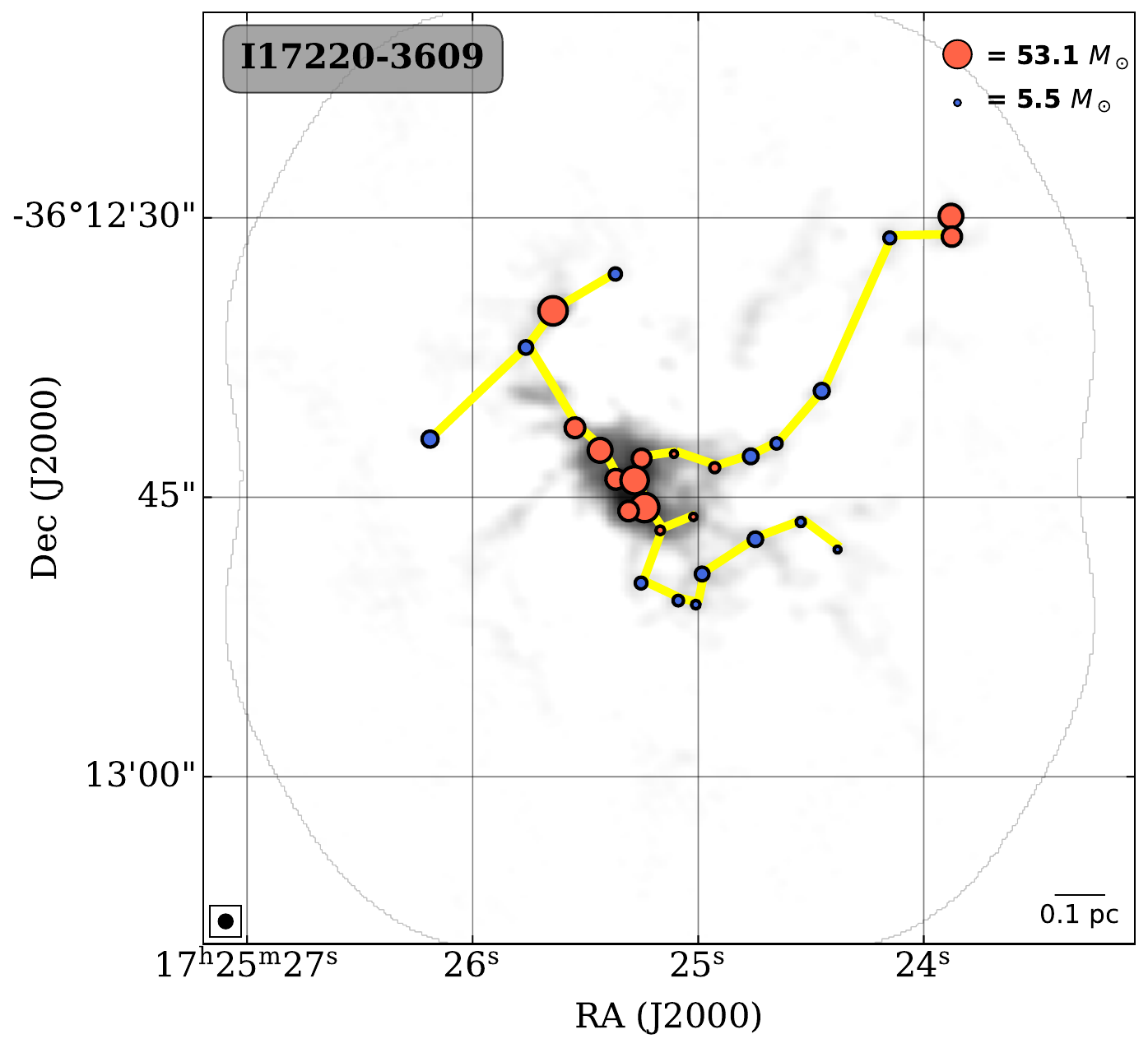}
\caption{MST results of ASSEMBLE protoclusters. Pre/proto-stellar cores are assgined with blue/red colors, with the size normalized by the square root of core mass $\sqrt{M_{\rm core}}$ and stretched within the mass range. Yellow segments connect the cores to minimize the sum of the lengths of segments. The beam size is shown on the lower left and the scale bar is on the right. \label{fig:assemble_mst}}
\end{figure*}

Minimum spanning tree (MST), first developed for astrophysical applications by \citet{Barrow1985MST}, has been applied to simulations \citep[e.g.,][]{Wu2017MST} and to observations \citep[][]{Wang2016Filament,Toth2017MST,Wang2021Software,Ge2022Filament}. In this paper, we use Prim's algorithm to find out the edges to form the tree including every node with the minimum sum of weights to form the MST. The Prim's algorithm starts with the single source node and later explores all the nodes adjacent to the source node with all the connecting edges. During the exploration, we choose the edges with the minimum weight and those which cannot cause a cycle. The edge weight is set to be the length between two vertices \citep{Prim1957Algorithm}. Therefore, MST determines a set of straight lines connecting a set of nodes (cores) that minimizes the sum of the lengths. Figures\,\ref{fig:assemble_mst} displays the MST results of the 11 ASSEMBLE protoclusters.

We also adopt the MST algorithm for the 12 core clusters in the ASHES pilot survey. To keep consistency when comparing with the ASSEMBLE results, the core catalog is updated using the same source extraction algorithm on the mock 0.87\,mm continuum data with the same array configuration and mosaicked coverage (see Appendix\,\ref{app:mock}). Compared to the original work by \citet{Sanhueza2019ASHES} who used the \texttt{astrodendro} for the 12m+ACA+TP combined data, our new core catalog is focused on the dense and concentrated structures.

\section{Mock Band-7 continuum images for ASHES clumps} \label{app:mock}

To compare the ASSEMBLE and ASHES results more directly, we simulate the Band-7 (0.87\,mm) observations following the ASSEMBLE project with the ASHES Band-6 (1.3\,mm) continuum data as input, to derive the mock continuum data. As the ASHES clumps are in their early stages, the free-free contamination especially at high-frequency bands (0.87 and 1.3\,mm) is negligible. In other words, the continuum emission mainly comes from dust gray-body emission. Therefore, the fluxes of 1.3\,mm and 0.87\,mm should follow the scaling relation as,

\begin{equation}
	\Gamma(T_{\rm dust}) \equiv \frac{F_{\rm 1.3\,mm}}{F_{\rm 0.87\,mm}} = \frac{\kappa_{\rm 1.3\,mm}}{\kappa_{\rm 0.87\,mm}} \frac{B_{\rm 1.3\,mm}(T_{\rm dust})}{B_{\rm 0.87\,mm}(T_{\rm dust})},
\end{equation}
where $\kappa_{\nu}$ is the opacity at certain frequency, $\kappa_{\rm 1.3\,mm}$ = 0.9\,cm$^2$\,g$^{-1}$ \citep{Sanhueza2019ASHES} and $\kappa_{\rm 0.87\,mm}=1.89$\,cm$^2$\,g$^{-1}$ \citep{Xu2023SDC335}. $B_{\nu}(T_{\rm dust})$ is the Planck function at corresponding frequency $\nu$ and dust temperature $T_{\rm dust}$. 

As the ASHES clumps are in their early stages and lack prominent central heating sources that induce temperature gradients, we assume that dust the temperature of the entire clump is uniform, i.e., the clump-averaged $T_{\rm dust}$ as listed in column (7) of \citet{Sanhueza2019ASHES}. Therefore, we simply adopt the same $T_{\rm dust}$ and then the same $\Gamma(T_{\rm dust})$ in one field to convert the flux density from the 1.3\,mm to 0.87\,mm continuum data. 

Afterwards, the flux-converted images are then cropped into fields with the same shape and size as ASSEMBLE data, to assure the same field of view where sources are extracted. The mock field is placed with both the major and minor axes aligned with those of the ASHES field with its center towards the densest part of the cluster. The 12 mock images are used as the basic input for source extraction algorithm \texttt{getsf} and the MST algorithm (Appendix\,\ref{app:mst}).


\clearpage
\bibliography{assemble}{}
\bibliographystyle{aasjournal}







\end{document}